\documentclass[a4paper,11pt,twoside]{book}

\usepackage{graphicx} 
\usepackage[english]{babel}
\usepackage{amsmath,amssymb}
\usepackage{natbib}
\usepackage[utf8x]{inputenc}
\usepackage{times}
\usepackage[dvips]{epsfig}
\usepackage[dvips]{graphicx}
\usepackage{textcomp}
\usepackage{rotating}
\usepackage{longtable}
\usepackage{pdfpages}
\usepackage{import}
\usepackage{appendix}
\usepackage{verbatim}
\usepackage{ulem}
\usepackage{lscape}
\usepackage{multirow}
\usepackage{hhline}
\usepackage[T1]{fontenc}
\usepackage{subfigure}
\usepackage{deluxetable} 
\usepackage{amsfonts}

\def\ltsim{\raise 2pt \hbox {$<$} \kern-1.1em \lower 4pt \hbox {$\sim$}}
\def\ltapprox{\raise 2pt \hbox {$<$} \kern-1.1em \lower 5pt \hbox {$\approx$}}
\def\gtsim{\raise 2pt \hbox {$>$} \kern-1.1em \lower 4pt \hbox {$\sim$}}
\def\gtapprox{\raise 2pt \hbox {$>$} \kern-1.1em \lower 5pt \hbox {$\approx$}}

\def\lapp{\ifmmode\stackrel{<}{_{\sim}}\else$\stackrel{<}{_{\sim}}$\fi}
\def\gapp{\ifmmode\stackrel{>}{_{\sim}}\else$\stackrel{<}{_{\sim}}$\fi}

\def\eps@scaling{.95}
\def\epsscale#1{\gdef\eps@scaling{#1}}
\def\plotone#1{\centering \leavevmode
    \epsfxsize=\eps@scaling\columnwidth \epsfbox{#1}}

\def\ltsima{$\; \buildrel < \over \sim \;$}
\def\lsim{\lower.5ex\hbox{\ltsima}}

\usepackage[top=4cm, bottom=3cm, left=3.5cm, right=3.5cm]{geometry}

\usepackage{fancyhdr}
\newcommand\arcsec{\mbox{$^{\prime\prime}$}}%
\newcommand\arcmin{\mbox{$^\prime$}}%

\def\aj{AJ}%
\def\actaa{Acta Astron.}%
\def\araa{ARA\&A}%
\def\apj{ApJ}%
\def\apjl{ApJL}%
\def\apjs{ApJS}%
\def\apss{Ap\&SS}%
\def\aap{A\&A}%
\def\aapr{A\&A~Rev.}%
\def\baas{BAAS}%
\def\mnras{MNRAS}%
\def\pasp{PASP}%
\def\nat{Nature}%
\def\procspie{Proc.~SPIE}%
\newcommand{\masyr}{${\rm mas\, yr^{-1}}$}

\begin{document}


  \thispagestyle{empty}   \hsize 140 mm
  \topmargin 0mm
  \enlargethispage*{3cm} 
\begin{center}
    
    {\Large \textbf{Alma Mater Studiorum}}\\\vspace{5mm}
    {\Large \textbf{Universit\`{a} degli Studi di Bologna}}\\ 
    \vspace{1mm}
    \rule{140mm}{0.2mm}\\
    \rule[2ex]{140mm}{0.5mm}
    \vspace{5mm}
    \large{DIPARTIMENTO DI FISICA E ASTRONOMIA\\      
           Dottorato di ricerca in Astronomia\\ 
            Ciclo XXVII\\

           \vspace{2.5cm}}
           
                  {\huge\textbf{COSMIC-LAB: Terzan 5 as a fossil remnant\\
                  \vspace{5mm} 
                  of the Galactic bulge formation epoch}}\\
                   		     
\end{center}

  \vspace{2.5  cm}
 
  \hspace{-6mm}
  \parbox[t]{50mm}{
    \begin{flushleft}
      {\large Dottorando: \vspace{1mm}\\
        \textbf{Davide Massari}}\\
    \end{flushleft}
  } \ \          
  \hspace{-5mm}
  \parbox[t]{90mm}{
    \vspace{0cm}
    \begin{flushright}
      {\large Relatore:\\ \vspace{1mm}         \textbf{Chiar.mo Prof. Francesco R. Ferraro}\\
        \vspace{0.3 cm}
        Co--Relatori: \\ \vspace{1mm}
        \textbf{Dr. Emanuele Dalessandro} \\
        \textbf{Dr. Alessio Mucciarelli} \\
        \textbf{Prof. Barbara Lanzoni} \\
        \textbf{Dr. Livia Origlia} \\
        \vspace{1 cm}
        Coordinatore: \\ \vspace{2mm}
        \textbf{Chiar.mo Prof. Lauro Moscardini}} \\
    \end{flushright}
  } \\

\begin{center}

    \vspace{1.5cm}
      {\large Esame finale anno 2014}\\
    \rule{140mm}{0.2mm}\\
    \rule[2ex]{140mm}{0.5mm}\\

       {\small         Settore Concorsuale: 02/C1 -- Astronomia, Astrofisica, Fisica della Terra e dei Pianeti\\
     Settore Scientifico-Disciplinare: FIS/05 -- Astronomia e Astrofisica\\}
         
          \end{center}

 \hsize 140 mm

\clearpage{\pagestyle{empty}\cleardoublepage}




\newpage
\mbox{ }
\thispagestyle{empty}

\vspace{6cm}
\begin{flushright}
{\sl A Cristiana}\\ 
   \vspace{2mm}
\end{flushright}


\newpage
\mbox{ }
\thispagestyle{empty}

\newpage
\mbox{ }
\thispagestyle{empty}

\vspace{4cm}
\begin{flushright}
{\sl ``Chi mira più alto, si differenzia più altamente; e 'l volgersi al gran libro della natura è il modo per alzar gli occhi.''}\\ 
   \vspace{2mm}
               {Galileo Galilei\ \ \  } 
\\
\end{flushright}


\newpage
\mbox{ }
\thispagestyle{empty}

\baselineskip 4ex

\newcommand{\ltae}{\raisebox{-0.6ex}{$\,\stackrel
{\raisebox{-.2ex}{$\textstyle <$}}{\sim}\,$}}
\newcommand{\gtae}{\raisebox{-0.6ex}{$\,\stackrel
{\raisebox{-.2ex}{$\textstyle >$}}{\sim}\,$}}

\baselineskip 4ex








\renewcommand{\thepage}{  }
\newpage

\pagenumbering{roman}

\markboth{\sc \ }{\sc Contents}
\tableofcontents
\clearpage{\pagestyle{empty}\cleardoublepage}




\setcounter{page}{1}

\renewcommand{\thepage}{\arabic{page}}

\clearpage

\markboth{Introduction}{Introduction}
\addcontentsline{toc}{chapter}{Introduction}
\chapter*{Introduction}

One of the most debated topic in modern astrophysics is the formation and
evolution of galaxy bulges.
According to some theoretical models, bulges may built up from the 
merger of substructures formed from the instability and fragmentation of a 
gaseous disk in the early phases of the evolution of a galaxy.
Such a scenario has been tested and confirmed by several numerical simulations
(e.g. \citealt{noguchi99, immeli04}), according to which the proto-disk
of a galaxy would locally fragment into massive clumps of gas, forming stars
at a very high star formation rate. Then, these clumps are forced to drift
towards the center of the galaxy because of dynamical friction and eventually
end up in merging together and form a bulge.
Such massive clumps may have been indeed observed at high redshift in the
chain and clumpy galaxies (see \citealt{elme08, dekel09}).
However, no confirmation about this scenario has been found for the closest 
and best studied bulge, the Galactic one. The only notable exception may be
the recent suggestion that the stellar system Terzan 5 could be the remnant
of one of these bulge pristine fragments (\citealt{f09}).

Terzan 5 has been historically cataloged as a globular cluster. It is located
in the bulge of the Galaxy, in a region of the sky strongly extincted by dust.
Its peculiar nature has remained hidden behind this dusty curtain until
adaptive-optics infrared observations revealed the presence of two well
separated red clumps in its color-magnitude diagram (\citealt{f09}).
A prompt spectroscopic follow-up demonstrated that such populations have very
different iron content, with a discrepancy as large as $\Delta$[Fe/H]$=0.5$ dex.
A more detailed study on a sample of 34 red giant branch stars further
revealed that the metal-poor component has a metallicity [Fe/H]$\simeq-0.27$
and is $\alpha$-enhanced, while the metal-rich population has an average 
[Fe/H]$\simeq+0.25$ and has solar-scaled $\alpha$-element abundance (\citealt{origlia}).
Moreover, no anti-correlations among light elements similar to those commonly observed 
in genuine Galactic globular clusters, have been found for Terzan 5.

All these features demonstrate that Terzan 5 is not a genuine globular cluster,
but a stellar system that experienced more complex star formation and
chemical enrichment histories. 
In fact, the [$\alpha$/Fe] vs. [Fe/H] trend shown by Terzan 5 populations indicates that
while the metal-poor component formed from a gas mainly enriched by type II 
supernovae, the metal-rich one has formed from a gas further polluted by
type Ia supernovae on a longer timescale. This means that the initial mass of the system
had to be large enough to retain the gas ejected by these violent explosions.
Moreover, the fact that the $\alpha$-elements
abundance starts to drop towards solar values at solar metallicity is very peculiar
and suggests a very high star formation rate.
All these chemical features are strikingly similar to those observed in only one other stellar system in the Galaxy: 
the Galactic bulge. Therefore, we believe that {\it Terzan 5 could be the remnant
of one of the massive clumps that contributed to form the Galactic bulge itself}.

Within this exciting scenario, we are carrying on a project aimed
at reconstructing the origin and the evolutionary history of Terzan 5.
To achieve this goal, a multi-fold approach is needed. First of all,
it is crucial to determine the
star formation history of Terzan 5 and thus to estimate the absolute ages of its 
populations via the Main-Sequence Turn-Off luminosity method. 
In fact, the color and magnitude separation of the two red clumps in the IR
color-magnitude diagram may suggest a younger age for the metal-rich component, but as argued
in \cite{dantona} also a difference in the helium content between the two
populations can explain the observed magnitude split, thus mitigating any age spread. 
The direct measure of any split in the Main-Sequence Turn-Off would definitely break such degeneracy. 
However, severe limitations to the detailed analysis of the
evolutionary sequences in the optical CMDs are introduced by the
strong contamination from the underlying bulge field population and by the
presence of large differential reddening. To face these problems,
we measured relative proper motions for Terzan 5 stars,
reaching several magnitudes below the Main-Sequence Turn-Off (see Chapter 2 of this
Thesis) and we built the highest-resolution extinction map ever constructed in the
direction of Terzan 5 (see Chapter 3).

The other crucial step toward a proper understanding of the nature and the evolutionary
history of Terzan 5 is the detailed study of its kinematical properties. We therefore collected
spectra for more than 1600 stars in the direction of the system. These have been used to
determine the chemical and kinematical properties of the surrounding bulge stars, as described 
in Chapter 4, and to build a
bulge-decontaminated metallicity distribution for Terzan 5 based on a very large number of stars. 
This allowed us to test (see Chapter 5 and 6) whether the actual metallicity 
distribution of Terzan 5 is bi/multi-modal (like that observed in massive systems
such as dwarf galaxies and suggesting a bursty star formation) or rather unimodally 
broad (thus mimicking a prolonged star formation).

The detailed photometric, spectroscopic and kinematical analysis of the
stellar populations of Terzan 5 is starting to shed new light on the
true nature of this fascinating system and will possibly allow us to test one of the most
promising scenarios about the formation of the Galactic bulge.  

\clearpage{\pagestyle{empty}\cleardoublepage}

\chapter{General context: the Galactic bulge}\label{chapintro}

More than half the light in the local Universe is found in spheroids
(e.g. \citealt{fuku98}). The Galactic bulge, that is the central structure of the Milky Way,
is the only spheroid where individual stars can be resolved and studied in detail.
The importance of this stellar system is therefore huge, and the understanding
of its formation and evolution is one of the fundamental goals of the modern astrophysics.

\begin{figure}[!htb]
 \centering%
 \includegraphics[scale=0.5]{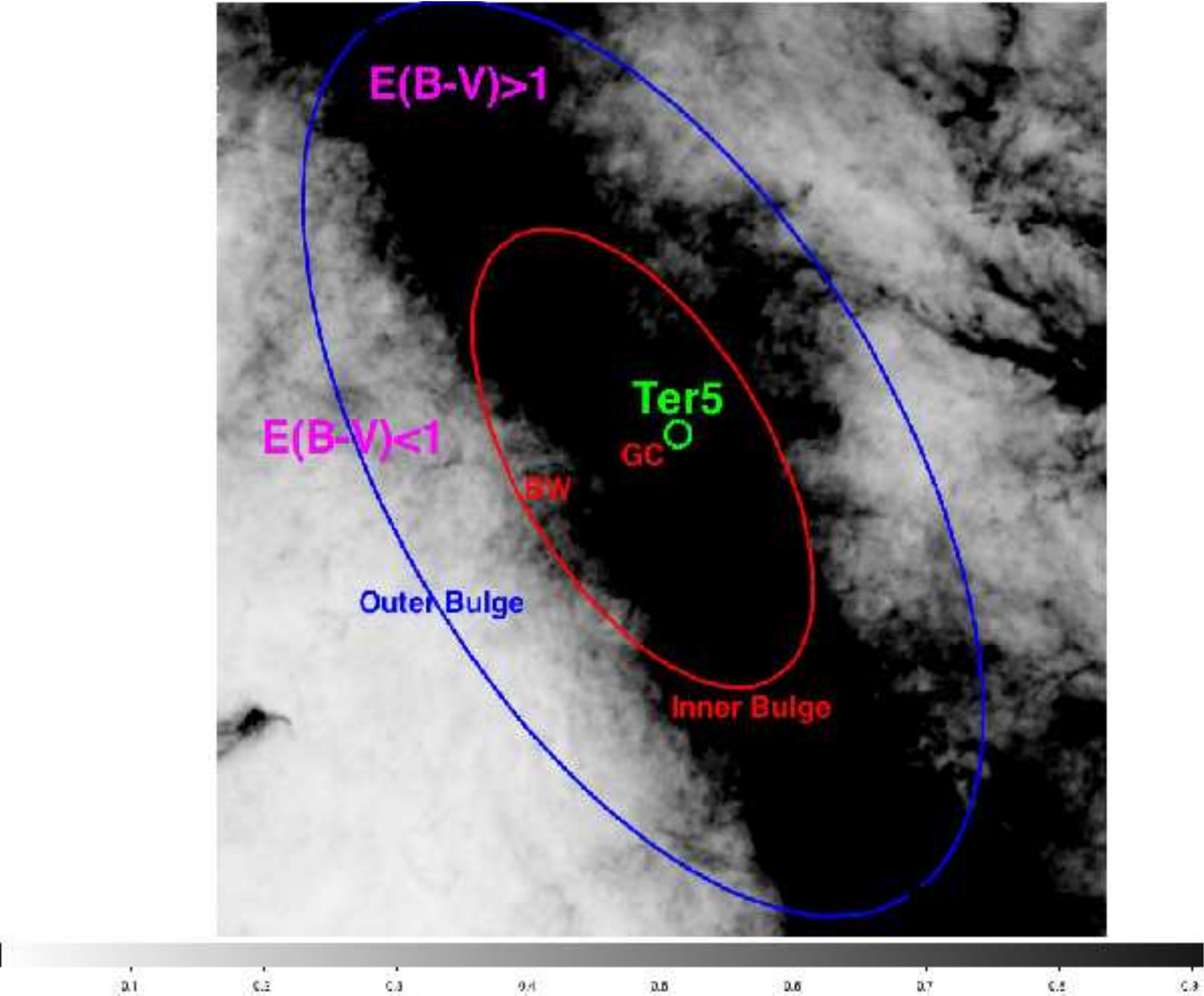}
 \caption{\small Reddening map of the central region of the Galaxy. Dark colors indicate large extinction,
with color excess E(B-V)$>1$ dex and vice-versa. The location of Terzan 5 is
marked in green, together with the extent of the outer bulge (blue ellipse) and of the inner bulge
(red ellipse). }
\label{bulgeredmap}
\end{figure}

Any precise definition of the Galactic bulge extent is somewhat arbitrary. Usually it is
defined as the region within the central 3 kpc of the Galaxy, whereas the central kpc is
often referred to as the inner bulge. It accounts for about
$20$\% of the total mass of the Milky Way and $25$\% of its bolometric luminosity.
Because of the strong and spatially varying extinction obscuring this region
(see the reddening map in Figure \ref{bulgeredmap}, taken from \citealt{schl98}), 
the bulge has always been very difficult to investigate, especially at optical wavelengths. 

A few fields are characterized by low extinction: these are the Baade's Window (at
Galactocentric coordinates l,b$=0.9$\textdegree,$-3.9$\textdegree), the Plaut's
Field (l,b$=0$\textdegree,$-8$\textdegree) and the Sagittarius I Field (l,b$=1.3$
\textdegree,$-2.65$\textdegree). For this reason, they have historically been the 
subject of the first photometric and spectroscopic studies on bulge stars (see, e.g.,
\citealt{baade51} for the discovery of RR Lyrae stars and \citealt{nassau58} for the
first detection of M giants in the direction of the Galactic center).
More recently, the advent of near-infrared (NIR) facilities allowed to
investigate the properties of the bulge also in the regions most affected by reddening,
and unveiled the complexity of its stellar populations.
Current and future large photometric and spectroscopic surveys 
in both optical and NIR bands will shed new light on this stellar system.
 
In the following Sections, a summary of the currently known properties of
the Galactic bulge will be provided, together with an overview of some of the models 
proposed to explain how it formed and evolved and the role of
the stellar system Terzan 5 in this context.

\section{The structure of the Galactic bulge}\label{structure}

The Galactic bulge is a triaxial, oblate system possibly composed of three bar structures:
a central massive bar, a long thin bar, and a nuclear bar. While the presence of the first main component
is well established, the existence of the other two, minor bars is still debated
(see e.g. \citealt{gmv12}).

\begin{figure}[!htb]
 \centering%
 \includegraphics[scale=0.9]{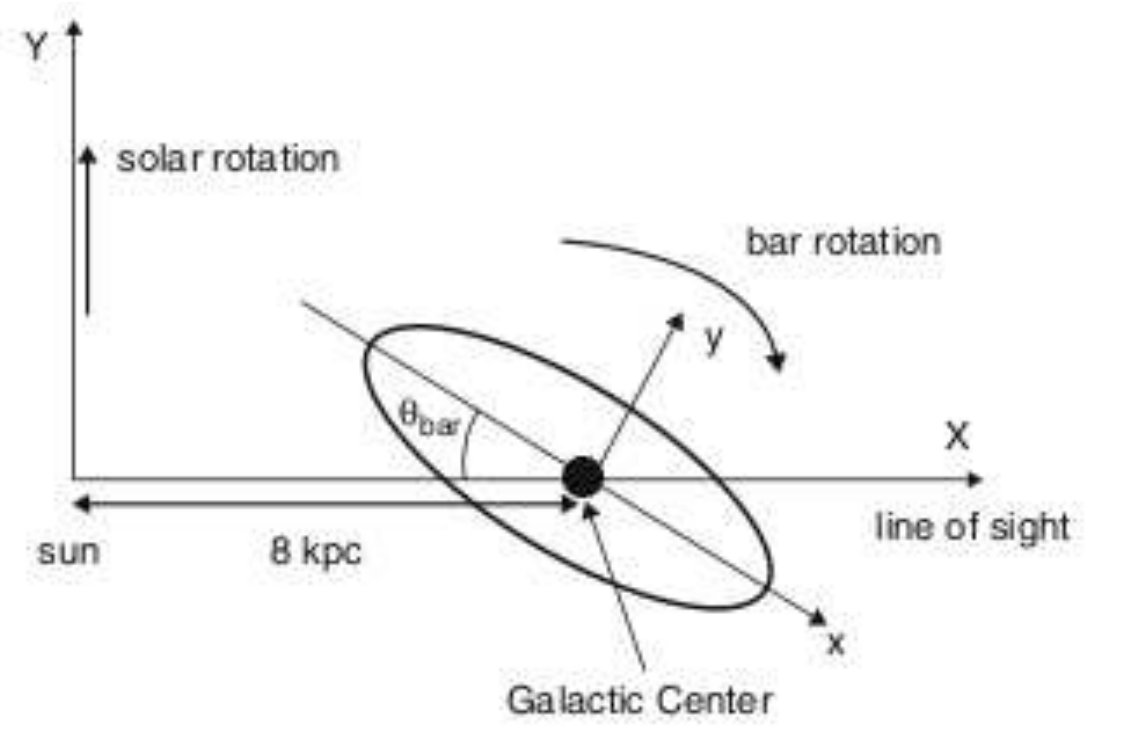}
 \caption{\small Shape and orientation of the main bar as it appears if the Galaxy is seen face-on. 
 The near side of the bar is at positive Galactic longitude. The clockwise rotation of the bar is also marked
 (Figure taken from \citealt{rich13}).}
\label{bar}
\end{figure}

The main component is a boxy bar, that accounts for 
most of the mass of the bulge itself, being as massive as about $2\times10^{10}$ M$_{\odot}$.
Its presence has been traced with different methods, from the kinematics of gas (\citealt{liszt80})
and planetary nebulae (\citealt{beau2000}), to star counts (e.g. \citealt{gonz}) and optical depth 
gradients in microlensing events (\citealt{zhao96}). The collected observables point toward the presence
of a bar which is $\sim2-3$ kpc in radius, with a vertical scale height of $\sim 300$ pc, an axis 
ratio of about $1:0.3:0.3$ and tilted by an angle of $\sim20-30$\textdegree 
with respect to the line of sight (\citealt{bab05, cao13}).
A sketch of the main bar structure is plotted in Figure \ref{bar}.

\begin{figure}[!htb]
 \centering%
 \includegraphics[scale=0.9]{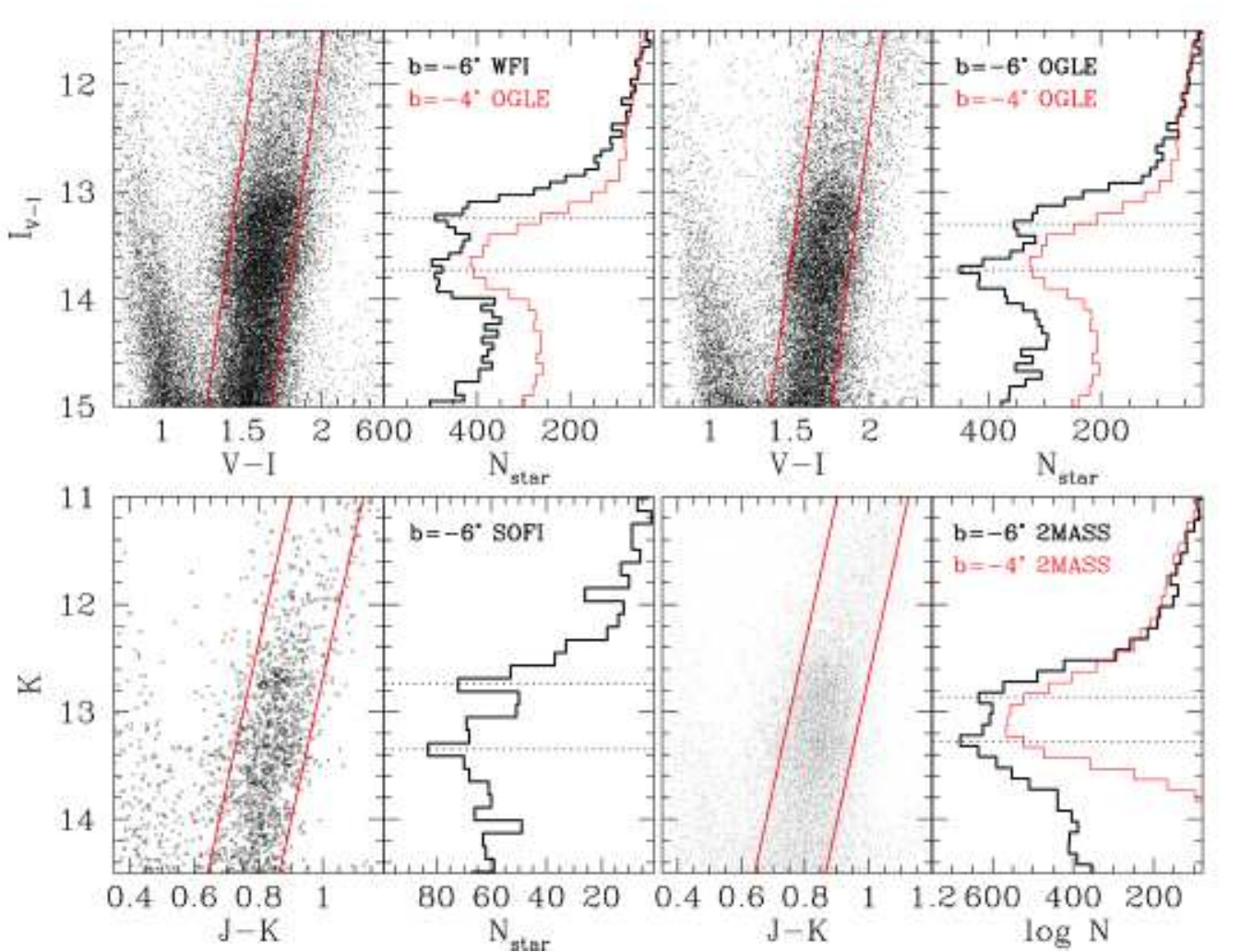}
 \caption{\small Optical (upper panels) and IR (lower panels) CMD of the bulge analyzed in \cite{mcz}.
 The luminosity functions for stars at Galactic latitude b$=-6$\textdegree~(black histograms) clearly show 
 two distinct peaks.}
\label{doublerc}
\end{figure}

Recent IR observations demonstrated that the color-magnitude diagram (CMD) of bulge stars shows a double
red clump (RC, see Figure \ref{doublerc}), clearly separated in magnitude (\citealt{mcz, nataf}). 
This feature has been interpreted in terms of a distance effect, due to a X-shaped distribution of stars in the bar. 
According to this interpretation, the bar would show its X-shape when seen tangentially, while it appears 
boxy/peanut-shaped when viewed from the Sun.

The stronger evidence for the presence of a \textquotedblleft long bar\textquotedblright comes from an
asymmetry in the disk star counts towards the Galactic center found by several authors 
(\citealt{hammer01, benj05}). According to these works, such a long bar would be thin, with a vertical
scale height of about $100$ pc, and seen with an inclination angle of 45\textdegree, thus almost
aligned with the main bar. However, such an asymmetry has been observed only in the first quadrant,
and its detection may be strongly biased by the presence of foreground disk stars.
Therefore, accurate proper motion data are necessary to disentangle this sub-structure from the main bar
and to firmly asses its actual existence.

An inner, nuclear bar has been also claimed, but its precise definition is more challenging, given both 
the strong extinction in the direction of the Galactic center and the contamination by the other sub-structures.
According to \cite{alard01}, this bar should have an inclination angle of $\sim70$\textdegree
and may be as massive as the long bar.

Finally the presence of any spheroidal component, and the amount of bulge mass enclosed in it, is still debated.
If we define the spheroidal component of the bulge as that formed either via hierarchical merging of substructures 
(\citealt{toomre77}) or from the monolithic collapse of primordial gas clouds (\citealt{eggen62}), then such
a component should be slowly rotating, mostly supported by random motions, and with surface brightness profiles
following a Sersic law $\mu(r)\sim r^{1/n}$, with Sersic index $n\sim4$. Recent kinematical observations (see Section \ref{kin})
ruled out any spheroidal component contributing for more than the $8\%$ of the bulge mass (\citealt{shen10}).
Moreover, measured values of the Sersic index are typically smaller, around 2-2.5 (\citealt{rich13}), closer
to what observed in the so-called pseudo-bulges (see \citealt{kk04}).
However, numerical simulations by \cite{saha12} demonstrated that the buckling instability of the bar may have spun 
up any possible spheroidal component, which would therefore be kinematically indistinguishable from the bar at the
present epoch.

\section{Bulge stellar populations}

In order to constrain the possible scenarios for the formation and evolution of the Galactic bulge, 
it is crucial to characterize its stellar populations in terms of age, chemistry and kinematics.

\subsection{Age}

\begin{figure}[!htb]
 \centering%
 \includegraphics[scale=1.1]{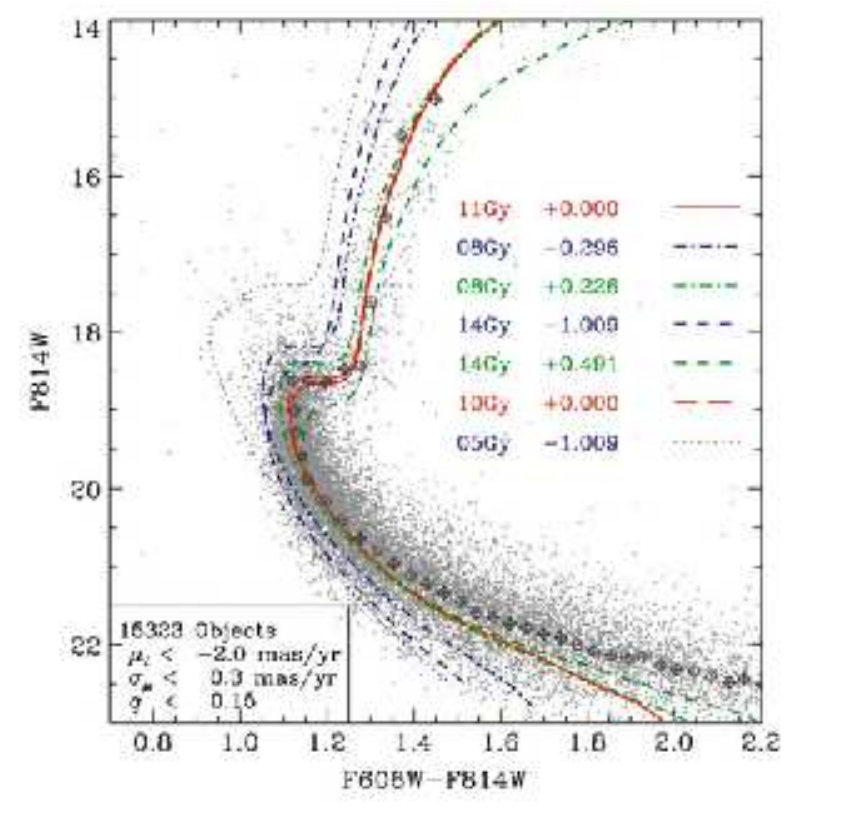}
 \caption{\small Proper-motion-selected optical CMD of the bulge stars in the Sagittarius I Field.
 The luminosity of the MSTO indicates that most of the population is old. The MS is in fact best-fitted
 by an 11 Gyr old isochrone, with very small room for satisfactory fits by younger models.}
\label{bulgecmd}
\end{figure}

The best way to determine the absolute age of a stellar population is by measuring the luminosity
of the Main-Sequence Turn-Off (MSTO) in the CMD. The CMD of the Galactic bulge is particularly difficult to
measure because of the strong contamination by disk field stars. When properly decontaminated by means
of statistical star counts (as in \citealt{zoccali03, valenti13}) or by proper motions analysis (\citealt{clark08},
see Figure \ref{bulgecmd}), the CMDs obtained in all the studies performed so far revealed that the bulge is 
predominantly old, at least as old as its globular clusters ($\gtrsim10$ Gyr). Moreover,
by considering also the population of Blue Straggler Stars (BSS) that in the CMD can mimic a younger component,
\cite{clark11} concluded that only $<3.4$\% of the bulge population can be younger than 5 Gyr.

The old age of the bulge is also confirmed by the discovery of RR Lyrae stars (e.g. \citealt{baade51, alcock98}), 
which are good tracers of old stellar populations.

Although the bulk of the bulge is old, several pieces of evidence have been found supporting the presence
of an intermediate-age population. This includes the discovery of long periods ($>400$ days) Mira variables,
that are associated with younger ages (\citealt{feast63}). However, a recent work on Mira
variables (\citealt{blo07}) compiled from the Optical Gravitational Lensing Experiment II survey 
(OGLE-II, \citealt{udalski2000}) concluded
that the majority of these stars lie in the innermost 50 pc of the bulge.

\begin{figure}[!htb]
 \centering%
 \includegraphics[scale=1.1]{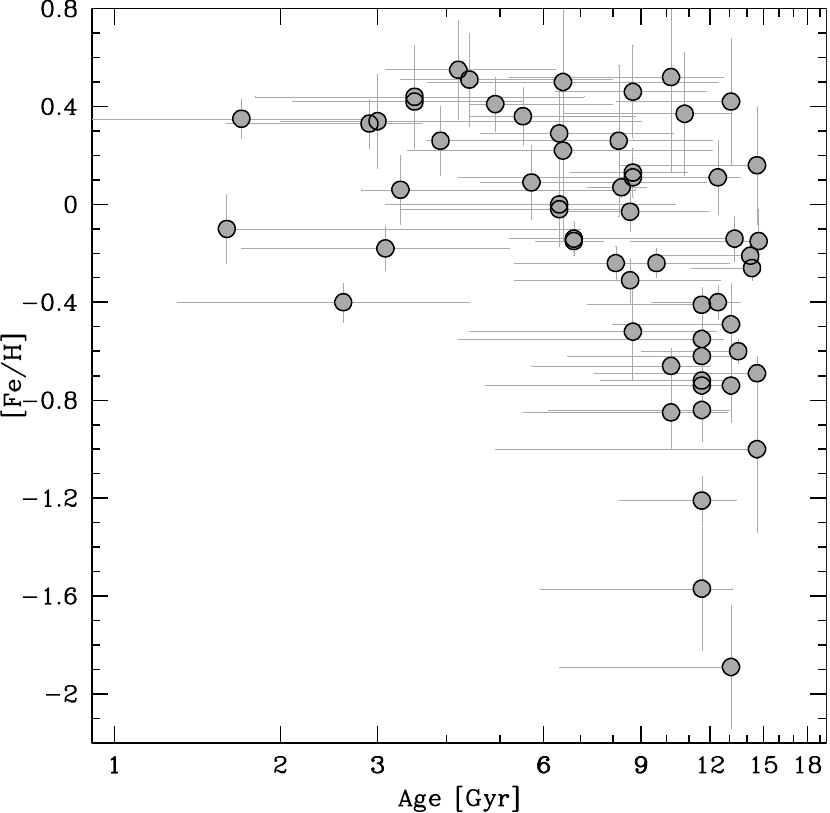}
 \caption{\small Age vs. metallicity relation for the sample of microlensed bulge dwarfs analyzed by
  \cite{bensby13}. An important fraction of these stars ($\sim25$\%) appear to be younger than 6 Gyr.}
\label{bensby}
\end{figure}

Microlensing is a very powerful approach to study the faint population of bulge dwarfs.
In fact, the strong magnification due to microlensing events can boost the magnitude of such
stars by up to 5 mag, thus making them very good targets for medium-high resolution spectroscopy.
From this kind of analysis, \cite{bensby13} found evidence for a quite large ($\sim25$\%) young
and metal-rich stellar component (see Figure \ref{bensby}), numerically inconsistent with the results 
drawn from the CMD analysis.
However, such studies are possibly affected by biases that favor the selection of metal-rich 
(and young) sources (\citealt{coh10}). Also, the precise age of these stars depends on the adopted
He-content (see e.g. \citealt{nataf12}).

Young and intermediate-age stellar populations are instead certainly present in the inner
100 pc of the bulge (\citealt{mezger99, vanloon, schuller06}).

\subsection{Chemistry}\label{chem}

The chemical composition of the Galactic bulge is a crucial information to constraint
its formation and evolutionary history, and to understand its connection with other
Galactic populations such as those in the disk and the halo.

The investigation of the bulge chemistry using medium and high-resolution spectroscopy
started a few decades ago. \cite{mcw94} were the first to measure abundances for a large
sample of K giants in the Baade's Window from high signal-to-noise (SNR), optical spectra with
resolution R$\sim17000$.
These authors found the iron abundances of bulge stars to span a wide range of values,
from [Fe/H]$\sim-1$ dex to [Fe/H]$\sim+0.5$ dex. Moreover, they found the $\alpha$-elements
to be enhanced relative to both the thick and the thin disk populations.

\begin{figure}[!htb]
 \centering
 \includegraphics[scale=1.4]{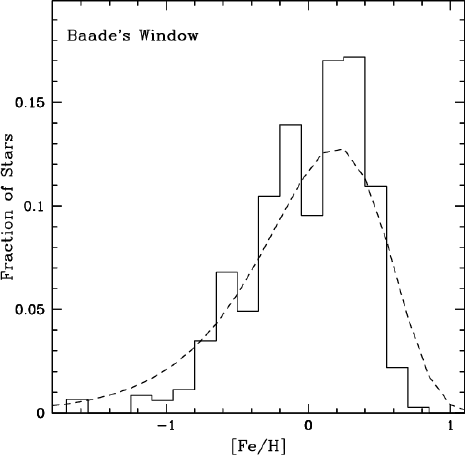}
 \caption{\small Metallicity distribution of the sample of bulge RC stars analyzed by \cite{zoccali08}. The distribution
peaks at solar values and displays a long metal-poor tail down to [Fe/H]$\sim-1.5$ dex, while reaching metallicity as high 
as [Fe/H]$\sim+0.5$ dex.}
\label{zoc08}
\end{figure}

These two main features have been confirmed by many subsequent works.
In fact, both optical (e.g. \citealt{ful06, zoccali08, johnson11, johnson12, johnson13, ness13, 
johnson14}) and infrared (e.g. \citealt{rich05, rich07, rich12}) studies, in different locations
of the bulge, demonstrated that the metallicity distribution of its stars peaks around
solar [Fe/H] (with a vertical gradient of about $-0.6$ dex kpc$^{-1}$, that likely flattens
at latitudes $|b|<4$\textdegree), and reaches iron abundances as high as $\sim+0.5$ dex with
a long metal-poor tail down to [Fe/H]$\sim-1.5$ dex (see for example Figure \ref{zoc08} 
taken from \citealt{zoccali08}). 

\begin{figure}[!htb]
 \includegraphics[scale=1.2]{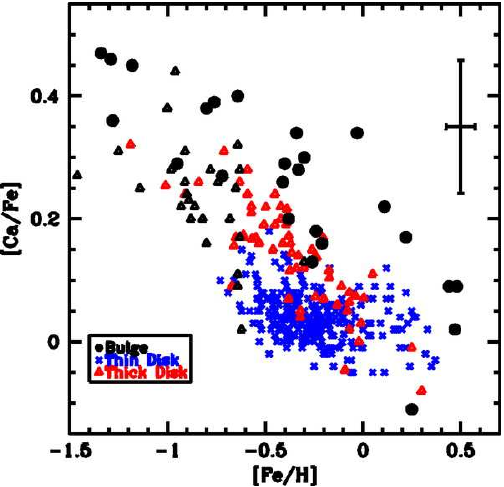}
 \includegraphics[scale=1.2]{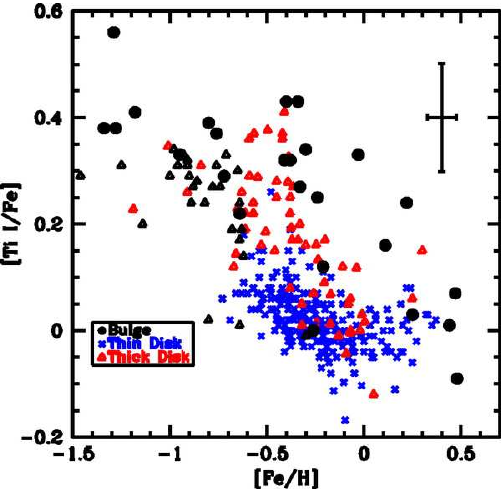}
 \caption{\small Examples of $\alpha$-enhancement for the sample of bulge stars (black dots) studied in \cite{ful07}. 
The left panel shows the behavior of Calcium, while Titanium is plotted in the right panel. In both cases the differences
among the bulge sample with respect to thin disk (blue crosses) and thick disk (red triangles) stars are evident,
with the bulge sample being more enhanced and with the knee point being located at higher metallicity.}
\label{cati}
\end{figure}

Some works found this
metallicity distribution to be multi-modal (\citealt{hill11, ness13}), with two main
components peaking at sub-solar ([Fe/H]$\sim-0.3$ dex) and super-solar ([Fe/H]$\sim+0.3$ dex) metallicity.
The $\alpha$-elements have generally been found to be enhanced with respect to iron at least up to solar [Fe/H]
and then progressively declining towards solar values of [$\alpha$/Fe] (see Figure \ref{cati}), with the location 
of the knee in the  [$\alpha$/Fe] {\it vs} [Fe/H] trend possibly occuring at different metallicities depending on the kind
of $\alpha$-element (\citealt{ful07, gonzalez11}). Such a behavior suggests that the stars
with [Fe/H]$<0$ dex probably formed from a gas mainly enriched by core collapse supernovae (ccSN) on very
short timescales and with a quite high star formation rate (SFR), while stars with super-solar
metallicity probably formed from a gas further polluted by SNIa.
Also, at odds with what is observed for iron abundances, no [$\alpha$/Fe] abundance gradient 
has been found in the bulge.

\begin{figure}[!htb]
 \centering
 \includegraphics[scale=1.6]{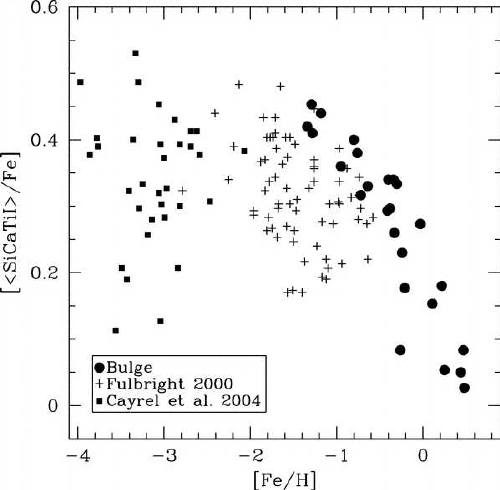}
 \caption{\small Summed abundances of the explosive $\alpha$-elements Si, Ca and Ti. The bulge stars (sample taken 
from \citealt{ful07} and plotted as black filled circles) clearly shows a smaller scatter than halo stars (other symbols,
from \citealt{ful2000, cayrel04}).}
\label{ful07_sicati}
\end{figure}

Given these abundance patterns, the bulge appears as a different population with respect to the
thin/thick disk and the halo in many respects.
First of all, the metallicity regime of the bulge is clearly much different with respect to that
of the halo (see Figure \ref{ful07_sicati}). Moreover, the rare bulge stars at [Fe/H]$<-1.3$ dex, which somewhat overlap the metal-intermediate
and poor halo populations, exhibit similar trends only in terms of \textquotedblleft explosive\textquotedblright~$\alpha$-elements 
(i.e. those produced during ccSN events, such as Ca, Si, Ti), but with a much
smaller scatter than what is observed in halo stars (\citealt{ful07}, see Figure \ref{ful07_sicati}).

\begin{figure}[!htb]
 \centering%
 \includegraphics[scale=1.1]{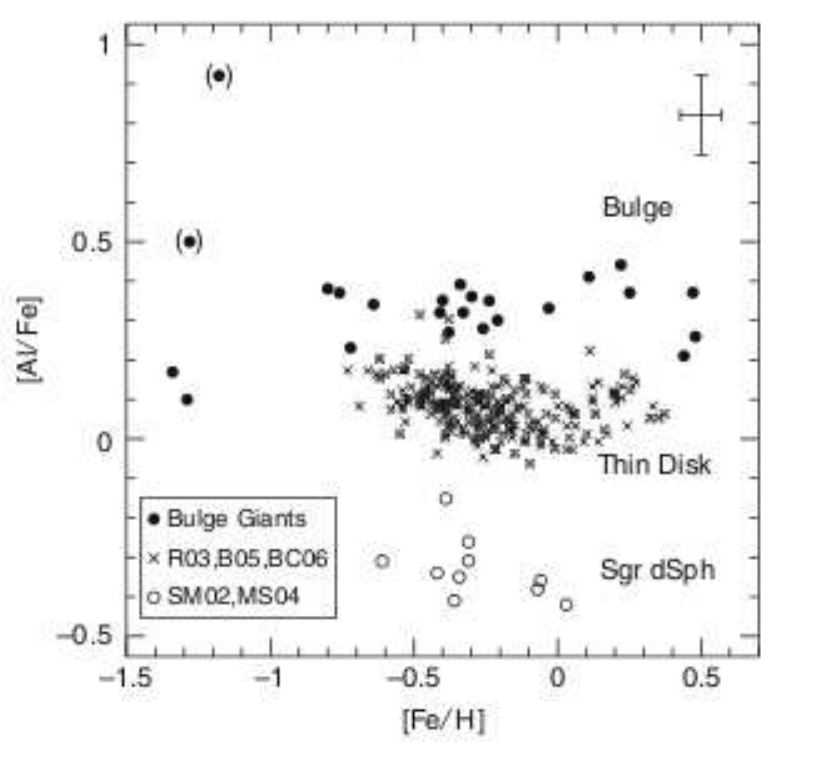}
 \caption{\small Comparison among [Al/Fe] vs [Fe/H] abundances for bulge, thin disk and dwarf galaxies stars.
Bulge stars (black filled circles) are clearly over-abundant in Al content with respect to the other samples.}
\label{alfe07}
\end{figure}

Secondly, the $\alpha$-element abundance pattern observed in the bulge is clearly different from
that of the thin disk (which stays almost constant at solar values, see Figure \ref{cati}), resulting from a more recent and
prolongated star formation. Also, the light odd element Al is an efficient tool 
in separating these populations, as clearly demonstrated in \cite{ful07} where Al in bulge stars
was found to be definitely enhanced with respect to both thin disk and dwarf galaxies (see Figure \ref{alfe07}).   

\begin{figure}[!htb]
 \centering%
 \includegraphics[scale=0.9]{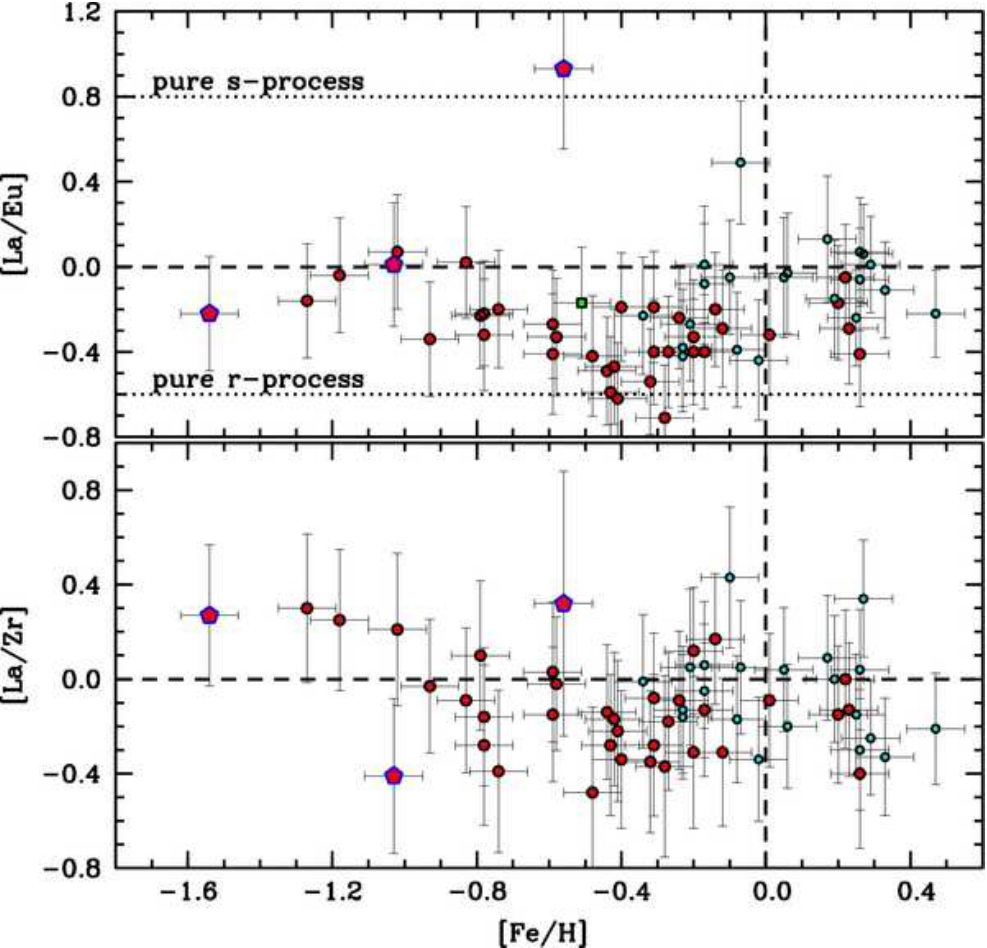}
 \caption{\small {\it Upper panel}: [La/Eu] vs. [Fe/H] for the bulge stars analyzed in \cite{johnson12}.
Red giant branch (RGB) stars are plotted in red, while RC stars are shown as cyan circles. 
The level of [La/Eu] is consistent with a pure r-process regime, indicating
an extremely fast enrichment.}
\label{laeu}
\end{figure}

Finally, the separation between bulge and thick disk stars in terms of $\alpha$-elements is less
neat and still disputed. Several works found the bulge giants to be more enhanced than thick disk giants. 
This has also been confirmed for a sample of microlensed dwarf stars by \cite{bensby13}. 
Only \cite{mele08, ryde09} and \cite{alvesbrito10} claim no or small differences between these 
two populations, especially in the [O/Fe] trend, but their samples are smaller and less statistically
significant. An important difference has been found
when measuring heavy neutron-capture element abundances, such as La and Eu.
In fact, bulge stars have low [La/Eu] over almost the entire range of 
metallicities (pure r-process regime, see \citealt{kap89}), this being consistent with a so fast enrichment 
that AGB stars had not enough time to pollute the star-forming gas with s-process (see the upper panel of 
Figure \ref{laeu} from \citealt{johnson12}). 
Instead, the thick disk has an higher [La/Eu], more compatible with a gas s-process polluted
on longer timescales.

\subsection{Kinematics}\label{kin}

The first studies of the bulge kinematics used neutral hydrogen (HI) gas as a tracer (\citealt{liszt80}).
From these works the first evidence of the presence of a bar was obtained (see Section \ref{structure}), 
with structural parameters close to those favored today (\citealt{binney91}).
Stellar tracers have been systematically used only later on, when multi-object spectrographs became
available to the community, allowing the measure of radial velocity for large samples of bulge stars.

In the recent years, two major radial velocity surveys of the outer bulge have been undertaken.

\begin{figure}[!htb]
 \centering%
 \includegraphics[scale=0.4]{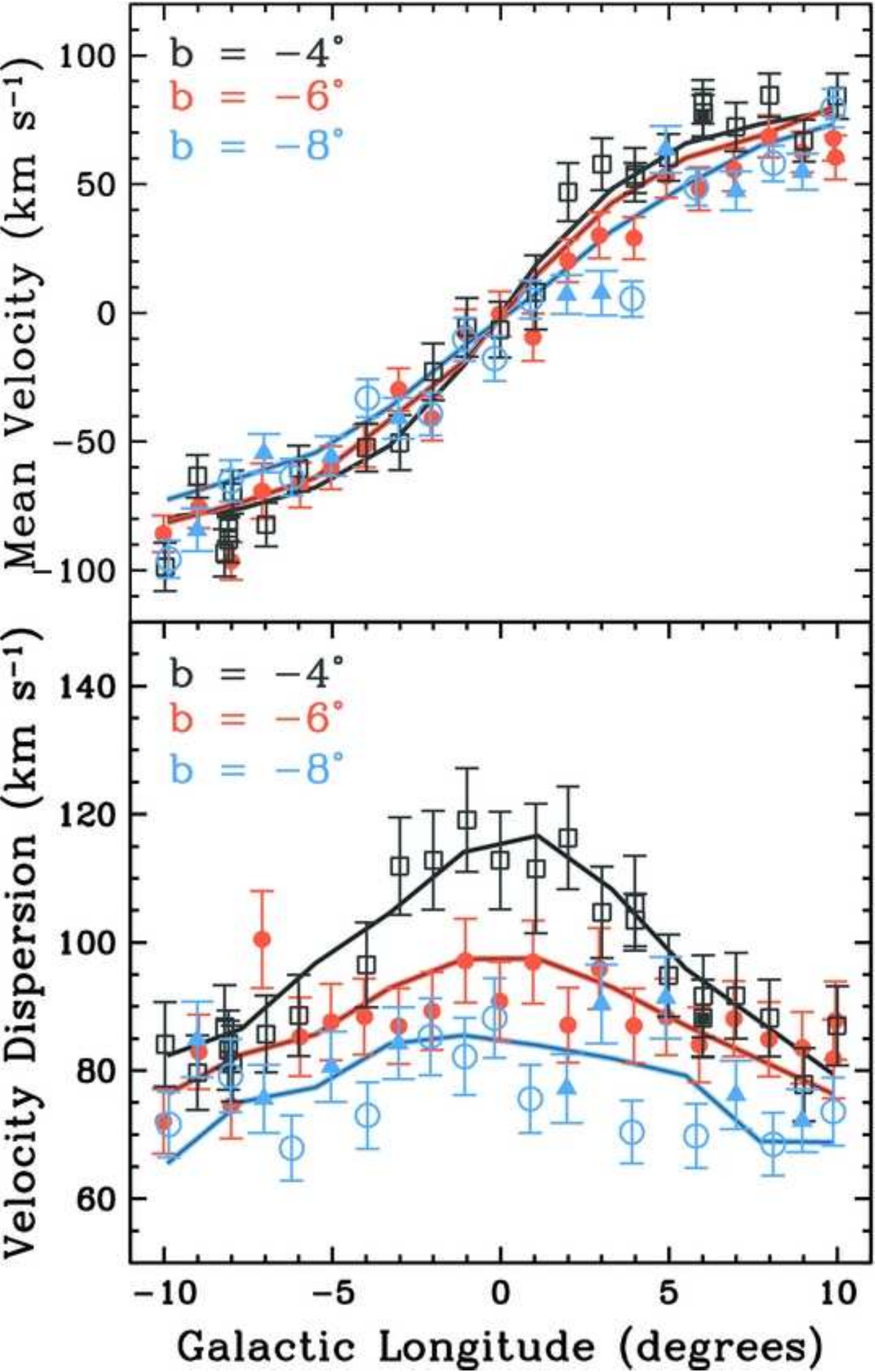}
 \caption{\small Radial velocity (upper panel) and velocity dispersion (lower panel) vs. Galactic longitude 
trends for the bulge fields studied in the BRAVA survey. The different colors indicate different Galactic
latitudes. The typical features of cylindrical rotation are evident in both cases.}
\label{kun12rot}
\end{figure}

($1$) The Bulge RAdial Velocity Assay (BRAVA, \citealt{rich07}) measured radial velocity for $4500$ M giants
located between $-10$\textdegree$<$l$<10$\textdegree and $-8$\textdegree$<$b$<-4$\textdegree, finding
that the bulge does not show a pure solid-body rotation but exhibits a cylindrical rotation (\citealt{howard08,
howard09, kunder12}, see Figure \ref{kun12rot}). 

According to these results, \cite{shen10} demonstrated using N-body simulations 
that the fraction of the bulge component in a non-barred configuration should be smaller than the 8\%.

\begin{figure}[!htb]
 \centering%
 \includegraphics[scale=0.99]{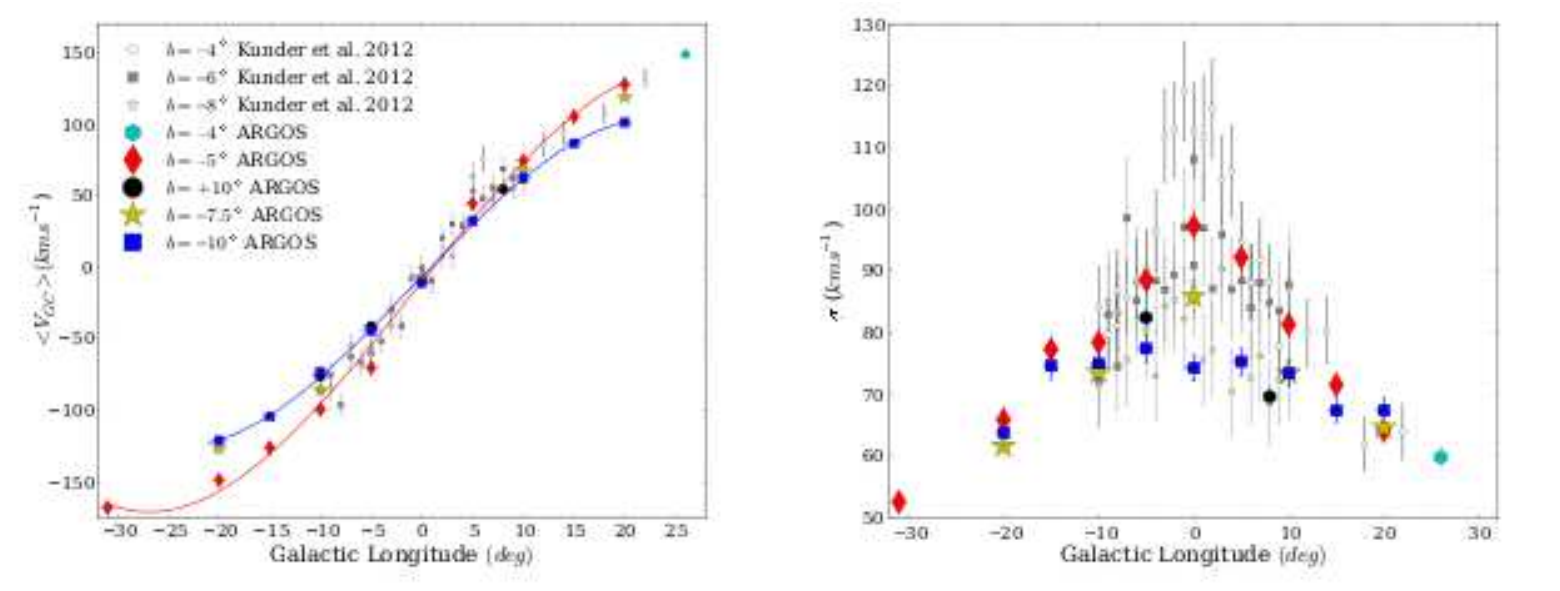}
 \caption{\small Same plots as in Figure \ref{kun12rot} but for the bulge fields surveyed by ARGOS.
The results of this survey (shown with different colored symbols depending on the latitude) agree well with the BRAVA
results (shown as grey symbols).}
\label{ness13rot}
\end{figure}

($2$) The Abundance and Radial velocity Galactic Origins Survey (ARGOS, \citealt{freeman13}) targeted more than
$28000$ stars in the bulge and in the inner disk measuring both radial velocity and chemical abundances
and found a rotation curve in good agreement with the BRAVA results (see Figure \ref{ness13rot}).

\begin{figure}[!htb]
 \centering%
 \includegraphics[scale=0.9]{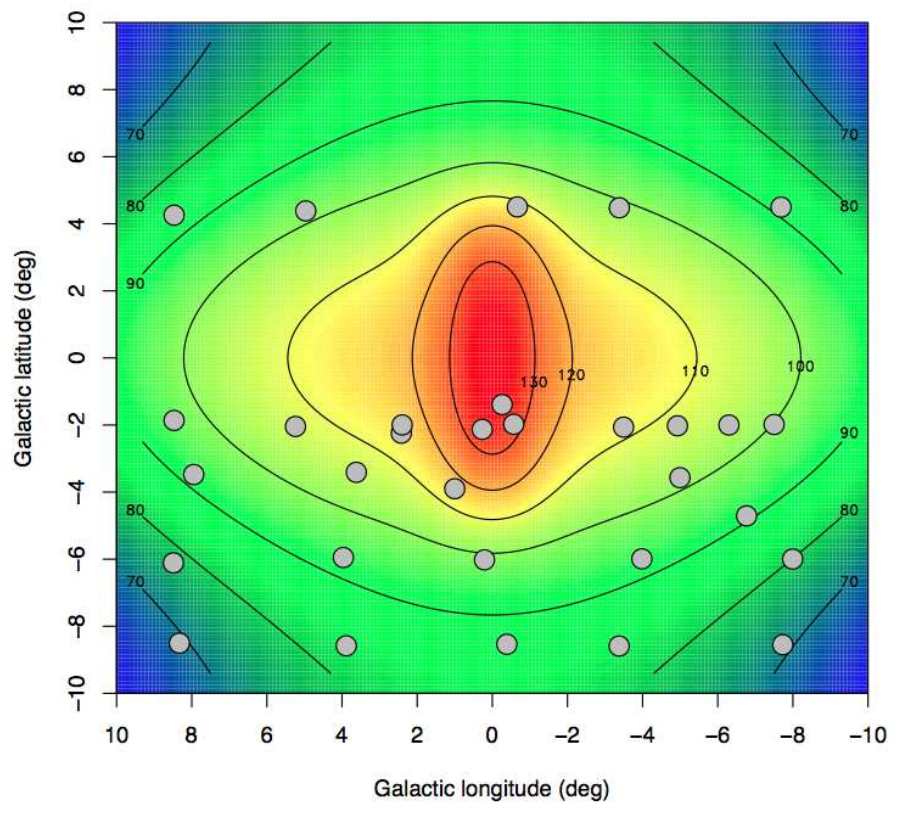}
 \caption{\small Color-coded velocity dispersion map obtained by the GIBS survey. A high velocity dispersion
peak is evident in the central region of the bulge, possibly indicating the presence of an over-density in the inner
250 pc of the Galaxy.}
\label{zoc14}
\end{figure}

Very recently the Giraffe Inner Bulge Survey (GIBS, \citealt{zoccali14}) has been carried out with
the aim of measuring radial velocities and chemical abundances for about $6500$ RC stars in the inner bulge.
The first results of the survey confirm the cylindrical rotation of the bulge also for a sample of K giants
at Galactic latitude b$=-2$\textdegree, in a region closer to the Galactic plane than probed
by the previous surveys, and found a velocity dispersion peak in the bulge central region, possibly indicating
the presence of an overdensity in the inner $250$ pc (see Figure \ref{zoc14}).

Other powerful approaches to study the kinematics of the bulge are the Fabry-Perot imaging in the
Ca II 8542 absorption line (see \citealt{rangwala09}) and proper motions.
In fact, the typical peak velocity dispersion of the bulge (of the order of $\sim100$ km$\,$s$^{-1}$) 
correspond to 2-3 \masyr in terms of proper motions.
Such a value is measurable in reasonable temporal baselines also with ground-based observations.
By using data from the OGLE-II survey,
\cite{ratten07} found that the proper-motion dispersion of bulge stars declines with increasing Galactic 
latitude and longitude.

Also, by combining radial velocity and proper motions it is possible to build the so-called velocity ellipsoid.
\cite{zhao94} found that for bulge stars a correlation between transverse proper motion and radial velocity
exists. This produces a velocity ellipsoid with a major axis which appears angled off of normal (see Figure \ref{vertex}).
Such a feature is called vertex deviation and appears to be related to stars with bar-like orbits 
(\citealt{zhao96, soto07}).



\begin{figure}[!htb]
 \centering%
 \includegraphics[scale=0.85]{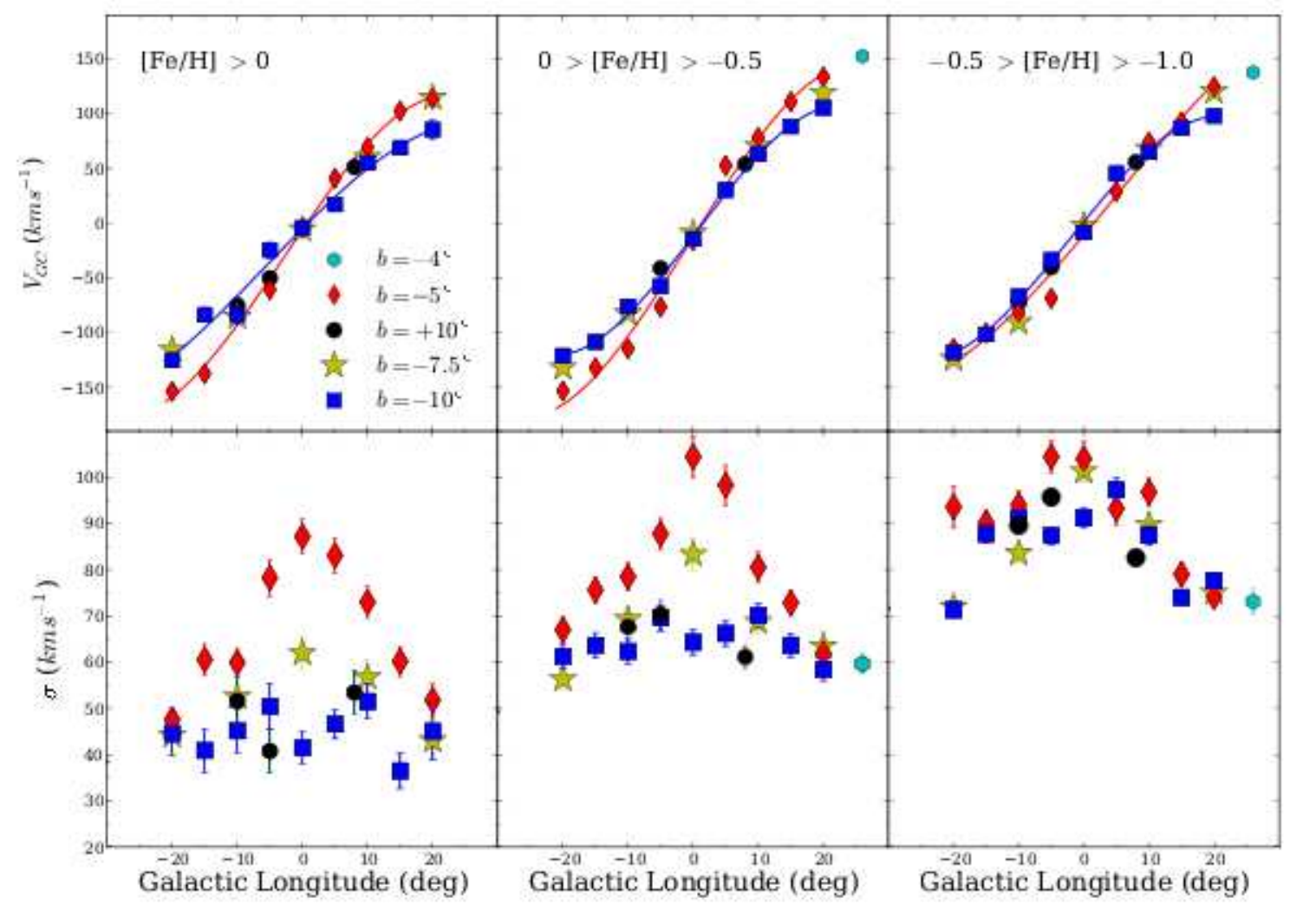}
 \caption{\small Same plots as in Figure \ref{ness13rot} but for three different bins of metallicity.
The sub and super-solar components (left and middle panels) show the same behavior, with the typical features 
of cylindrical rotation. On the other end, the most metal-poor component (right panel) shows a kinematic
more compatible with that of a slowly rotating spheroidal.}
\label{chemokin}
\end{figure}

\begin{figure}[!htb]
 \includegraphics[scale=0.33]{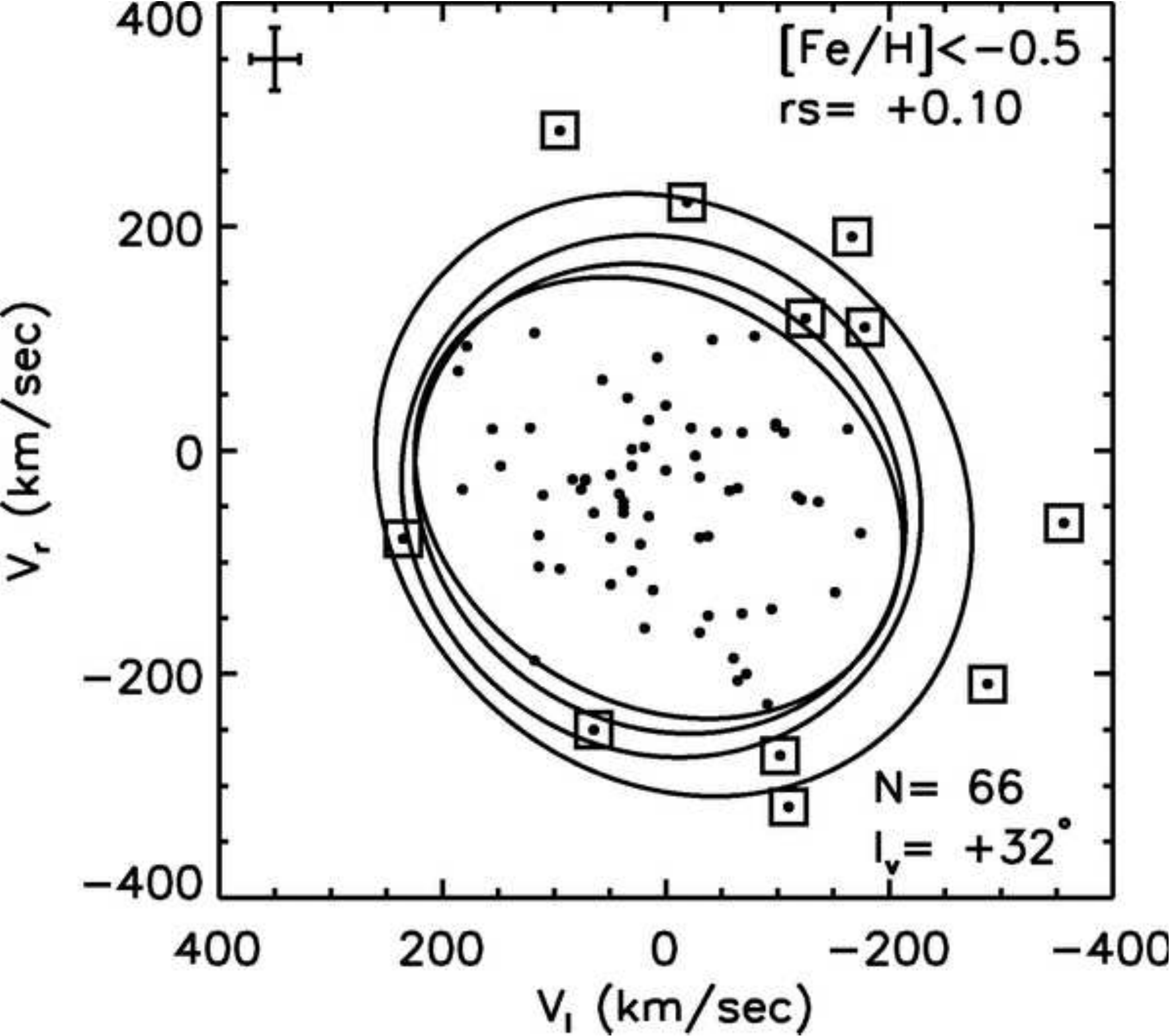}
 \includegraphics[scale=0.33]{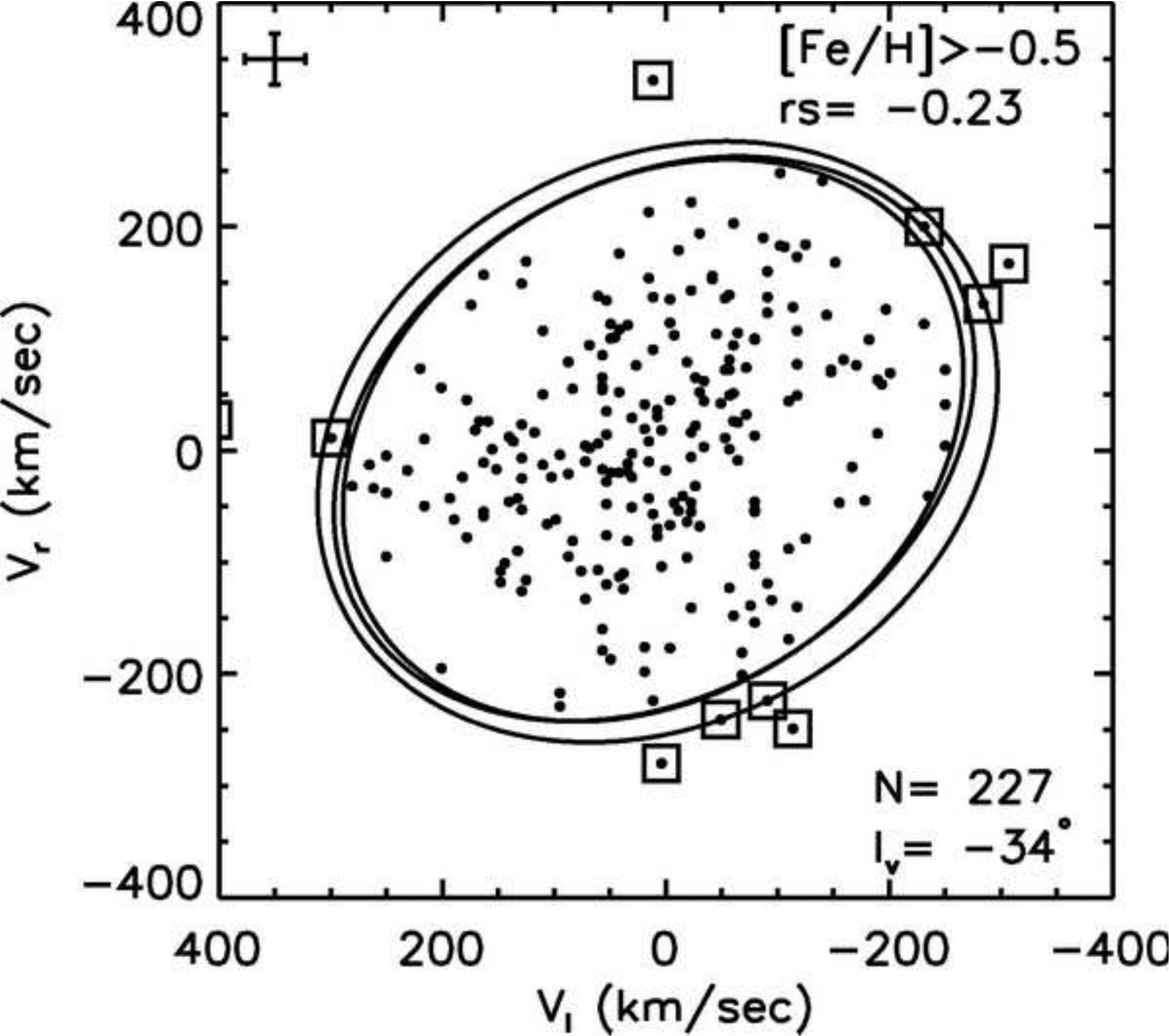}
 \caption{\small Example of vertex deviation (from \citealt{soto07}). In the left panel, stars with [Fe/H]$<-0.5$ dex 
do not show evidence of correlation among the velocity ellipsoid components. In the right panel, stars with [Fe/H]$>-0.5$ dex
display a correlation between radial velocity and the transverse component of proper motion. This causes the velocity ellipsoid
to have a preferential orientation which is defined as vertex deviation.}
\label{vertex}
\end{figure}

The combination of chemical and kinematical information gives other important clues to understand
the complexity of the bulge populations.
\cite{ness13kin} decomposed the metallicity distribution
of the bulge in several Gaussian components and distinguished two main populations
with a rotating bar kinematics, one peaking at [Fe/H]$\simeq-0.25$ dex and being $\alpha$-enhanced and the 
other peaking at [Fe/H]$\simeq+0.15$ and kinematically colder. On the other hand, they found only a small fraction 
(5\%) of metal-poor stars ([Fe/H]$<-1$ dex) showing a kinematics typical of a slowly rotating spheroidal, with a velocity
dispersion not changing with longitude. These kinematical features for the three components are summarized in Figure \ref{chemokin},
where both the rotation curve and the velocity dispersion trend with respect to the Galactic longitude are shown
(different colors represent samples at different Galactic latitude).

Finally, \cite{zhao94, soto07, babu10} observed a vertex deviation in the velocity ellipsoid 
only for stars with [Fe/H]$>-0.5$ dex (see Figure \ref{vertex}), thus indicating that most of the bar 
population should be more metal-rich than this value.

\section{Bulge formation and evolution theories}



The old age of the bulge stars and the observed chemical patterns indicate that the bulge formed
early and rapidly, from a gas mainly enriched by ccSN (\citealt{matteucci90, matteucci99}).  
Chemical evolutionary models have been able to reproduce the observed chemistry by requiring
a formation timescale smaller than 1 Gyr (\citealt{ballero07, cescutti11}).
Such properties, together with the presence of a vertical abundance gradient, are naturally
accounted for by a dissipative collapse model (\citealt{eggen62}).

On the other hand the stellar kinematics of the bulge, which appears peanut/X-shaped and which is
cylindrically rotating, is consistent with a purely dynamical evolution of a disk buckling into
a bar (\citealt{shen10, saha12}). However, these authors argued that the presence of a bar would 
have spun up to cylindrical rotation any classical bulge component, thus making it kinematically indistinguishable.
\cite{mvp13} demonstrated that a bulge formed from the instability of the disk would show
a vertical abundance gradient if the original disk were characterized by a radial abundance gradient.

Another interesting hypothesis (see e.g. \citealt{immeli04, carollo07, elme08}) is that bulges form
at high redshift (z$\sim2-3$) via a combination of disk
instabilities and mergers of giant clumps (\citealt{elme08, dekel09}) on short dynamical timescales.
These giant clumps would form stars rapidly and with very high SFR (thus producing stars that today would appear
old and with the same chemistry as that observed in the bulge) and then would dynamically evolve
towards the center of the galaxy. There, they would merge together and their stars would develop 
the typical kinematics of a bar.

The only way to directly test this latter scenario is to find the relics of such massive clumps,
which may still be orbiting in the bulge of the host galaxy.
\citet[][hereafter F09]{f09} may have discovered the first remnant of one of these objects in our Galaxy: the bulge stellar 
system Terzan 5.

\section{Terzan 5}

\begin{figure}[!htb]
 \centering%
 \includegraphics[scale=1.1]{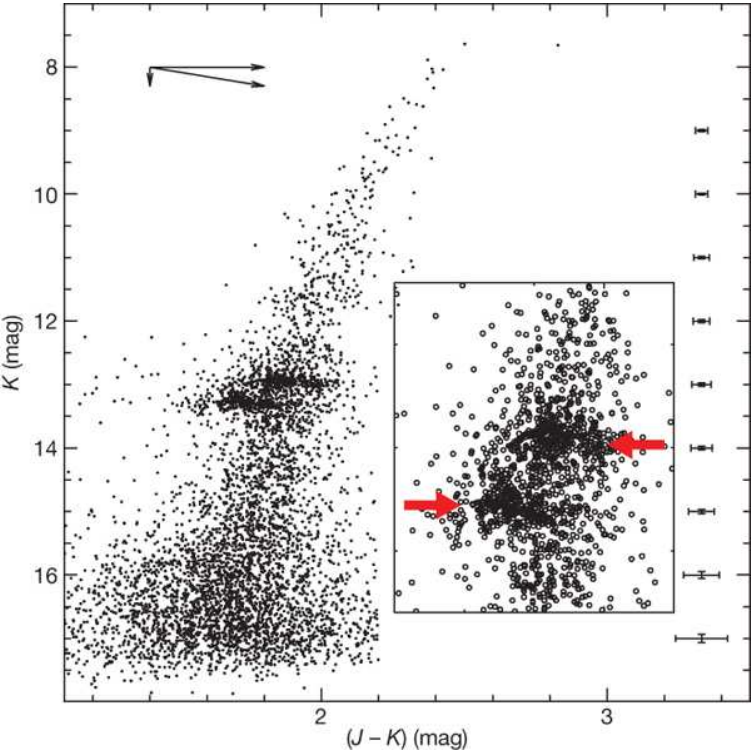}
 \caption{\small IR CMD of Terzan 5. The inset highlights the presence of the two RCs, well separated in color
and magnitude, in a direction perpendicular to that of the reddening vector (shown as a black arrow in the top-left 
corner of the Figure). Photometric errors are also marked on the right side of the plot.}
\label{f09cmd}
\end{figure}

Terzan 5 is a stellar system commonly catalogued as an old (\citealt{ortolani01}) globular cluster
(GC), located in the bulge of our Galaxy 
(its Galactic coordinates are $l=3.8395$\textdegree, $b=1.6868$\textdegree).
The distance (5.9 kpc, see also \citealt{ortolani07}) and reddening (E(B-V)$=2.38$,
see also \citealt{barbuy}) we adopt in this Thesis are from \cite{valenti07}\footnote{These values
have been obtained by using the method described in \cite{ferraro06} with the parameters defined
in \cite{valenti04}.}
F09 discovered the presence of two distinct sub-populations,
which define two RCs clearly separated in luminosity and color in the
$(K,J-K)$ CMD (Figure \ref{f09cmd}) obtained through observations taken with the
Multi-Conjugate Adaptive Optics Demonstrator (MAD) mounted at the Very Large Telescope (VLT).

\begin{figure}[!htb]
 \centering%
 \includegraphics[scale=1.4]{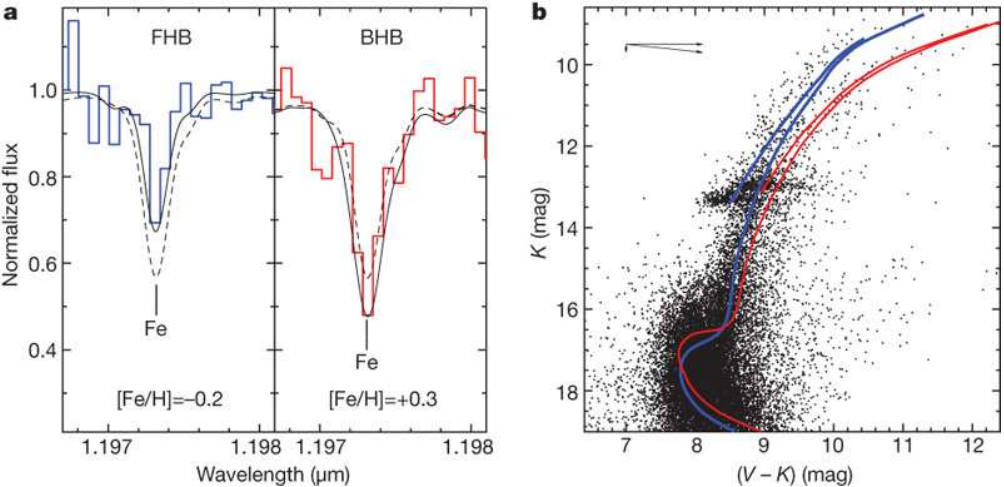}
 \caption{\small {\it Left panel:} Combined spectra measured for three faint RC (FHB) and three bright RC (BHB) stars in Terzan 5.
Solid lines correspond to the best-fit synthetic spectra computes at the marked metallicities ([Fe/H]$\simeq-0.2$ dex and
[Fe/H]$\simeq+0.3$ dex). The dashed lines show the spectra expected by assuming the reverse metallicities ([Fe/H]$\simeq0.3$ dex
for the FHB and [Fe/H]$\simeq-0.2$ dex for the BHB).
{\it Right panel:} isochrone fitting of the IR CMD of Terzan 5. The metal-poor component is best fitted with an old ($12$ Gyr) isochrone,
while the metal-rich population is well fitted by a younger isochrone ($6$ Gyr old). All the theoretical models are
taken from the BaSTI database (\citealt{pie04, pie06}).}
\label{f09iron}
\end{figure}

The analysis of high-resolution IR spectra obtained with NIRSPEC at the Keck II
telescope, demonstrated that the two populations have significantly
different iron content (see the left panel of Figure \ref{f09iron}): 
the bright RC (at $K = 12.85$) is
populated by a quite metal rich component ([Fe/H]$\simeq +0.3$),
while the faint clump (at $K = 13.15$) corresponds to a relatively metal
poor population at [Fe/H]$\simeq -0.2$. Before this discovery,
such a large difference in the iron content ($\Delta$ [Fe/H]$>0.5$
dex) was found only in $\omega$ Centauri, a GC-like system in
the Galactic halo, now believed
to be the remnant of a dwarf galaxy accreted by the Milky Way.

\begin{figure}[!htb]
 \includegraphics[scale=2.1]{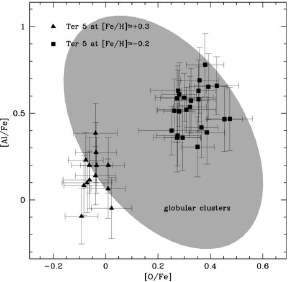}
 \includegraphics[scale=1.65]{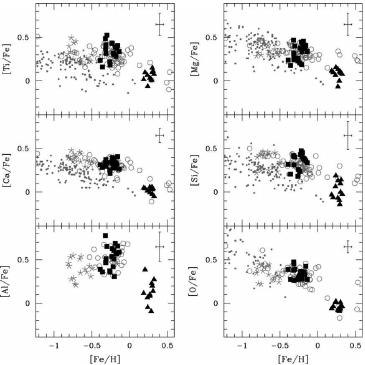}
 \caption{\small Spectroscopic screening of Terzan 5 giants from \cite{origlia}. {\it Left panel:}
the light elements measured in Terzan 5 stars (solid circles and triangles) clearly do not show 
the typical anti-correlation commonly observed in genuine GCs (the grey
shaded area). {\it Right panel:} all the measured $\alpha$-elements for Terzan 5 targets (black filled symbols) 
follow the same trend observed for bulge stars (grey symbols) in the [$\alpha$/Fe] vs [Fe/H] plot.}
\label{o11fig}
\end{figure}

\citet[][hereafter O11]{origlia} presented a detailed study of the abundance patterns of
Terzan 5, demonstrating that (1) the abundances of light elements
(like O, Mg, and Al) measured in both the sub-populations do not
follow the typical anti-correlations observed in genuine GCs (\citealt{carretta09, muccia09}, 
see left panel of Figure \ref{o11fig}); (2) the
overall iron abundance and the $\alpha-$enhancement of the metal poor
component demonstrate that it formed from a gas mainly enriched by
Type II supernovae (SNII) on a short timescale, while the progenitor
gas of the metal rich component was further polluted by SNIa on longer
timescales; (3) these chemical patterns are strikingly similar to
those measured in the bulge field stars (see the right panel of Figure \ref{o11fig}), 
with the $\alpha$-elements being enhanced up to solar metallicity and then
progressively decreasing towards solar values (see Section \ref{chem}).

There is also another interesting chemical
similarity between Terzan~5 and the bulge stellar population. In fact, as already
described in Section \ref{chem}, the latter
shows a metallicity distribution with two major peaks at sub-Solar and
super-Solar [Fe/H], very similar to the metallicities of the two populations 
discovered in Terzan 5. Chemical abundances of bulge dwarf stars
from microlensing experiments \citep[see e.g.][and references
therein]{coh10,bensby13} also suggest the presence of two populations,
a sub-Solar and old one with [$\alpha$/Fe] enhancement, and a possibly
younger, more metal-rich one with decreasing [$\alpha$/Fe] enhancement
for increasing [Fe/H].

Each Terzan 5 component shows a small internal metallicity spread and the most
metal-rich population is also more centrally concentrated (F09, \citealt[][hereafter L10]{l10}).

All these observational results demonstrate that Terzan 5 is not a genuine
GC, but a stellar system with a more complex star formation and
chemical enrichment history. Indeed it is likely to have been much more
massive in the past than today (with a mass of at least a few
$10^7-10^8 M_\odot$, while its current value is $\sim 10^6
M_\odot$; L10), thus to retain the high-velocity gas ejected
by violent SN explosions. Moreover, it likely formed and 
evolved in strict connection with its present-day
environment (the bulge), thus suggesting {\it the possibility
that it is the relic of one of the pristine fragments that contributed
to form the Galactic bulge itself}.  In this context, also the
extraordinary population of millisecond pulsars (MSPs) observed in
Terzan 5 can find a natural explanation. 
In fact, the system hosts 34 MSPs. This amounts to $\sim25\%$ of the
entire sample of MSPs known to date in Galactic GCs \citep[][; see
the updated list at {\tt www.naic.edu/$\sim$pfreire/GCpsr.html}]{ransom}.
In order to account for the observed chemical abundance patterns, a large number of SNII 
is required. These SNII are expected to have produced
a large population of neutron stars, mostly retained by the deep
potential well of the massive {\it proto}-Terzan 5.  The large
collisional rate of this system (Verbunt \& Hut, L10) may also have favored
the formation of binary systems containing neutron stars and promoted
the re-cycling process responsible for the production of the large MSP
population now observed in Terzan 5.

\clearpage{\pagestyle{empty}\cleardoublepage}

\chapter{{HST} Relative Proper Motions of Terzan 5}\label{chaptermoti}

As already discussed in Chapter \ref{chapintro}, the analysis of the evolutionary
sequences in the optical CMD of Terzan 5 is extremely difficult because of
the large differential extinction and the strong contamination
of the underlying bulge population and foreground sources.

In this Chapter the issue of field contamination is addressed.
In general, the most efficient way to decontaminate CMDs from
non-member stars is the determination of accurate stellar proper
motions (PMs). To this aim, we analyzed two epochs of high-resolution {\it Hubble
Space Telescope (HST)}
images obtaining relative PMs (i.e. PMs of stars in the {\it HST} field of view 
with respect to the average motion of Terzan 5) for more than $100,000$ stars
reaching {\it m}$_{{\rm F606W}}\simeq28$, i.e. about 3 magnitudes below the MSTO.
This allowed us to define a method to reliably select cluster member stars
and discard foreground and background sources.  

Two additional applications of the technique used in this Thesis 
are also presented in the Appendix. 

\section{Observations}\label{datamoti}

In order to measure the PMs in the direction of Terzan~5 we used two
{\it HST} high-resolution data sets acquired with the Wide Field Channel (WFC) 
of the Advanced Camera for Survey (ACS). The WFC/ACS is made up of two 
$2048 \times 4096$ pixel detectors with a pixel scale of $\sim0.05
\arcsec\,$pixel$^{-1}$ and separated by a gap of about 50 pixels, for
a total field of view (FoV) of $\sim200\arcsec \times 200\arcsec$.
The data set used as first epoch was obtained under
GO-9799 (PI:\ Rich). It consists of two deep exposure images, one
in the F606W filter and the other in the F814W filter (with exposure times of $340$ s), 
and one short exposure ($10$ s) image in the F814W filter, 
taken on September 9, 2003.  

The second-epoch data set is composed of data obtained
through GO-12933 (PI:\ Ferraro). This program consists of several
deep images taken both with the WFC/ACS in the F606W and F814W
filters, and with the IR channel of the Wide
Field Camera 3 (WFC3) in the F110W and F160W filters.  The WFC3
IR camera is made of a single $1024 \times 1024$ pixel detector.  
Its pixel scale is $\sim0.13
\arcsec\,$pixel$^{-1}$ and the total FoV is $123\arcsec \times
136\arcsec$.  Because of the larger pixel size of the WFC3 IR detector
(it is almost three time the size of the WFC/ACS pixel) and
given that the Full Width Half Maximum (FWHM) of the Point Spread Function (PSF)
in the IR is larger than in the optical bands,
we used only the WFC/ACS images for the PM determination. 
The sample used consists of $5\times365\,$s images in F606W and $5\times365\,$s images in
F814W, with one short exposure image per filter ($50$ s and $10$ s, respectively). 
These observations were taken on August 18, 2013, therefore the two available data
sets provide a temporal baseline of $\sim 9.927$ yrs.

\section{Relative Proper Motions}\label{motirel}

The techniques applied in the present work have been developed in the
context of the HSTPROMO collaboration (\citealt{hstpromo, bellini14}\footnote{For details see
HSTPROMO home page at http://www.stsci.edu/~marel/hstpromo.html}), of which
I am currently member. HSTPROMO
aims at improving our understanding of the dynamical evolution
of stars, stellar clusters and galaxies in the nearby Universe through
measurement and interpretation of proper motions.

The analysis has been performed on \_FLC images, which have been flat-fielded,
bias-subtracted and corrected for Charge Transfer Efficiency (CTE) losses by 
the pre-reduction pipeline with the pixel-based correction 
described in \cite{jaybedin10} and \cite{ubedajay}.
The main data-reduction procedures we used are described in detail in
\cite{jayking06}. Here we provide only a brief description of the main
steps of the analysis.  The first step consists in the photometric reduction
of each individual exposure of the two epochs with the publicly available program
\texttt{img2xym$\_$WFC.09$\times$10}.  This program uses a pre-determined model of
spatially varying PSFs plus a single time-dependent perturbation PSF
(to account for focus changes or spacecraft breathing). The final output
is a catalog with instrumental positions and magnitudes for a sample of
sources above a given flux threshold in each exposure.
Star positions were then corrected in each catalog for
geometric distortion, by means of the solution provided by \cite{jayacs}.

\begin{figure}[!htb]
 \centering%
 \includegraphics[scale=0.6]{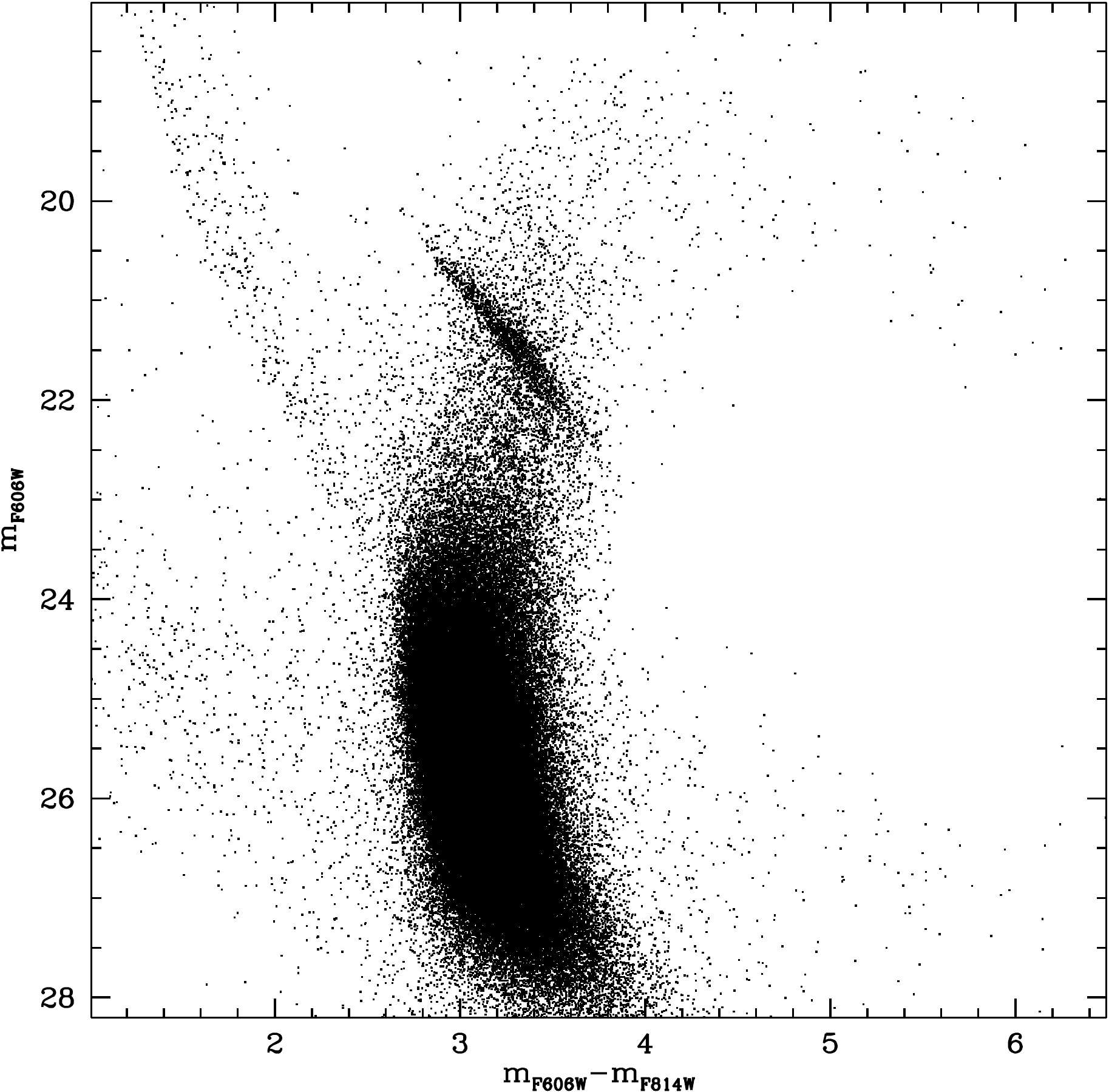}
 \caption{\small ({\it m}$_{{\rm F606W}}$, {\it m}$_{{\rm F606W}}-${\it m}$_{{\rm F814W}}$) 
CMD of Terzan 5. All the cluster evolutionary sequences are distorted because of differential
reddening effect. A bright, blue sequence is clearly separated from the cluster sequences and
it is likely traced by young field stars.}
\label{vvimoti}
\end{figure}

To check the quality of our photometry, we built the WFC/ACS ({\it m}$_{{\rm F606W}}$,
{\it m}$_{{\rm F606W}}-${\it m}$_{{\rm F814W}}$) CMD of Terzan~5.  The
F606W and the F814W samples were constructed by selecting stars in common 
among at least 3 out of 5 deep-single-exposure catalogs.
The CMD resulting from these two samples is shown in Figure
\ref{vvimoti}. The instrumental magnitudes have been calibrated onto the
VEGAMAG system using aperture corrections and zeropoints reported
in the WFC3 web
page\footnote{http://www.stsci.edu/hst/wfc3/phot\_zp\_lbn.}.
As it is evident from Figure \ref{vvimoti},
the evolutionary sequences of Terzan 5 are strongly affected
by differential reddening, however they can be identified well in the
CMD obtained.  
The MS extends for almost $4$ magnitudes below the TO. 
A blue sequence is visible at {\it m}$_{{\rm
F606W}}<23$ mag and ({\it m}$_{{\rm F606W}}-${\it m}$_{{\rm
F814W}})<2.6$ mag and it remains well separated from the cluster
RGB. This sequence is likely populated by young field stars.

The second step in measuring relative PMs is to astrometrically relate
each exposure to a distortion-free reference frame, which from now on
we will refer to as the {\it master frame}. 
Since no high-resolution photometry other than that coming from
these data sets is available, we defined as master frame the catalog obtained
from the combination of all the second-epoch single-exposure catalogs
corrected for geometric distortions. In this way, the master frame is composed
only of stars with at least $10$ position measurements (5 for each filter).
We then applied a counter-clockwise rotation of $88.84$\textdegree~
in order to give to the master frame the same orientation as the absolute
reference frame, here defined by the Two Micron All Sky Survey (2MASS) catalog (see L10).

We then transformed the measured position of each star in each exposure into 
the master frame by means of a six-parameter linear transformation 
based on the positions of several hundreds reference stars. 
Such reference stars are the stars with respect to which our PMs would be computed.
For convenience, we chose to compute all PMs relative to the mean motion of the
cluster. Therefore our reference list is composed of stars which are likely cluster
members. These are initially selected on the basis of their
location on the CMD, including in the list only well-measured and
unsaturated stars. Then, for each star in each catalog, we computed
the position on the master frame using a transformation based on only 
the closest 50 reference stars.
To maximize the accuracy of these transformations we
treated each chip of our exposures separately, in order to avoid spurious
effects related to the presence of the gap.

\begin{figure}[!htb]
 \centering%
 \includegraphics[scale=0.45]{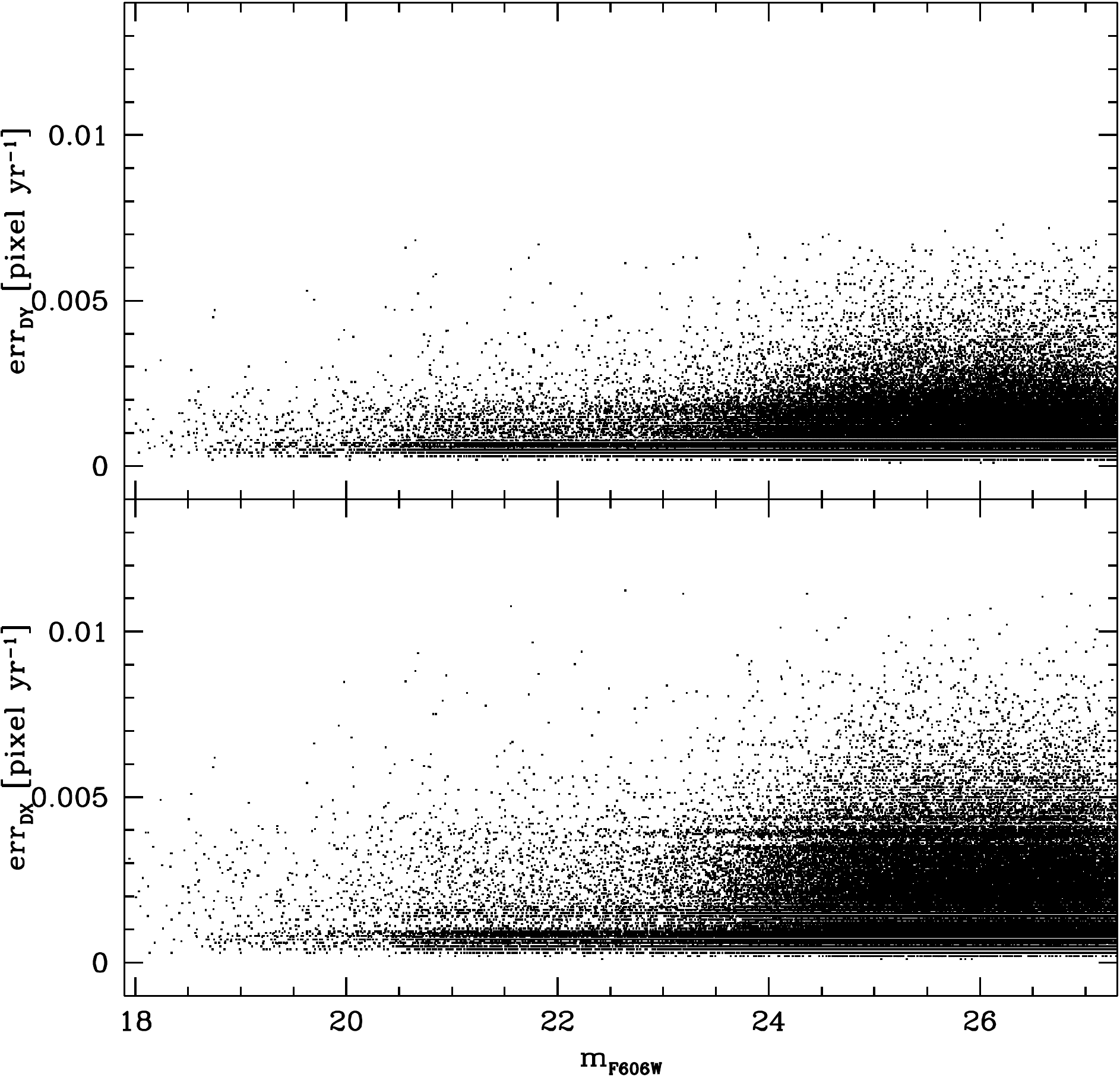}
 \caption{\small Uncertainties in the X and Y displacements in units of pixel yr$^{-1}$.
Bright, well measured stars have displacement errors typically smaller than 0.002
pixel yr$^{-1}$ in each coordinate, i.e. smaller than 0.1 \masyr.}
\label{errmoti}
\end{figure}

At the end of the procedure, for each star we have up to $3$ first-epoch
position measurements and up to $12$ second-epoch positions on the master frame. 
However, stars brighter than {\it m}$_{{\rm
F606W}}\simeq17.9$ saturate in the long exposures.
Therefore, for the brightest stars we have only $1$ first-epoch and $2$
second epoch positions. To estimate the relative motion of each star we adopted a
3$\sigma$-clipping algorithm and computed the median X and Y positions of
each star in the first and in the second epoch. The difference between the two 
median positions gives the star's X and Y displacements in $\Delta {\rm T}=9.927$ years.  
To determine the displacements of bright stars with only one or two positional measurements, 
we adopted either the single or the mean X and Y position values, respectively.
The errors in each direction and within each epoch ($\sigma_{1,2}^{{\rm X,Y}}$) 
were computed as:
\begin{equation}
{\rm rms_{1,2}^{{\rm X,Y}}}/\sqrt{N_{1,2}}
\end{equation}
where rms$_{1,2}$ is the rms of the positional residuals about the median value, 
and N$_{1,2}$ is the number of measurements. Therefore, the error in each PM-component 
associated to each star is simply the sum in quadrature between first- and
second-epoch errors:\ $\sigma_{\rm
  PM}^{{\rm X}}=\sqrt{(\sigma_1^{{\rm X}})^{2}+(\sigma_2^{{\rm X}})^{2}}/\Delta {\rm T}$
and $\sigma_{\rm
  PM}^{{\rm Y}}=\sqrt{(\sigma_1^{{\rm Y}})^{2}+(\sigma_2^{{\rm Y}})^{2}}/\Delta {\rm T}$.
The error associated to the PM of the brightest stars measured only in the short
exposures were computed by adopting as positional uncertainties the typical errors
determined in the long exposure catalogs at the same {\it instrumental} magnitude.

\begin{figure}[!htb]
 \centering%
 \includegraphics[scale=0.5]{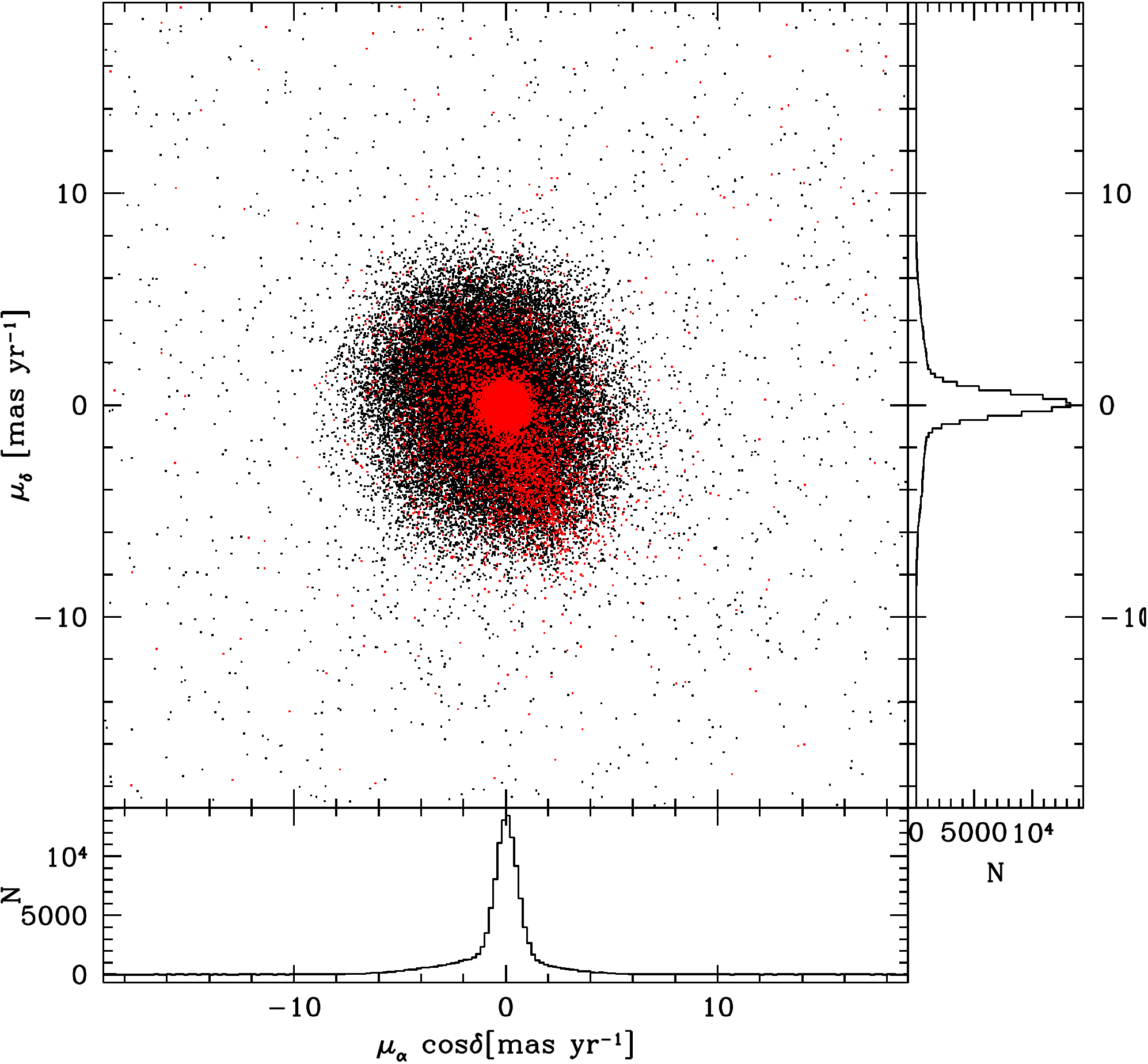}
 \caption{\small Vector Point Diagram (VPD) of the PMs measured for the $127,172$
stars (black dots) in the direction of Terzan 5. Their distributions in the
Right Ascension and Declination axis are shown in the histograms in the bottom and
right panel, respectively. PMs measured for stars with m$_{F606W}<24$ mag are shown
as red dots. At least two components are visible: the first showing a symmetric
distribution centered around the origin (corresponding to the bulk of Terzan 5
member stars) and an asymmetric structure roughly centered at (2.5, -5) \masyr.}
\label{vpdmoti}
\end{figure}

By selecting stars on the basis of this first PM determination,
we repeated the entire procedure the number of times needed to make the
number of stars in the reference list stable, i.e. with variations smaller than
2-3\%.  To be conservative, for all
the unsaturated stars in the deep exposures we
decided to build the final PM catalog taking into account only the
123$\,$172 stars having at least 2 position measurements in each
epoch. The typical error as a function of magnitude is shown in Figure ~\ref{errmoti}.
For well-exposed stars it is smaller than 0.002
pixel yr$^{-1}$ in each coordinate, i.e. smaller than 0.1 \masyr. 
Faint stars or stars with only few epochs
measurements show larger errors, but always smaller than 0.01 pixel yr$^{-1}$.

\begin{figure}[!htb]
 \centering%
 \includegraphics[scale=0.5]{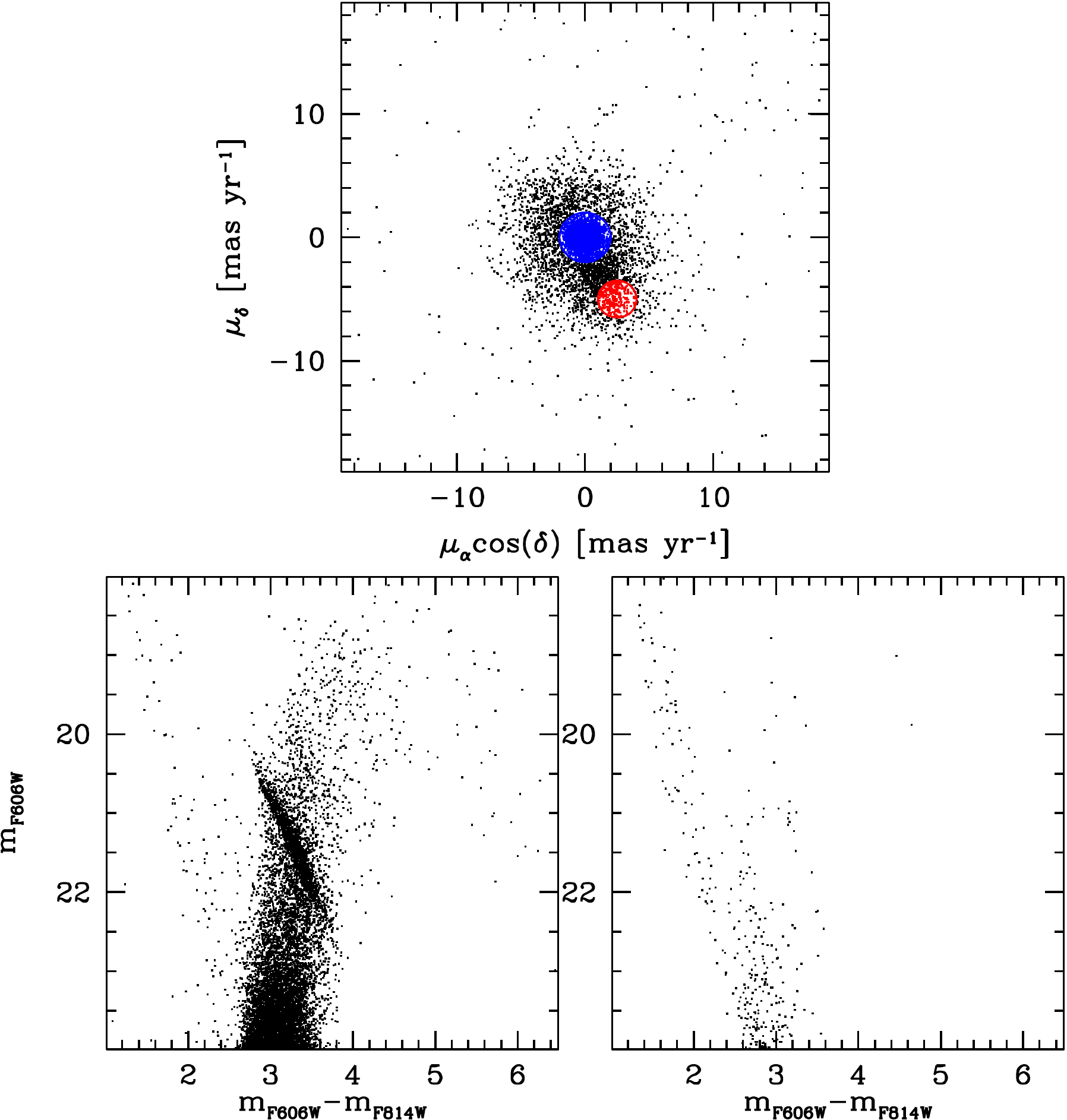}
 \caption{\small {\it Upper panel}: VPD of the stars brighter than m$_{F606W}=24$ mag. Stars belonging
to the symmetric component of likely members are plotted in blue, while those belonging to the
asymmetric component centered at (2.5, -5) \masyr are plotted in red. {\it Left-lower panel}: CMD described
by VPD-selected likely member stars. with the exception of few contaminating stars, only the cluster
evolutionary sequences are visible. {\it Right-lower panel}: CMD described by stars belonging to the VPD
asymmetric feature. This population is dominated by young foreground disk stars along the blue plum.}
\label{two_moti}
\end{figure}

We converted the PMs into absolute units  (\masyr) by multiplying the measured
displacements by the pixel scale of the master frame ($0.05 \arcsec/$pixel) 
and dividing by the temporal
baseline ($9.927$ yr). Since the master frame is already oriented
according to the equatorial coordinate system, the X PM-component
corresponds to that projected along (negative) Right Ascension
($-\mu_{\alpha}\cos\delta$), while the Y PM-component corresponds to that along
Declination ($\mu_{\delta}$).  The output of this analysis is
summarized in Figure \ref{vpdmoti}, where we show
the Vector Point Diagram (VPD) for all the stars with a measured PM.

By selecting stars with {\it m}$_{{\rm F606W}}<24$ (red points in Figure \ref{vpdmoti}), 
which typically have the most accurate PMs (red points in Figure \ref{errmoti}), the VPD clearly shows 
at least two components. One is a symmetric distribution centered around the
origin, corresponding to Terzan 5 
member stars. The bulk of this population is confined within a circle of radius 2\masyr.
The other is an asymmetric structure approximately centered around the coordinate
(2.5, -5) \masyr in the VPD.
The location of these two components in the CMD clearly reveal their nature (see Figure \ref{two_moti}).
In fact, while the stars of the first component (shown as blue dots in the VPD) describe in the CMD
the evolutionary sequences of the cluster (left-lower panel), the stars belonging to the
asymmetric component (red dots) correspond in the CMD to the blue plum (right-lower panel) essentially 
populated by young disk stars in the foreground of Terzan 5.

\begin{figure}[!htb]
 \centering%
 \includegraphics[scale=0.45]{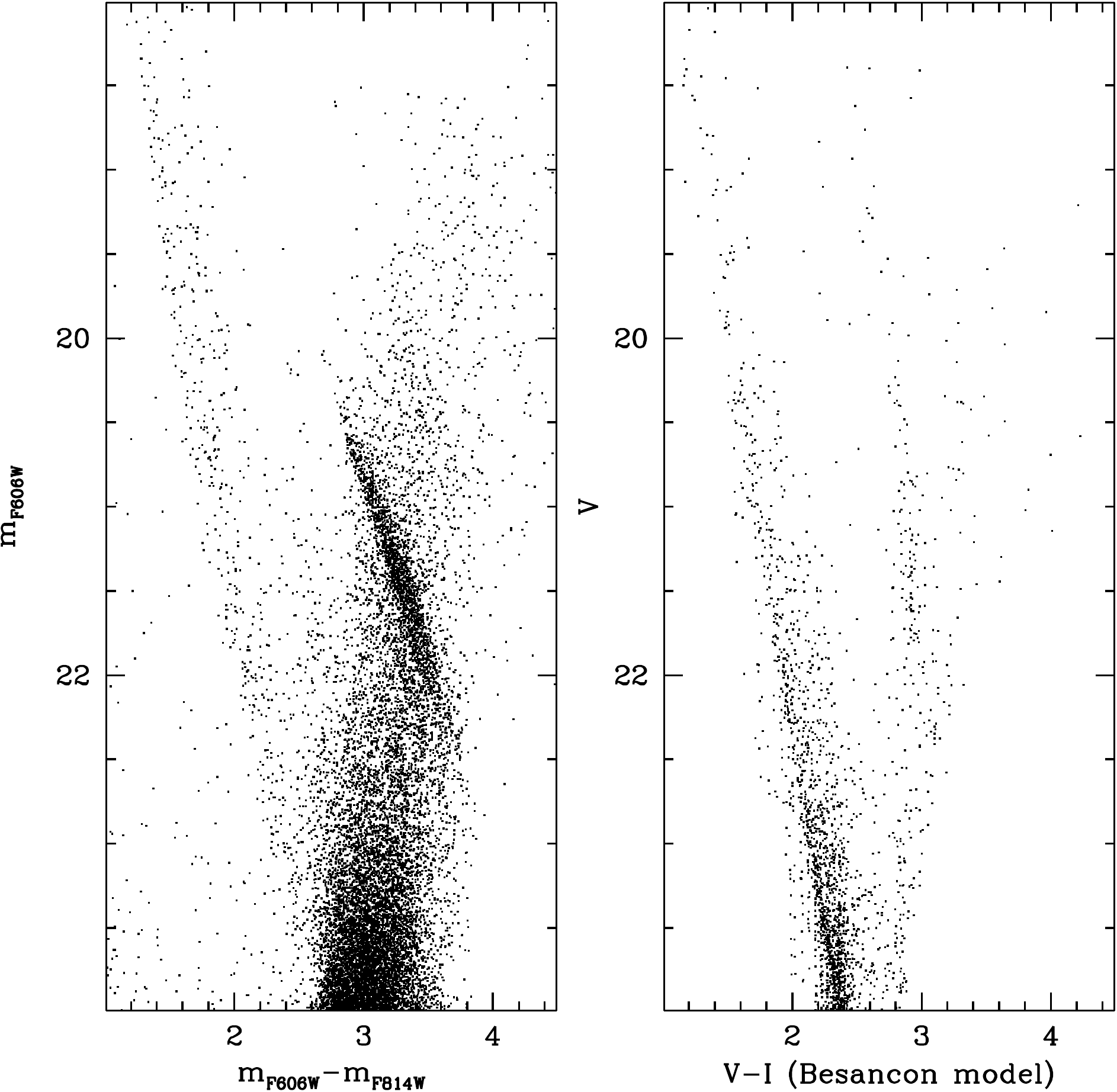}
 \caption{\small Comparison between the ACS optical CMD and that predicted by a simulation of
the Besan\c{c}on model including only Galactic disk stars younger than 7 Gyr. Such a comparison
clearly demonstrates that the bluer sequence in the CMD of Terzan 5, already identified as composed
of field stars by the PMs, correspond to the MS of foreground Disk stars. }
\label{vvibes}
\end{figure}

Such a conclusion is confirmed by the comparison with the prediction of the Besan\c{c}on model (\citealt{robin}) 
for a field centered at the coordinates of Terzan 5 and having the same size as that of the WFC/ACS and with
only young (t$_{age}<7$ Gyr) Galactic disk stars (see Figure \ref{vvibes}).

\section{Analysis of the PM-selected CMD}

In this Section we analyze the CMDs obtained with different data sets and filters after an appropriate
decontamination from non-member stars performed on the basis of the measured PMs.

\subsection{Optical CMD}\label{motiottici}

\begin{figure}[!htb]
 \centering%
 \includegraphics[scale=0.5]{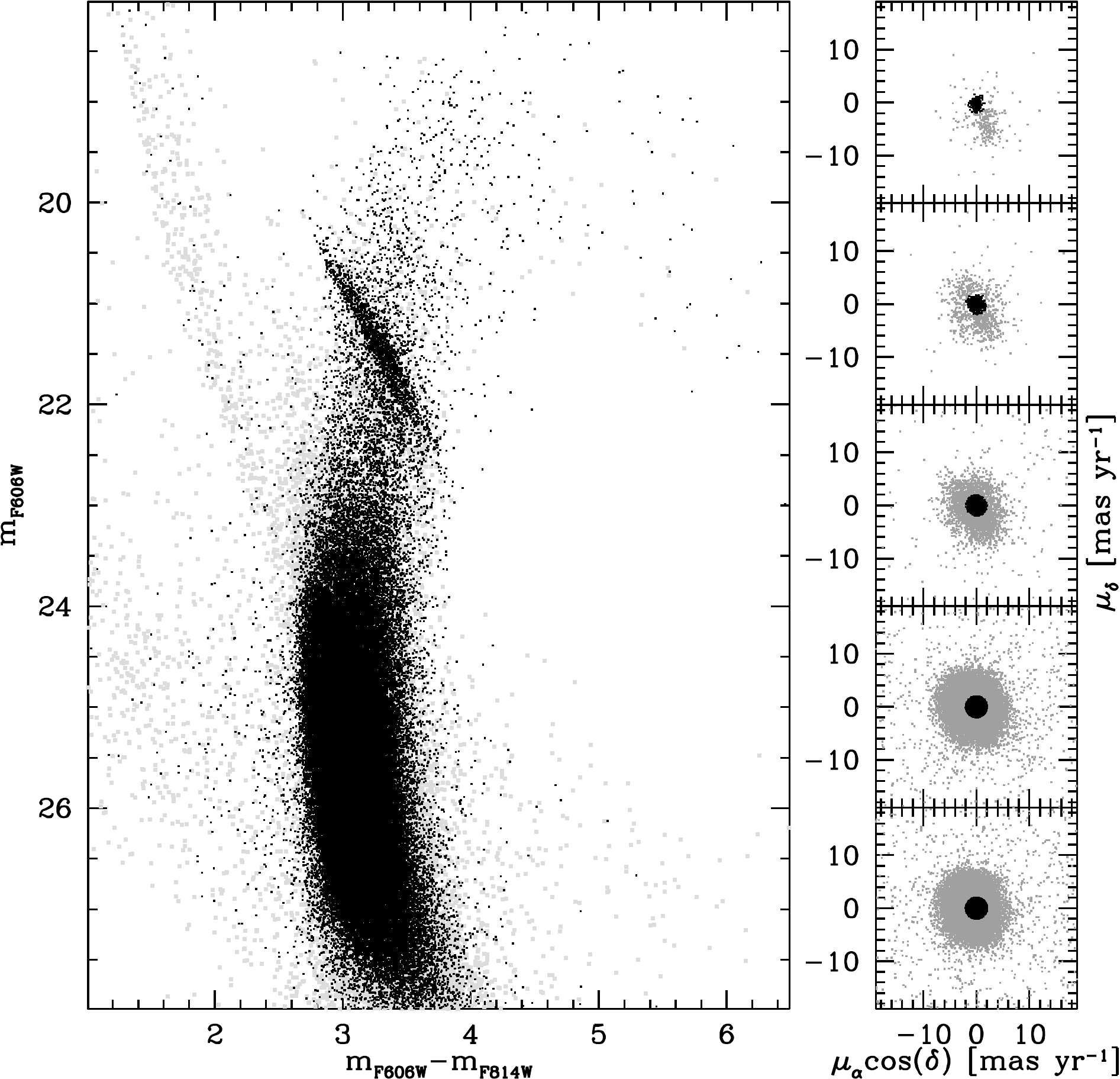}
 \caption{\small {\it Left panel:} optical CMD of Terzan 5. Stars selected as likely members
are shown as black dots, while sources excluded by the selection are plotted in grey. 
The measured relative PMs are efficient in decontaminating the CMD and only
cluster evolutionary sequences survive the selection criterium. {\it Right panels:} magnitude-binned
VPDs for all the stars in the optical catalogue. Each bin has a size of 2 mag. Sources are
color coded as in the left panel, and the red circles mark the selection of asymmetric feature.}
\label{vviclean}
\end{figure}

The first analysis is performed on the optical ({\it m}$_{{\rm F606W}}$, 
{\it m}$_{{\rm F606W}}-${\it m}$_{{\rm F814W}}$) CMD of Terzan 5 obtained with the 
WFC/ACS dataset described in Section \ref{datamoti}.
Since, as shown in Figure \ref{vpdmoti}, the bulk of the PMs in the VPD is concentrated 
within a circle of radius $2$ \masyr and we expect it to be dominated by likely 
member stars, we adopted this as member selection criterium. 
We therefore considered all the sources lying outside such a circle as non-member stars.
The CMD obtained from such a selection is shown in Figure \ref{vviclean}.

In the Figure, member stars are plotted as black dots, while non members stars are shown
in grey. The selection applied leaves in the CMD only stars
clearly belonging to the cluster evolutionary sequences while excluding most of the
outliers. A small degree of contamination is still present, probably because the distribution
of field stars (mainly bulge stars) in the VPD overlaps that of Terzan 5 members.  However, we can conclude that
the PMs analysis performed in this Thesis is efficient in decontaminating the CMD from foreground and
background sources in Terzan 5. 


Since the method described to identify and reject foreground and background stellar sources in the direction
of Terzan 5 is reliable and works well, in the following we will analyze the IR CMD
obtained in F09 by means of observations taken with the MAD camera in order to check whether all 
the properties discovered in that CMD hold after the PM decontamination.

\subsection{MAD Infrared CMD}

With the aim of studying the large population of MSPs hosted by Terzan 5, F09 exploited
the great capability of the multi-conjugate adaptive optics system MAD mounted at the VLT.
The obtained (K, J-K) IR CMD revealed the presence of two well separated RCs (see Chapter 1). 
Since their separation in both color and magnitude is perpendicular with
respect to the direction of the reddening vector, it has been interpreted as a genuine feature due
to the presence of two sub-populations with different
properties in terms of age, metallicity or helium content. A prompt spectroscopic follow-up
demonstrated that the two populations have at least different metallicities, with stars belonging
to the faint RC having [Fe/H]$\simeq-0.3$ dex and those belonging to the bright RC with [Fe/H]$\simeq+0.3$.

This spectroscopic follow-up also revealed that the samples of stars belonging to the two RCs have the same
radial velocity, corresponding to the systemic velocity of Terzan 5. Moreover, the photometric study
performed in L10 showed that the two RCs also share the same center of gravity. Finally,
the bright RC was also found to be more centrally concentrated than the faint RC. All these findings
clearly suggest that the two populations are members of Terzan 5 and exclude that they are the result
of a superposition in the sky (in this case the bright RC should be the closer and thus less concentrated
population, at odds with what is observed). 

\begin{figure}[!htb]
 \centering%
 \includegraphics[scale=0.5]{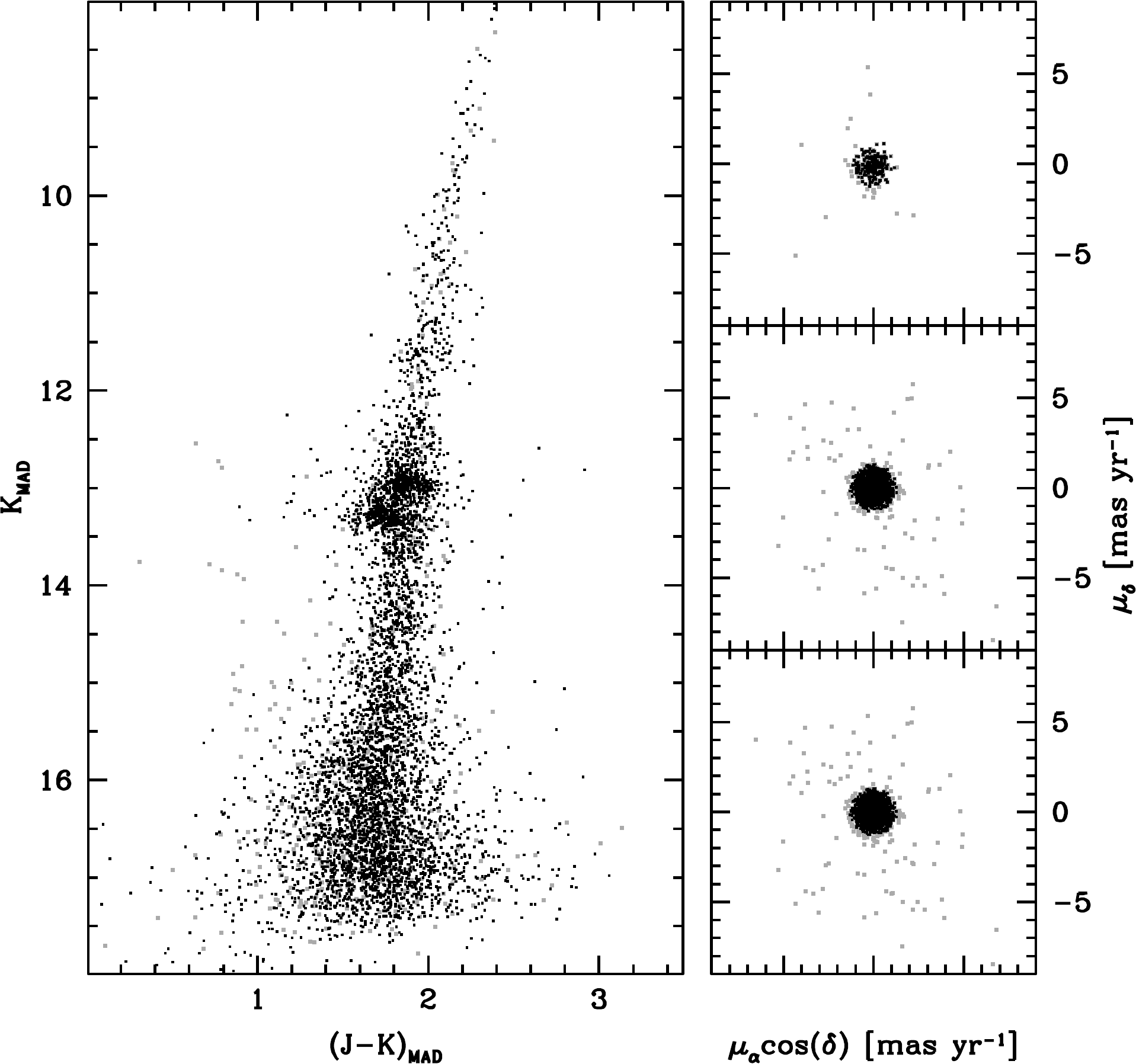}
 \caption{\small {\it Left panel:} IR CMD of Terzan 5 obtained from MAD observations. The PM-selected
member stars are shown as black dots, while the sources excluded as grey dots. Despite the stringent
selection, the presence of the two RC is evident. {\it Right panels:} magnitude-binned VPDs (each magnitude 
bin has a size of 3 mag). Stars are color-coded as in the left panel.}
\label{kjkclean}
\end{figure}

Despite this evidence, the membership of the two populations has been questioned in some
works (see for example \citealt{willman}). To solve this issue, our PM analysis is applied 
to the IR CMD. The result is summarized in Figure \ref{kjkclean}. 

The left panel shows the IR CMD
of Terzan 5 after the decontamination (black dots), with the sources excluded by the PM selection
shown as grey dots. Such a selection is even more stringent in this case than that adopted in 
Section \ref{motiottici}. In fact, as shown in the magnitude-binned VPDs in the right panels, 
we excluded all the stars with a total PM vector 
$\mu=\sqrt{(\mu_{\alpha}\cos\delta)^{2}+(\mu_{\delta})^{2}}>1.3$ \masyr.
The overall smaller dispersion should not be interpreted as an intrinsic feature but rather as
due to the fact that the IR CMD is less deep than the optical one and faint stars with more uncertain PM
are not present.
Despite such a tighter selection, the decontaminated CMD clearly exhibits the two well separated RCs.

\begin{figure}[!htb]
 \centering%
 \includegraphics[scale=0.5]{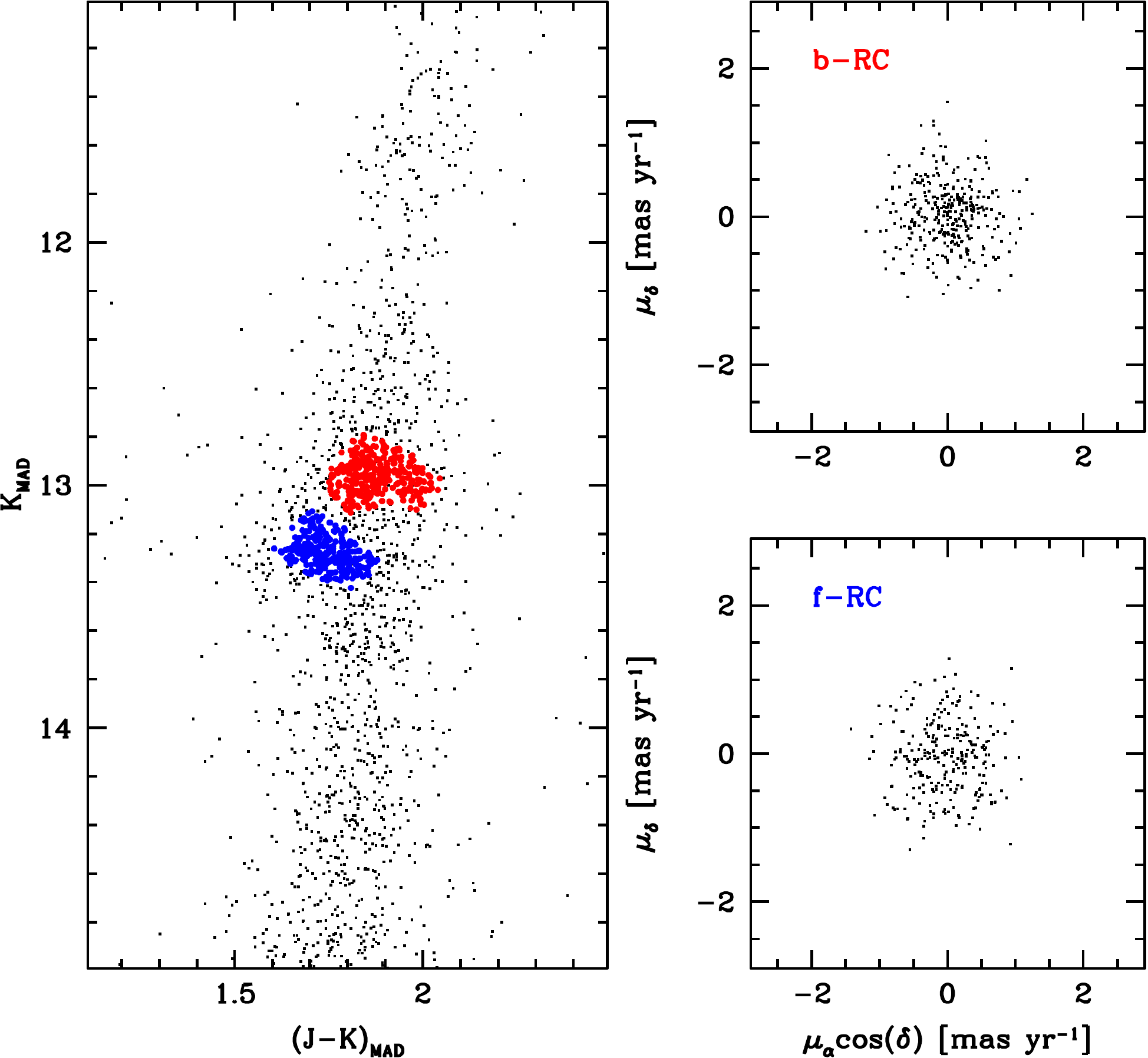}
 \caption{\small {\it Left panel:} IR-MAD CMD of Terzan 5 zoomed in the RC region. Faint-RC stars are plotted
in blue, while bright-RC stars are plotted in red. {\it Right panels:} VPDs of b-RC (upper panel) and f-RC
(lower panel) stars: in both cases stars lie within the 2\masyr circle adopted as membership selection
criterium. No clearcut differences between the two distributions are visible.}
\label{vpd2pops}
\end{figure}

Figure \ref{vpd2pops} shows the VPDs of stars properly selected in the two RCs.
As can be seen, the two distributions appear quite symmetric, both showing a small (1.3\masyr)
dispersion around the origin.
However we underline that the accuracy of these PMs is not sufficient to reveal
possible intrinsic differences in the kinematics of the two populations. At this stage, we can only
conclude that they are not distinguishable in terms of cluster membership, being both well within
the adopted VPD-members selection criterium. 

Overall the number of contaminants in the IR CMD is much smaller than what observed in the optical case.
However this is somehow expected, given that this photometry comes from a smaller FoV in the very
central region of the system, where the cluster population is supposed to dominate.

Therefore, the relative PMs measured in this Chapter
and applied to the MAD CMD definitely demonstrate that {\it both the populations
discovered in Terzan 5 are clearly member of the system}.

\section{Conclusions}

In this Chapter we addressed the issue of field contamination in the photometric analysis of Terzan 5.
We performed a relative PM analysis to separate fore- and background stars from those belonging
to the system. In particular, we used the approach developed in the context of the HSTPROMO collaboration to
measure relative PMs for 123$\,$172 optical sources found in the direction of Terzan 5.

The resulting VPD shows a dominant component distributed symmetrically around the origin of
the diagram and extending out to about $2$ \masyr, and two minor and sparser components, one
of them being clearly asymmetric and preferentially located in the fourth quadrant.

From the analysis of the optical CMD, we demonstrated that the measured PMs are efficient
in decontaminating the photometry from non-member populations, and by comparing our photometry
with that predicted by the Besan\c{c}on model at the Terzan 5 coordinates we showed that the
asymmetric feature observed in the VPD is likely to be populated by young Galactic disk stars.

Finally, the analysis of the PM-selected IR CMD showed that the two populations discovered
in Terzan 5 are present in the CMD even after a stringent decontamination. Such an evidence,
coupled with the features already discovered in previous works, 
demonstrates that both the populations in Terzan 5 are genuine members of the system.

\clearpage{\pagestyle{empty}\cleardoublepage}

\chapter{High resolution reddening map in the direction of the stellar system Terzan 5}\label{chapred}

Severe limitations to a detailed analysis of the
evolutionary sequences in the CMDs of Terzan 5 are introduced by the
presence of large differential reddening. To face this problem 
we build the highest-resolution extinction map ever constructed in the
direction of this system. This is the subject of the present Chapter.

In particular, we used optical images acquired with the {\it HST}
to construct an extinction map in the direction
of Terzan 5 which has a spatial resolution of $8\arcsec \times
8\arcsec$, over a total FoV of $200\arcsec\times
200\arcsec$. The absorption clouds show a patchy structure on a
typical scale of $20\arcsec$ and extinction variations as large as
$\delta E(B-V)\sim 0.67$ mag.  These correspond to an absolute color
excess ranging from $E(B-V)=2.15$ mag, up to 2.82 mag.  After the correction
for differential reddening, two distinct red giant branches become
clearly visible in the optical color magnitude diagram of Terzan 5 and we
verified that they well correspond to the two sub-populations with different iron
abundances recently discovered in this system.

All the details of this study are described in \cite{massari}.

\section{Differential reddening correction} 
\label{obsdata}

\subsection{The data-set}
\begin{figure}[!htp]
  \centering
  \includegraphics[scale=0.5]{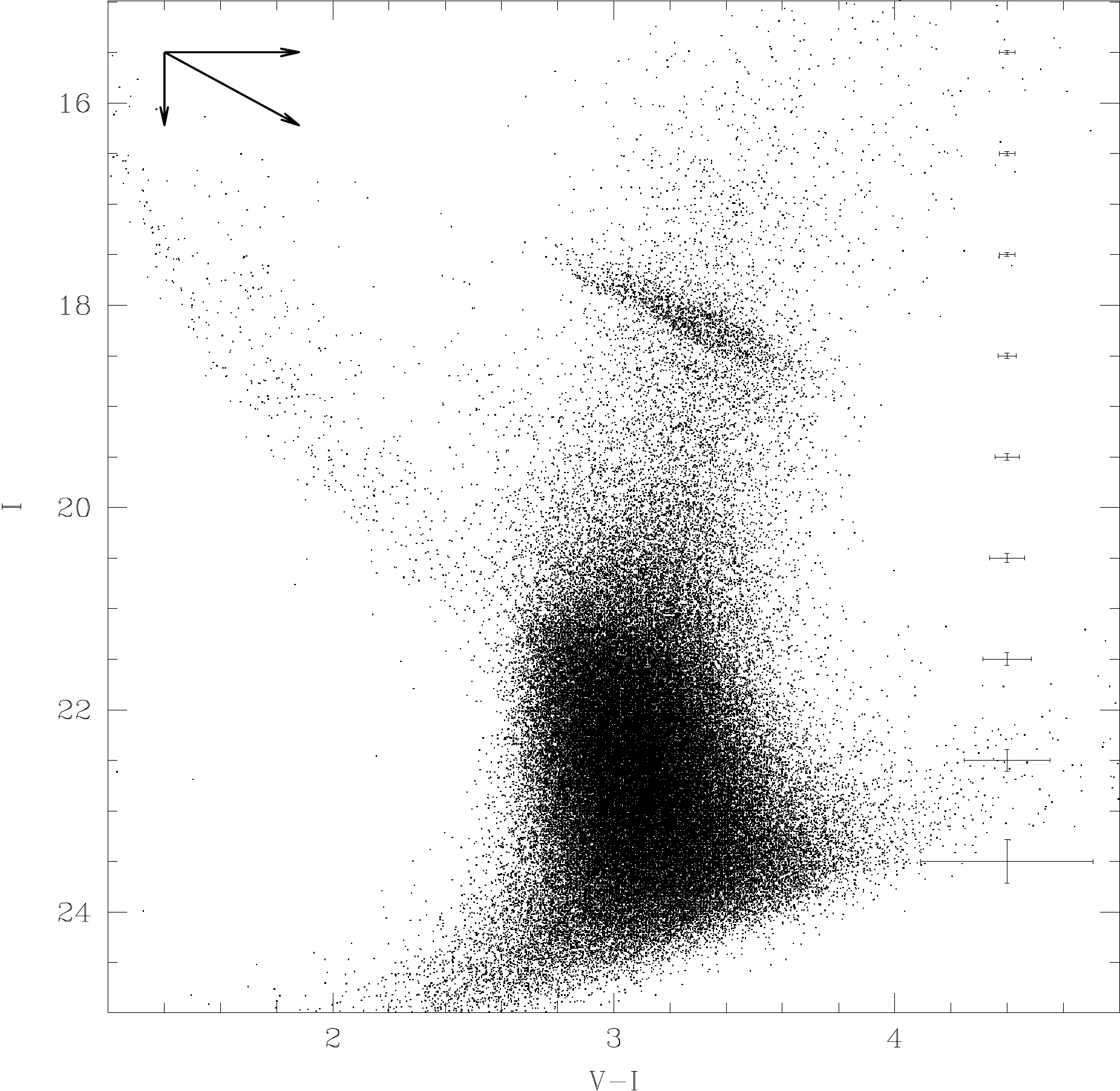}
  \caption{\small $(I,V-I)$ CMD of the $\sim127,000$ stars measured
    in the ACS FoV. The photometric errors at different magnitude levels 
    are shown.  
    Note how the distortions of the
    evolutionary sequences (in particular the RCs) 
    follow the reddening vector, shown in the
    upper left corner.}
\label{red}
\end{figure}

The photometric data used in this work consist of a set of
high-resolution images obtained with the WFC of
the ACS on board the {\it HST} (GO-9799, see F09 and L10).  
The WFC/ACS camera has a FoV of $\sim200\arcsec \times 200\arcsec$ with a plate-scale of
  0.05\arcsec/pixel. Both F606W (hereafter $V$) and F814W ($I$)
  magnitudes are available for a sample of about 127,000 stars. The
  magnitudes were calibrated on the VEGAMAG photometric system by
  using the prescriptions and zero points by \cite{sirianni}.  The
  final catalog was placed onto the 2MASS
  absolute astrometric system by following the standard procedure
  discussed in previous works (e.g., L10).  The $(I,V-I)$ CMD
  shown in Figure \ref{red} clearly demonstrates the difficulty of
  studying the evolutionary sequences in the optical plane,
  because of the broadening and distortion induced by differential
  reddening. In particular, the RGB is anomalously wide ($\Delta (V-I)\sim0.8$ mag)
  and the two RCs appear highly stretched along the reddening
  vector.

\subsection{The method}
The method here adopted to compute the differential reddening within
the ACS FoV is similar to those already used in the
literature \citep[see e.g.,][] {mcz, nataf}. Briefly, the amount of
reddening is evaluated from the shift along the reddening vector
needed to match a given (reddened) evolutionary sequence to the
reference one, which is selected as the least affected by the
extinction.  Thus, the first step of this procedure is to define the
reddening vector in the considered CMD.  It is well known that the
extinction $A_\lambda$ varies as a function of the wavelength
$\lambda$, and the shape of the extinction curve is commonly described
by the parameter $R_\lambda=A_\lambda/E(B-V)$. In order to determine
the value of $R_\lambda$ at the reference wavelengths of the F606W and
F814W filters ($\lambda_V=595.8$ and $\lambda_I=808.7$ nm,
respectively; see {\tt http://etc.stsci.edu/etcstatic/users\_guide}),
we adopted the equations 1, 3a and 3b of \citet{cardelli}, obtaining $R_V=2.83$ and
$R_I=1.82$.  With these values we then computed the reddening vector
shown in Figure \ref{red}. A close inspection
of the CMD shows that the direction
of the distortions along the RCs and the RGB is well aligned with the
reddening vector.
  
\begin{figure}[!htp]
  \centering
  \includegraphics[scale=0.5]{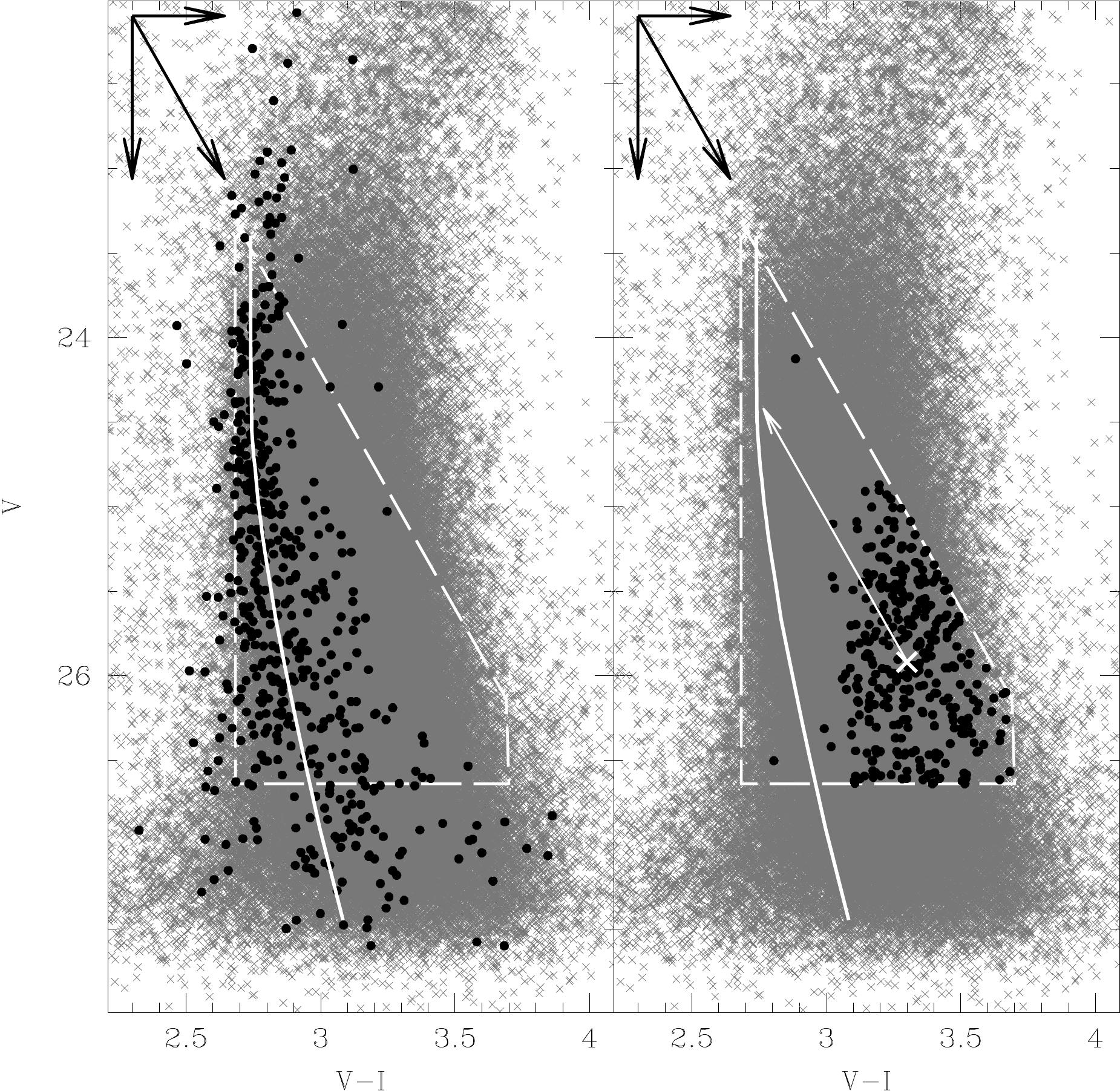}
  \caption{$(V,V-I)$ CMD of Terzan 5 zoomed in the MS region (grey
    crosses). The white dashed lines (same in both panels) delimit the
    selection box for the stars used for the
    computation of the differential reddening correction in each cell.
    The reddening vector is shown in the upper left
    corners of the diagrams. {\it Left panel--} The stars in
    the least extincted (bluest) cell are marked as
    black dots and their best-fit is shown as
    a white solid line.  {\it Right panel--} The mean color and
    magnitude of the stars selected in the $i$-th cell (black dots)
    define the {\it equivalent cell-point} ($\langle V-I\rangle$,
    $\langle V \rangle$), marked as a white cross.  The color
    excess of the cell is obtained by quantifying the shift
    needed to project this point onto the reference line along the
    reddening vector.}
\label{method}
\end{figure}

As second step, the ACS FoV has been divided into a regular
grid of $m\times n$ cells.  The cell size has been chosen small enough
to provide the highest possible spatial resolution, while guaranteeing
the sampling of a sufficient number of stars to properly define the
evolutionary sequences in the CMD. In order to maximize the number of
stars sampled in each cell, we used the Main Sequence.  
After several experiments varying the cell size, we defined a
grid of 25\texttimes25 cells, corresponding to a resolution
of $8.0\arcsec\times 8.0\arcsec$. In order to minimize spurious
effects due to photometric errors and to avoid non-member stars, we considered 
only stars brighter than $V=26.6$ and with $2.7<(V-I)<3.7$ colors.  
We also set
the upper edge of the CMD selection box as the line running parallel to
the reddening vector (see Figure \ref{method}). With these
prescriptions the number of stars typically sampled in each cell is
larger than 60, even at large distance from the cluster center.

The accurate inspection of the MS population in each cell allowed us
to identify the one with the lowest extinction (i.e. where the MS
population shows the bluest average color): it is located in the
South-East region of the cluster at a distance $r\simeq80\arcsec$ from
its center. The stars in this
cell are shown in the left panel of Figure \ref{method} and those enclosed
in the selection box have been used as reference sequence for
evaluating the differential reddening in each cell.  As a "guide line"
of this sequence we used an isochrone of 12 Gyr and metallicity $Z=0.01$
\citep[from][]{marigo,girardi} suitably shifted to best-fit the MS
star distribution (see the heavy white line in Figure \ref{method}).

For each cell of the grid we determined the mean $\langle
V-I\rangle$ color and $\langle V\rangle$ magnitude. A
sigma-clipping rejection at 2-$\sigma$ has been adopted to minimize
the contribution of Galactic disc stars (typically much
bluer than those of Terzan~5) and any other interloper. Each cell is
then described by the ($\langle V-I\rangle, \langle V\rangle$)
color-magnitude pair, which defines the {\it
  equivalent cell-point} in the CMD (as an example, see the cross
marked in the right panel of Figure \ref{method}).  The relative color
excess of each $i$-th cell, $\delta[E(V-I)]_i$, is estimated by
quantifying the shift needed to move the {\it equivalent cell-point}
onto the reference sequence along the reddening vector (see
the right panel of Figure \ref{method}). From the value of
$\delta[E(V-I)]_i$, the corresponding $\delta[E(B-V)]_i$ is easily
computed using the relation
 \begin{equation}
 \delta [E(B-V)]_i=\frac{\delta [E(V-I)]_i}{(R_V-R_I)},
\end{equation}
where $i=1,m\times n$ and $m\times n=625$ is the total number of cells in
our grid.  The $V$ and $I$ magnitudes of all stars
in the $i$-th cell are then corrected by using the derived
$\delta[E(B-V)]_i$ and a new CMD is built.  The whole procedure is
iteratively repeated and a residual $\delta [E(B-V)]_i$ is calculated
after each iteration. The process stops when the difference in the color
excess between two subsequent steps becomes negligible
($\lesssim0.02$ mag). The final value of the relative color excess in each
cell $\delta [E(B-V)]_i$ is thus given by the sum over all the
iterative steps.  For robustness, we applied this
procedure in both the $(I,V-I)$ and $(V,V-I)$ planes.  The
difference between the two estimates turned out to be always smaller
than $\sim0.01$ mag and the average of the two measures was then
adopted as the final estimate of the differential reddening in each
cell.

\subsection{Error estimate and caveats}
\label{error}

Our estimate of the error associated to the color-excess in each cell
is based on the method described by {\citealp{vonbraun} (see also
  \citealp{alonso}).  We considered the uncertainty on the mean color
  of the $i$-th cell as the main source of error on the value of
  $\delta [E(B-V)]_i$. This latter was then computed as the ratio
  between the 1-$\sigma$ dispersion of the mean color and the
  parameter $a=\cos(180-\theta)$, where $\theta$ is the angle between
  the reddening vector and the color-axis.  Geometrically,
  this is equivalent to measure the difference between the values of
  $\delta [E(B-V)]_i$ of the first and last contact-points
  of the color error-bar when moved along the reddening vector
  to match the reference line.  We did not consider the
  error on the mean magnitude because, since the reference line is
  almost vertical, its contribution is negligible.  Following these
  prescriptions we obtain a typical formal error of about $0.03$ mag
  on each color excess value $\delta [E(B-V)]_i$.


A potential problem with this procedure to quantify the
differential reddening of Terzan 5 is the presence of
two stellar populations with distinct iron abundances.  Indeed, the metal-rich
population is expected to be systematically redder than the metal-poor one in
the CMD, and we therefore expect that at least a fraction of stars
with redder colors along the MS are genuine metal-rich objects, and not metal-poor
stars affected by a larger extinction.  However, by using the
\cite{girardi} isochrones, the expected intrinsic difference in the
$(V-I)$ color between the metal-rich and metal-poor populations is only $\delta
(V-I)\sim0.05$ mag.  Moreover, the metal-rich population has been found to be
more centrally segregated than the metal-poor one (F09, L10). Hence, we
expect the former to become progressively negligible with increasing
radial distance from the cluster center. On the other hand, the
uncertainties due to the photometric errors are dominant in the
central region of the system, where the two populations are comparable
in number.  Finally, the use of average values for the color and
magnitude in each cell ($\langle V-I\rangle$ and $\langle V\rangle$),
with the addition of a sigma-clipping rejection algorithm, should
reduce the effect of contamination by metal-rich stars.  Thus, an overall
error of $0.05$ mag on the color excesses $\delta [E(B-V)]_i$ is
conservatively adopted to take into account any possible residual
effects due to the presence of a double population in Terzan 5.

\section{Results}
\label{results}
 
The final differential reddening map in the direction of Terzan 5 is
shown in Figure \ref{map}, with lighter colors indicating less
obscured regions and the center of gravity and core radius (r$_{c}$, see L10)
also marked for reference.  We find that, within the area covered by
the WFC/ACS, the color excess variations can be as large as $\delta
E(B-V)=0.67$ mag.  This is consistent with the value of 0.69 mag estimated
by \cite{ortolani96} from the elongation of the RC.  The
obscuring clouds appear to be structured in two main dusty patches:
the first one is located in the North-Western corner of the map at
$30\arcsec-35\arcsec$ from the center, with an average
differential extinction $\delta E(B-V)>0.4$ mag and a peak value of
0.67 mag.  The second one is placed in the South-Eastern corner, with
typical values of $\delta E(B-V)\sim0.3$ mag.  These two
regions seem to be connected by a bridge-like structure with
$\delta E(B-V)\gtrsim 0.2-0.3$ mag.
 
\begin{figure}[!htb]
 \centering
 \includegraphics[height=11cm,width=13cm]{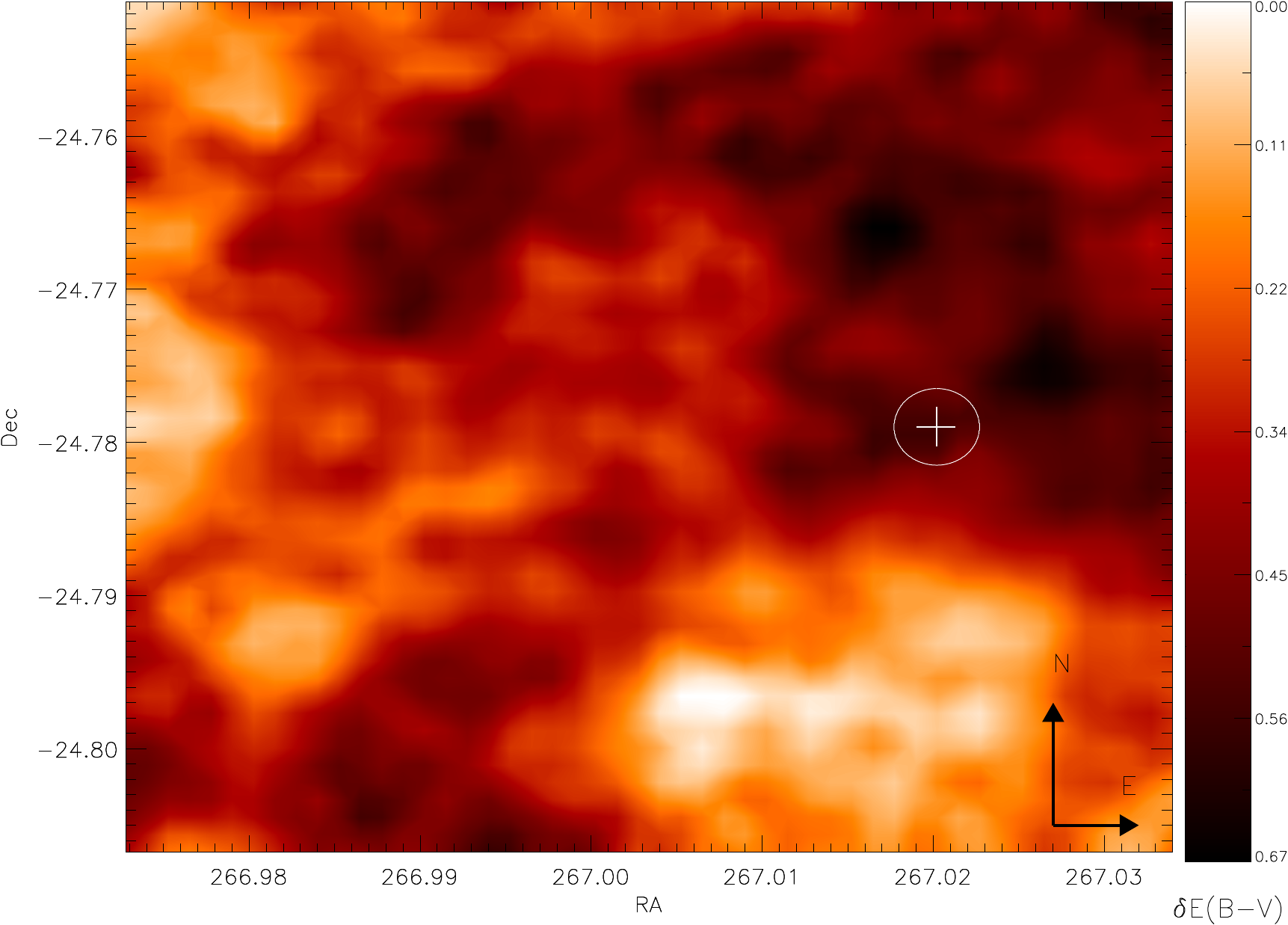}
 \caption{\small Reddening map of the WFC/ACS FoV
   ($200\arcsec\times200\arcsec$) in the direction of Terzan 5. The
   differential color excess $\delta E(B-V)$ ranges between zero (lightest)
   and 0.67 (darkest).  The gravity center and core
   radius of Terzan 5 (L10) are marked for reference as
   white cross and circle, respectively.}
\label{map}
\end{figure}

We used this map to correct our photometric catalogue. Figure
\ref{cmd} shows the comparison between the observed (left panel) and
the differential-reddening corrected (right panel) CMDs in the
$(V,V-I)$ plane. After the correction, both the color extension of the RC and the RGB width are
significantly reduced by 40\% and $>50$\%, respectively, and V magnitudes
become $\sim0.5$ mag brighter. 
To properly quantify the effect of such a correction on
the MS width, we selected the stars along an almost vertical portion
of MS and compared their color distributions before and after the
correction.  To this end, we selected stars with $25<V<25.5$ in the
observed CMD, and 0.5 mag brighter in the
corrected one (see the dashed lines in Figure \ref{cmd}).  The
result is shown in the bottom panels of the figure. Before the
correction the MS color distribution is well represented by a Gaussian
with a dispersion $\sigma=0.18$, significantly larger than the
photometric error at this magnitude level ($\sigma_{phot}\sim0.13$).
Instead, the intrinsic width of the corrected MS is well reproduced by
the convolution of two Gaussian functions separated by 0.05 mag in
color, with a ratio of 1.6 between their amplitudes, and each one
having $\sigma=0.13$ equal to the photometric error. Such
a color separation corresponds to what expected for two stellar
populations with metallicities equal to those measured in Terzan 5
(see Sect. \ref{error}).  The adopted ratio between the amplitudes
corresponds to the number counts ratio between metal-poor and metal-rich populations
(L10). Hence, these two
Gaussian functions correspond to the two sub-populations at different
metallicities observed in Terzan 5.  Note that the corrected MS color
distribution shows an asymmetry
toward the redder side, which is more pronounced in
the center of the system and decreases at progressively
larger distances. The highest amplitude Gaussian (corresponding to the
metal-poor population) is unable to properly account for this feature, while
the convolution with the reddest and lowest amplitude Gaussian
(corresponding to the metal-rich population, which is observed to decrease in
number with increasing distance from the center) provides an excellent
match.

\begin{figure}[!htb]
 \centering
 \includegraphics[scale=0.5]{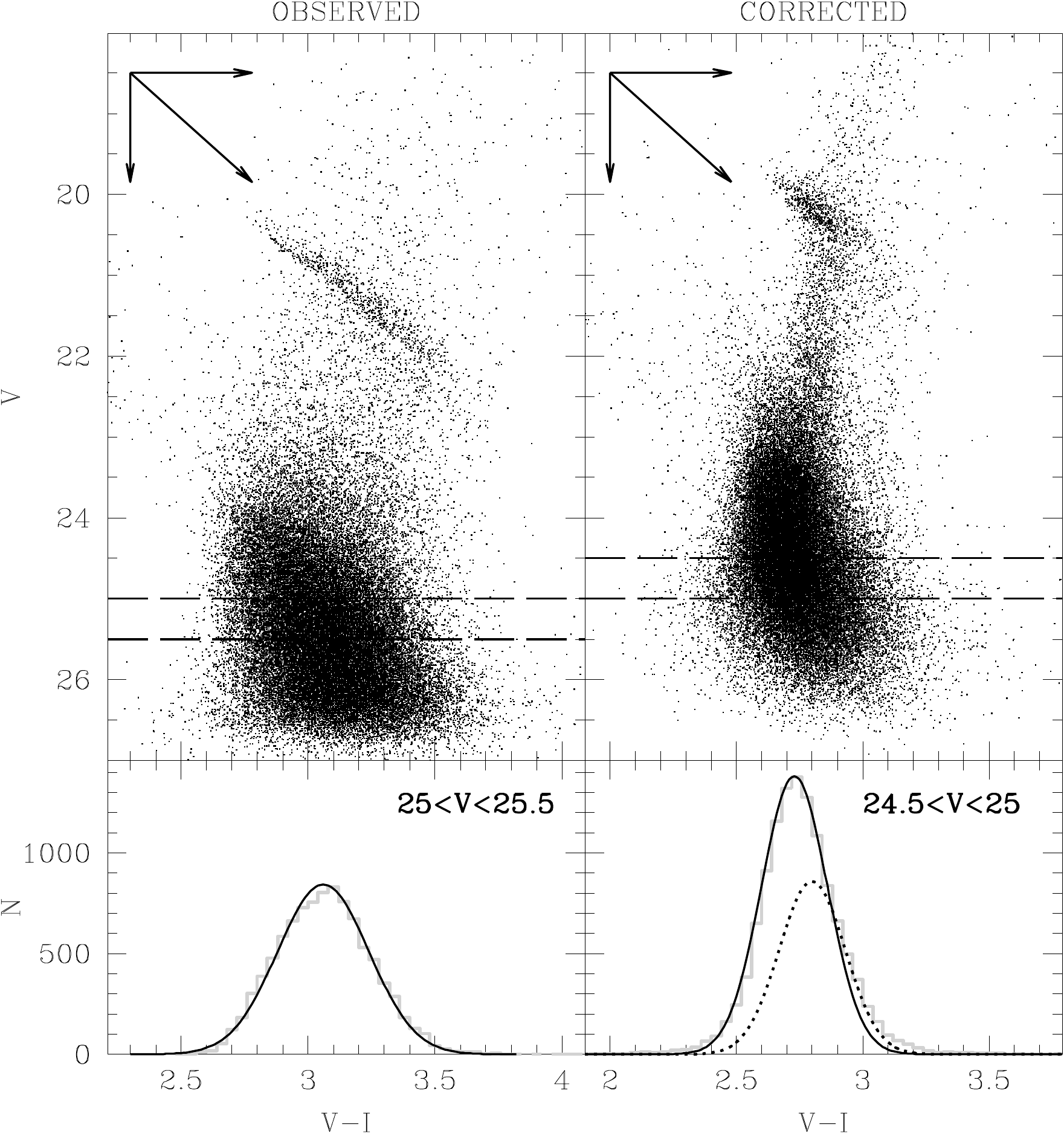}
 \caption{\small Comparison between the optical CMDs of Terzan 5
   before (left panel) and after (right panel) the differential
   reddening correction.  Only stars located at a distance
   $20\arcsec<r<80\arcsec$ are plotted for sake of clarity. All the
   sequences in the corrected CMD are much less stretched along the
   reddening vector.  The bottom panels show the color
   distributions (grey histograms) for a nearly vertical portion of
   MS at $25<V<25.5$ in the observed CMD, and at
   $24.5<V<25$ in the corrected one (see the
   dashed lines in the two upper panels).  Before the correction, the
   color distribution is well represented by a Gaussian with
   $\sigma=0.18$ (while the photometric error is
   $\sigma_{phot}\sim0.13$). After the correction, the distribution is well fitted
   by the convolution of two
   Gaussian functions with $\sigma=0.13$, separated by 0.05 mag in color and with an amplitude ratio of
   1.6. The solid Gaussian
   corresponds to the metal-poor population of Terzan 5,
   while the dotted one represents the metal-rich component (Sect. \ref{results}).}
\label{cmd}
\end{figure}

The derived reddening correction was also applied to the $(K,V-K)$ CMD
obtained from the combination of the ACS and
near-infrared data (see F09). Figure \ref{rgbs} shows the corrected CMD 
with two well separated RGB sub-populations and the two
distinct RCs.  The ratio between the number of stars counted
along the two RGBs is $\sim1.5$, in very good agreement
with the value from the RCs (see above and L10).

\begin{figure}[!htb]
 \centering%
 \includegraphics[scale=0.5]{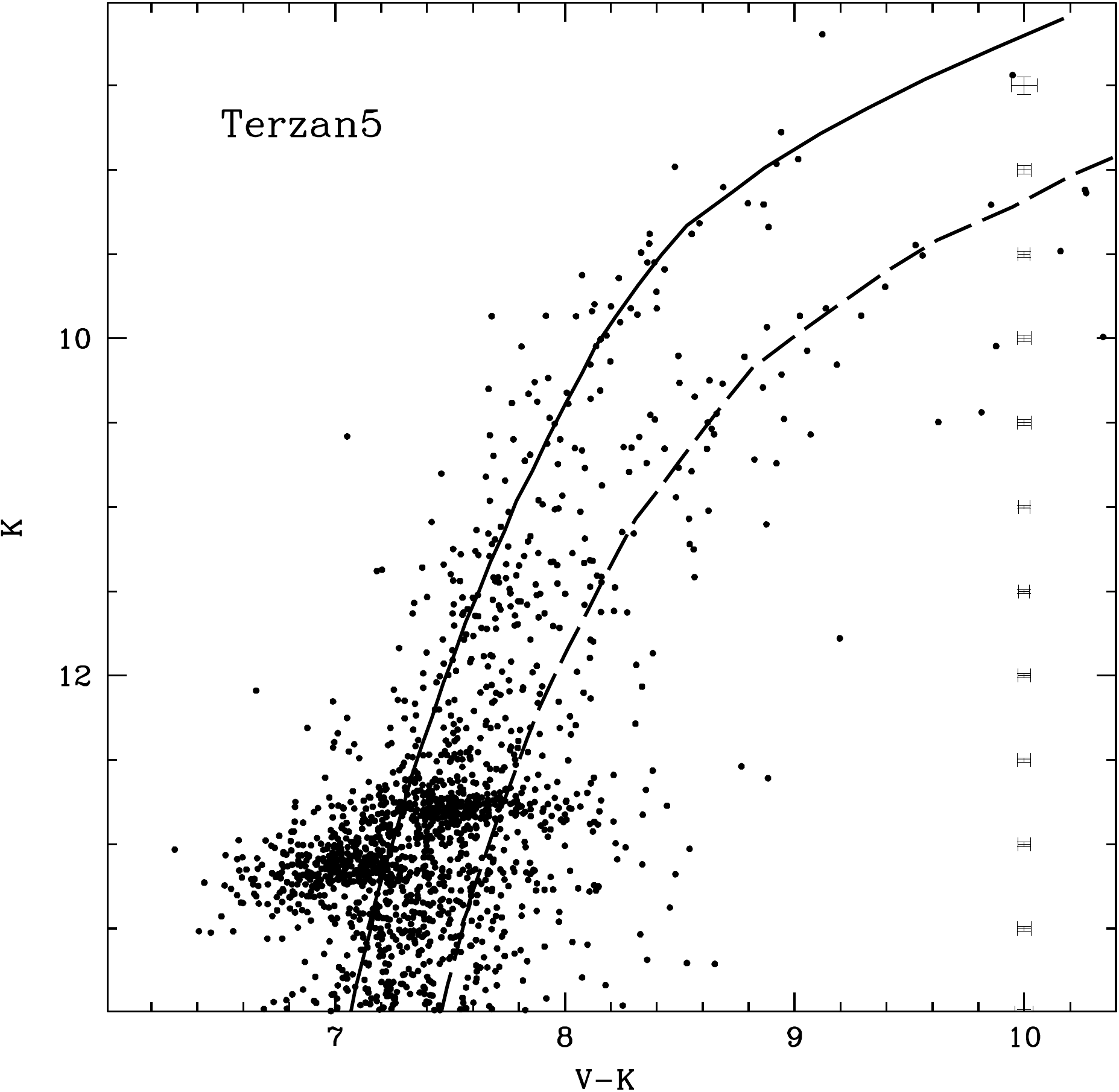}
 \caption{\small Brightest portion of the differential reddening corrected
  $(K,V-K)$ CMD of Terzan 5, with error bars also reported.  
   Beside the two RCs, also two well separated RGBs are clearly
   distinguishable. The solid and dashed lines correspond to the mean
   ridge lines of the metal-poor and the metal-rich sub-populations, respectively. }
\label{rgbs}
\end{figure}

The differential reddening corrected CMD can be finally used to
estimate the absolute color excess in the direction of Terzan
5.\footnote{A free tool providing the color excess values at any
  coordinate within the WFC/ACS FoV can be found at the web site {\tt
    http://www.cosmic-lab.eu/Cosmic-Lab/products}} Different values of
$E(B-V)$ are provided in the literature, ranging from 1.65 (estimated by
\citealt{az} from the strength of an interstellar band at 8621\AA), up
to 2.19 and 2.39, derived from optical or infrared photometry
\citep[][]{barbuy,cohn,valenti07}. However, all these estimates are
average values and do not take into account the presence of
differential reddening.  Here, instead, we want to build a
2-dimensional map of the absolute reddening and, to this end, we
shifted the corrected $(V,V-I)$ CMD of Terzan 5 along the reddening
direction until it matches the CMD of 47 Tucanae, adopted as reference
cluster since it is metal-rich, low extincted and with a well-determined distance modulus.  
In particular we
looked for the best match between the RC of the metal-poor population
of Terzan 5 and the RC of 47 Tucanae.  We adopted the color
excess $E(B-V)=0.04$ and the distance modulus $\mu_0=13.32$ for 47
Tucanae \citep[from][]{ferr99}, and $\mu_0=13.87$ for Terzan 5
\citep{valenti07}.
From \citet{girardi} model, in the $(V,V-I)$
plane the RC of 47 Tucanae turns out to be 0.02 mag brighter and 0.03 mag bluer
than the metal-poor one of Terzan 5 because of a difference in their metallicity
([Fe/H]$=-0.70$ for 47 Tucanae and 
[Fe/H]$=-0.27$ for the metal-poor population of Terzan 5; see \citealt{ferr99} and
O11, respectively).
Taking into account these slight differences, a
nice match of the two RCs is obtained by
adopting $E(B-V)=2.15$ mag. Since the corrected CMD is, by construction,
referred to the bluest cell, the absolute color excess within the
WFC/ACS FoV varies from $E(B-V)=2.15$ up to
$E(B-V)=2.82$ mag. 
In order to check the reliability of these estimates, we compared it with
the values found by \citet{gonz} from the Vista Variable in the Via Lactea
survey. In a $2\arcmin \times 2\arcmin$ region centered on Terzan 5, 
these authors found an extinction $A_{K}=0.80$ mag.
Using \citet{cardelli} coefficients to convert $E(B-V)$ to $A_{K}$, our estimate varies from
$A_{K}=0.75$ to $A_{K}=0.98$ mag, in nice agreement with \citet{gonz} result.

Moreover, we looked for a possible correlation between the color excess and the 
dispersion measures for 34 MSPs of Terzan 5 studied by \cite{dm}. 
In this case
we did not find a strong correlation, probably because mostly 
(75\%) of the MSP sample is situated within the inner $20\arcsec$ of the system, where the
estimate of $E(B-V)$ is more uncertain (see Sect. \ref{error}).

\clearpage{\pagestyle{empty}\cleardoublepage}

\chapter{Chemical and kinematical properties of Galactic bulge stars surrounding the 
stellar system Terzan~5}\label{chapfield}

In order to investigate the kinematical and chemical properties of
Terzan 5, we have collected spectra for more than 1600 stars in its
direction. In this Chapter we focus on the properties of the bulge field
population surrounding the system, with the aim of providing crucial
information for further studies of both Terzan 5 and the bulge
itself. In fact, this is a statistically significant sample of field
stars which can be used to decontaminate the population of Terzan 5
from non-members.  Moreover, it is one of the few large
samples of bulge stars spectroscopically investigated at low and
positive latitudes (b$=+1.7$\textdegree), thus allowing interesting
comparisons with other well studied bulge regions.

The complete description of this analysis can be found in \cite{massari14a}.

\section{The sample}
\label{obs}
This study is based on a sample of 1608 stars within a radius of 800\arcsec
from the center of Terzan 5 ($\alpha_{{\rm J}2000}=17^{\rm
  h}:48^{\rm m}:4^{\rm s\!.}85$, $\delta_{{\rm
    J}2000}=-24^{\circ}:46\arcmin:44\arcsec6$; see F09, L10).
While the overall
survey will be described in a forthcoming paper (Francesco
Ferraro et al. in preparation), in this work we focus on a sub-sample
of stars representative of the field population surrounding Terzan
5. Given the value of the tidal radius of the system
(r$_{t}\simeq300\arcsec$; L10, \citealt{miocchi13}), we conservatively
selected as genuine field population members all the targets more
distant than $400\arcsec$ from the center of Terzan 5.  This
sub-sample is composed of $615$ stars belonging to two different
datasets obtained with FLAMES \citep{pasquini} at the ESO-VLT and with DEIMOS (\citealt{faber}) at the Keck II
Telescope.  Each target has been selected from the ESO-WFI optical
catalog described in L10, along the brightest portion of the
RGB, with magnitudes brighter than $I<18.5$.  In
order to avoid contamination from other sources, in the selection
process of the spectroscopic targets we avoided stars with bright
neighbors (I$_{neighbor}<{\rm I_{star}}+1.0$) within a distance of
2\arcsec.  The spatial distribution of the entire sample is shown in
Figure \ref{map_f}, where the selected field members are shown as large
filled circles.

\begin{figure}
\plotone{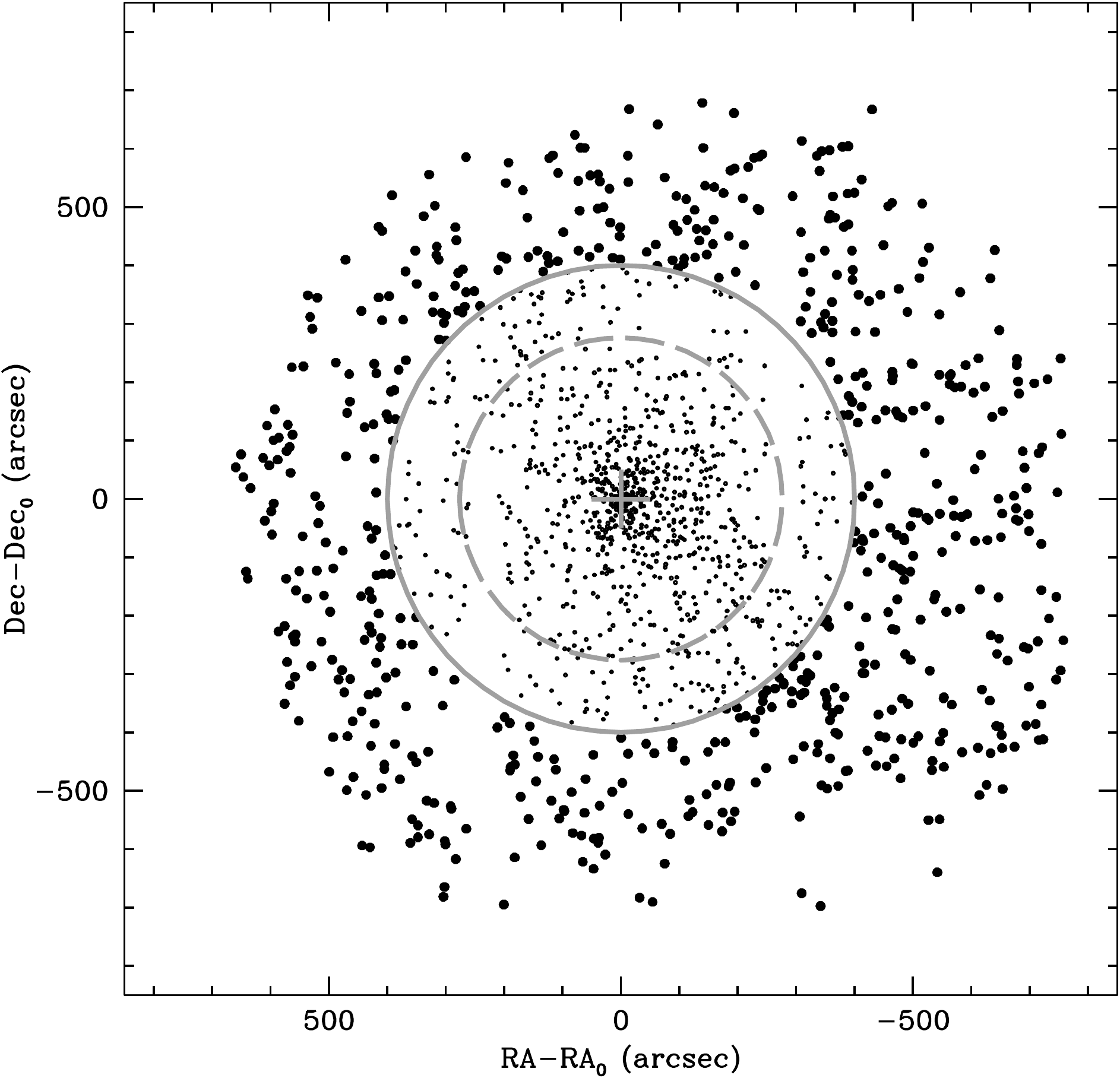}
\caption{\small Spatial distribution of all the targets in our
  spectroscopic survey in the direction of Terzan 5. The center of
  gravity and tidal radius ($r_t\simeq 300\arcsec$) of Terzan 5 (from
  L10) are marked with a gray cross and a gray dashed
  circle, respectively. The targets discussed in this work (shown as
  filled black circles) are all located at more than 400\arcsec ~from
  the center (gray solid circle), well beyond the Terzan 5 tidal
  radius.}
\label{map_f}
\end{figure}

The FLAMES dataset was collected under three different programs:
087.D-0716(B), PI: Ferraro; 087.D-0748(A), PI: Lovisi; and
283.D-5027(A), PI: Ferraro.  All the spectra were acquired in the
GIRAFFE/MEDUSA mode, allowing the allocation of 132 fibers across a
$25\arcmin$ diameter FoV in a single pointing. We used
the GIRAFFE setup HR21, with a resolving power of 16200 and a spectral
coverage ranging from 8484 \AA{} to 9001 \AA{}.  This grating was
chosen because it includes the prominent Ca~II triplet lines, which
are excellent features to measure radial velocities also in relatively
low signal-to-noise ratio (SNR) spectra. Other metal lines (mainly
Fe~I) lie in this spectral range, thus allowing a direct measurement
of the iron abundance.  Multiple exposures (with integration times
ranging from 1500 to 3000 s, according to the magnitude of the
targets) were secured for the majority of the stars, in order to reach
SNR$\sim$30 even for the faintest ($I\sim 18.5$) targets.  The data
reduction was performed with the FLAMES-GIRAFFE
pipeline\footnote{http://www.eso.org/sci/software/pipelines/},
including bias-subtraction, flat-field correction, wavelength
calibration with a standard Th-Ar lamp, re-sampling at a constant
pixel-size and extraction of one-dimensional spectra.  Since a correct
sky subtraction is particularly crucial in this spectral range
(because of the large number of O$_{2}$ and OH emission lines), 15-20
fibers were used to measure the sky in each exposure. Then a master
sky spectrum was obtained from the median of these spectra and it was
subtracted from the target spectra.  Finally, all the spectra were
shifted to zero-velocity and in the case of multiple exposures they
were co-added.

The DEIMOS dataset was acquired using the 1200 line/mm grating coupled
with the GG495 and GG550 order-blocking filters, covering the
$\sim$6500-9500 \AA{} spectral range at a resolution of R$\sim$7000 at
$\sim$8500 \AA{}.  The DEIMOS FoV is $16$\arcmin$\times5$\arcmin,
allowing the allocation of more than 100 slits in a single mask.  The
observations were performed with an exposure time of 600 s, securing
SNR$\sim$50-60 spectra for the brightest targets and achieving
SNR$\sim$15-20 for the faintest ones (I$\sim17$).  The spectra have
been reduced by means of the package developed for an optimal
reduction and extraction of DEIMOS spectra and described in
\citet{ibata11}.

\section{Radial velocities}
\label{vrad}
Radial velocities ($v_{\rm rad}$) for the target stars were measured
by cross-correlating the observed spectra with a template of known
velocity, following the procedure described in \citet{tonry} and
implemented in the FXCOR software under IRAF.  As templates we adopted
synthetic spectra computed with the SYNTHE code \citep{sbordone04}.
For most of the stars the cross-correlation procedure is performed in
the spectral region $\sim$8490-8700 \AA{}, including the prominent
Ca~II triplet lines that can be well detected also in noisy spectra.
For some very cool stars, strong TiO molecular
bands dominate this spectral region, preventing any reliable
measurement of the Ca features (Figure \ref{tiosel} shows the
comparison between two FLAMES spectra, with and without strong
molecular bands). In these cases, the radial velocity was measured
from the TiO lines by considering only the spectral region around the
TiO bandhead at $\sim$8860 \AA{} and by using as template a synthetic
spectrum including all these features.  Because several stars show
both the Ca~II triplet lines and weak TiO bandheads, for some of them
the radial velocity has been measured independently using the two
spectral regions. We always found an excellent agreement between the
two measurements, thus ruling out possible offsets between the two
$v_{\rm rad}$ diagnostics.

\begin{figure}
\plotone{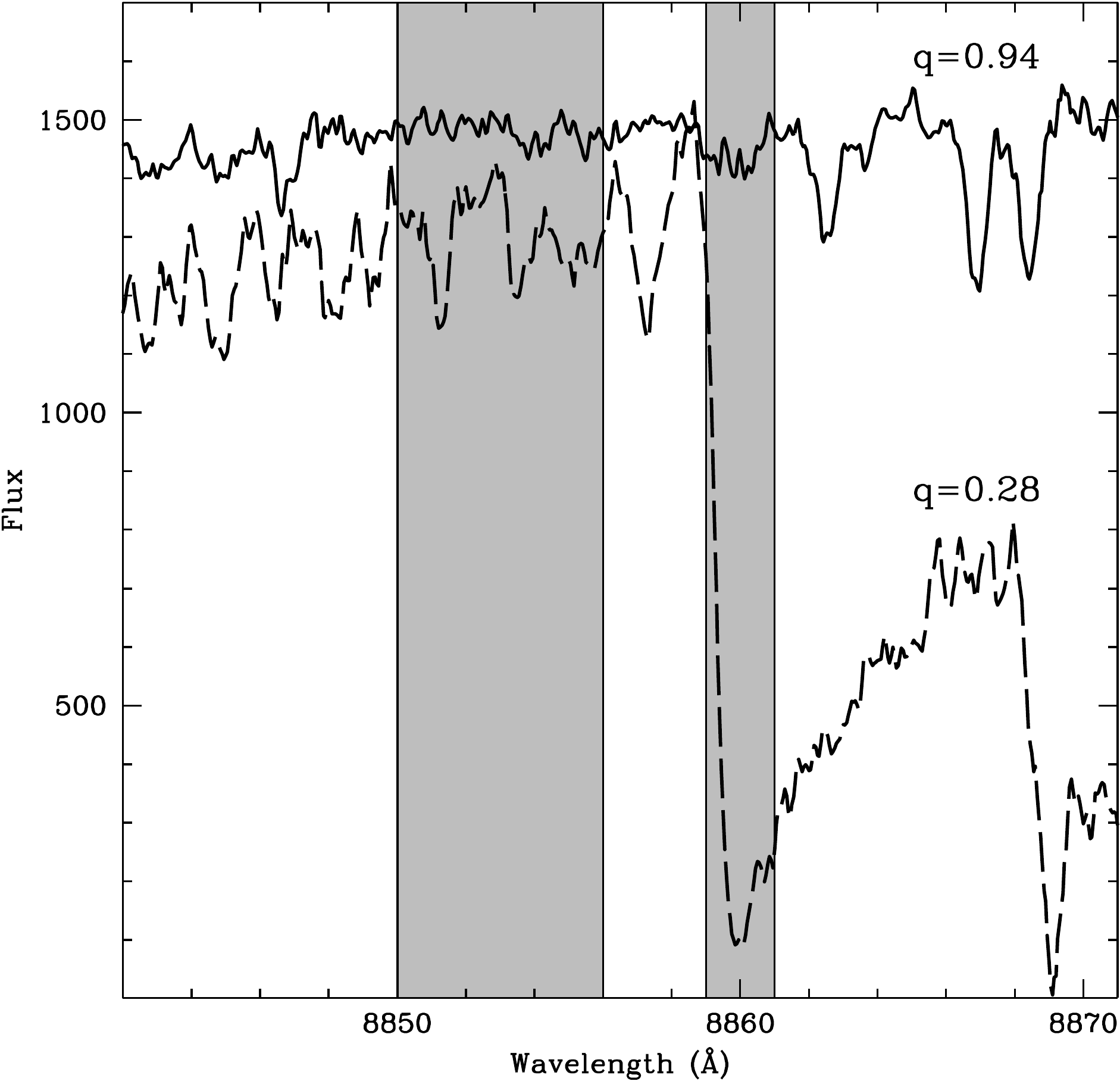}
\caption{\small Examples of spectra poorly (solid line) and
  severely (dashed line) affected by the TiO molecular bands
  ($\lambda>8860$ \AA).  The gray regions highlight the wavelength
  ranges adopted to compute the $q$-parameter defined in the text
  (Section \ref{iron}).  For these two spectra very different values
  of $q$ have been obtained: $q=0.94$ and $q=0.28$ for the solid and
  dashed spectrum, respectively. According to the adopted selection
  criterion (q$>0.6$), the iron abundance has not been computed from
  the latter.  }
\label{tiosel}
\end{figure} 

For the FLAMES dataset, where multiple exposures were secured for most
of the stars, radial velocities were obtained from each exposure
independently. The final radial velocity is computed as the weighted
mean of the individual velocities (each corrected for its own
heliocentric velocity), by using the formal errors provided by FXCOR
as weights.  For the DEIMOS spectra, we checked for possible velocity
offsets due to the mis-centering of the target within the slit
\citep[see the discussion about this effect in][]{simon}, through the
cross-correlation of the A telluric band (7600-7630 \AA{}). We found
these offsets to be of the order of few km$\,$s$^{-1}$.  The
uncertainty on the determination of this correction (always smaller
than 1 km$\,$s$^{-1}$) has been added in quadrature to that provided
by FXCOR.  The typical final error on our measured v$_{{\rm rad}}$ is
$\sim1.0$ km$\,$s$^{-1}$.

The distribution of the measured $v_{\rm rad}$ for the 615 targets is
shown in   Figure \ref{vraddist}.  It ranges from
$-264.0$ km s$^{-1}$ to $+303.9$ km s$^{-1}$.  By using a
Maximum-Likelihood procedure, we find that the Gaussian function that
best describes the distribution has mean $\langle v_{\rm
  rad}\rangle=21.0\pm4.6$ km s$^{-1}$ and $\sigma_{v}=113.0\pm2.7$ km
s$^{-1}$.
We converted radial velocities to Galactocentric velocities ($v_{\rm
  GC}$) by correcting for the Solar reflex motion (220 km s$^{-1}$;
\citealt{kerrlyn86}) and assuming as peculiar velocity of the Sun in
the direction (l,b)$=$(53\textdegree,25\textdegree) $v=18.0$ km
s$^{-1}$ \citep{schonrich}. The conversion equation is then:
\begin{equation}
 v_{{\rm GC}}=v_{{\rm rad}}+220[\sin(l)\cos(b)]+18[\sin(b)\sin(25)+\cos(l)\cos(25)\cos(l-53)],
\end{equation}
where velocities are in km s$^{-1}$, and (l,b)=(3.8\textdegree,
1.7\textdegree) is the location of Terzan 5.
The Galactocentric velocity distribution estimated in this way peaks at $<v_{{\rm GC}}>=47.7\pm4.6$ km$\,$s$^{-1}$.

\begin{figure}
\plotone{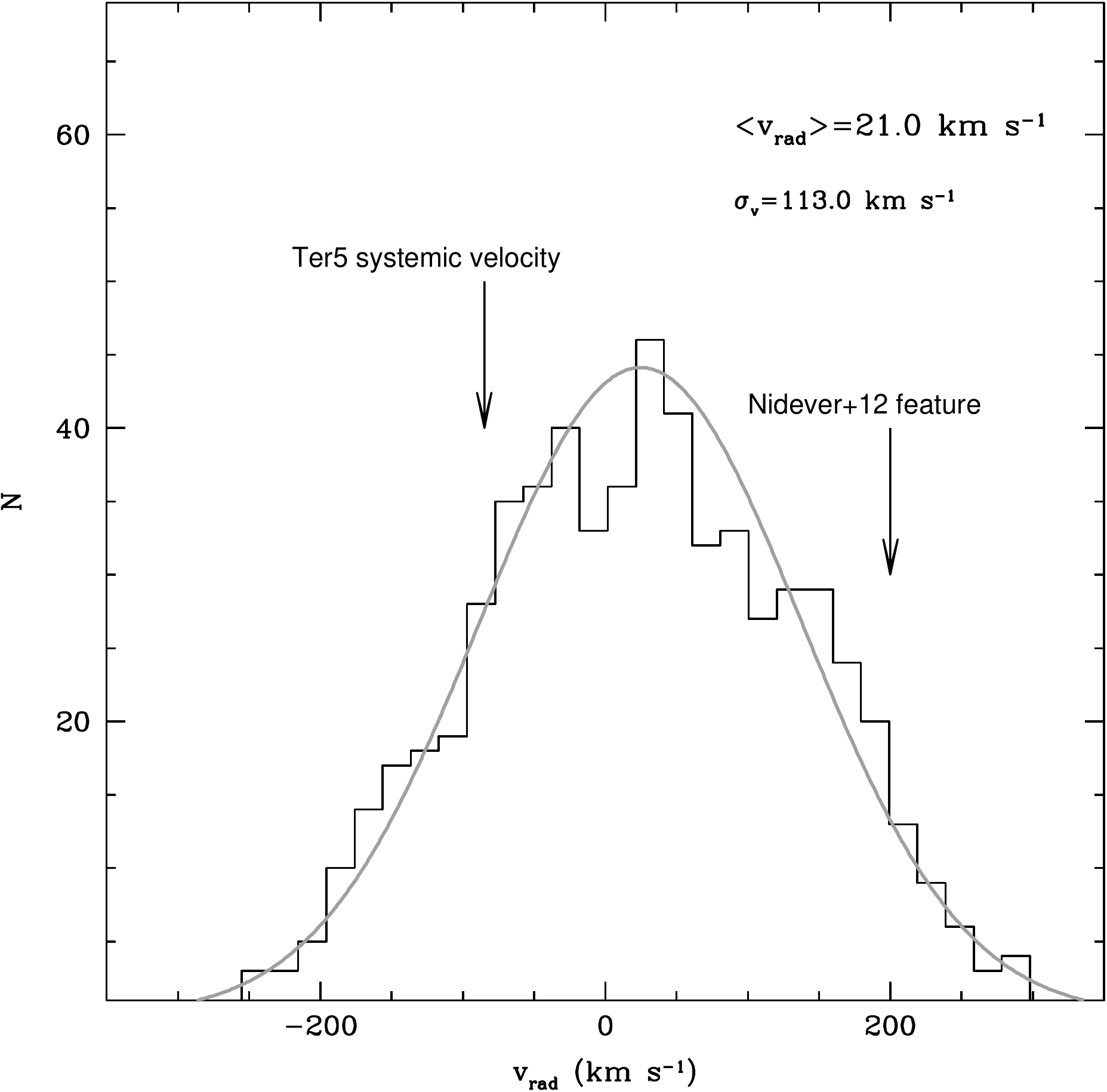}
\caption{\small Radial velocity distribution for the 615 spectroscopic
  targets at $r>400\arcsec$. The mean value and dispersion are
  indicated. The bin size (20 km$\,$s$^{-1}$) is the same as that
  adopted by \cite{nidever}, but in our case no high velocity
  sub-components are found. The systemic velocity of Terzan 5 
(v$_{{\rm rad}}\simeq-83$ km$\,$s$^{-1}$) and the location of the subcomponent
found in \cite{nidever} (v$_{{\rm rad}}\simeq200$ km$\,$s$^{-1}$) are also marked 
with the black arrows for sake of comparison.}
\label{vraddist}
\end{figure} 

This value turns out to be in good agreement with the values found in the context of three recent kinematic surveys of the Galactic
bulge: the BRAVA survey (\citealt{rich07}), the ARGOS survey (\citealt{freeman13}) and the GIBS survey (\citealt{zoccali14}).
In fact, in fields located close to the Galactic Plane and with Galactic longitude 
as similar as possible to ours, \citet[][see also Kunder et al. 2012]{howard08} found $<v_{{\rm GC}}>=53.0\pm10.3$ km$\,$s$^{-1}$ at 
(l,b)=(4\textdegree,-3.5\textdegree) for BRAVA, \cite{ness13kin} found $<v_{{\rm GC}}>=44.4\pm3.8$ km$\,$s$^{-1}$ 
at (l,b)=(5\textdegree,-5\textdegree) for ARGOS and \cite{zoccali14} obtained $<v_{{\rm GC}}>=55.9\pm3.9$ km$\,$s$^{-1}$ 
at (l,b)=(3\textdegree,-2\textdegree) for GIBS.
Thus, all the measurements agree within the errors with our result.
As for the velocity dispersion, our estimate ($\sigma_{v}=113\pm2.7$ km$\,$s$^{-1}$) agrees well with the result of 
\cite{kunder12}, who found $\sigma_{v}=106$ km$\,$s$^{-1}$, and with that of \cite{zoccali08} who measured
$\sigma_{v}=112.5\pm6.4$ km$\,$s$^{-1}$. Instead, it is larger than that quoted 
in \cite{ness13kin} $\sigma_{v}=92.2\pm2.7$ km$\,$s$^{-1}$. Since their field is the farthest from ours among
those selected for the comparison, we ascribe such a difference to the different location in the bulge.
As a further check, we used the Besan\c{c}on Galactic model
(\citealt{robin}) to simulate a field with the same size of our
photometric sample (i.e., the WFI FoV) around the location of Terzan
5, and we selected all the bulge stars lying within the same color and
magnitude limits of our sample. The velocity dispersion for these
simulated stars is $\sigma_{v}=119$ km s$^{-1}$, in agreement with
our estimate.

It is worth mentioning that \cite{nidever} identified a high velocity
($v_{{\rm rad}}\sim$200 km s$^{-1}$) sub-component that accounts for about
10\% of their entire sample of $\sim4700$ bulge stars.  Such a feature
has been found in eight fields located at
-4.3\textdegree$<$b$<2$\textdegree and
4\textdegree$<$l$<14$\textdegree.  \cite{nidever} suggest that such a
high-velocity feature may correspond to stars in the Galactic bar
which have been missed by other surveys because of the low latitude of
the sampled fields.  However, as is evident in Figure \ref{vraddist}
(where we adopted the same bin-size used in Fig. 2 of \cite{nidever}
for sake of comparison), we do not find neither high-velocity peak
nor isolated substructures in our sample, despite its low latitude.  
In fact, the skewness calculated for our distribution is $-0.02$, 
clearly demonstrating its symmetry.  Also \cite{zoccali14} did not find 
any significant peak at such large velocity in the recent GIBS survey.

\section{Metallicities}\label{iron}

For a sub-sample of $284$ stars we were able to also derive
metallicity. As already pointed out in Section \ref{vrad},
spectra of cool giants are affected by the presence of prominent TiO molecular
bands. These bands make particularly uncertain the determination of
the continuum level.  While this effect has no consequences on the
determination of radial velocities, it could critically affects the
metallicity estimate. Therefore we limited the metallicity analysis
only to stars whose spectra suffer from little contamination from the
TiO bands. In order to properly evaluate the impact of the TiO bands
in the considered wavelength range, we performed a detailed analysis
of a large set of synthetic spectra and we defined a parameter $q$ as
the ratio between the flux of the deepest feature of the TiO bandhead
at 8860 \AA{} (computed as the minimum value in the spectral range
8859.5 \AA{}$<\lambda<$8861 \AA{}) and the continuum level measured
with an iterative 3 $\sigma$-clipping procedure in the adjacent
spectral range, 8850\AA{}$<\lambda<$8856 \AA{} (see the shaded regions
of Figure \ref{tiosel}).  

We found that the continuum level of synthetic spectra for 
stars with $q>0.8$ is slightly ($<2\%$) affected by TiO bands over 
the entire spectral range, while for stars with $0.6<q<0.8$ the region
marginally ($<5\%$) affected by the contamination is confined between 8680
\AA{} and 8850 \AA. Instead, stars with $q<0.6$ have no useful spectral
ranges (where at least one of the Fe~I in our linelist falls) 
with TiO contamination weak enough to allow a 
reliable chemical analysis. We therefore analyzed
targets with $q>0.8$ (counting 126 objects) using the full linelist
(see Section \ref{analysis}), while for targets with $0.6<q<0.8$ (158
objects) only a sub-set of atomic lines lying in the safe spectral
range $8680-8850$ \AA{} has been adopted.  All targets with $q<0.6$
(329 stars) have been excluded from the metallicity analysis.  Hence
the metallicity analysis is limited to 284 stars (corresponding to
$\sim 46\%$ of the entire sample observed in the spectroscopic
survey).  In Section \ref{results_f} we discuss the impact of this
selection on the results of the analysis.

\subsection{Atmospheric parameters}\label{atmpar}

We derived effective temperatures (T$_{{\rm eff}}$) and gravities ($\log$ $g$) photometrically. 
In order to minimize the effect of differential reddening we used the 2MASS catalog, correcting the (K, J-K) CMD
for differential extinction according to our new wide-field reddening map, shown in Figure \ref{redmap}.
This was obtained by applying to the optical WFI catalog the same procedure described in Chapter \ref{chapred}. 
Because of the large incompleteness at the MS level we were forced to use red clump stars as reference.
Since these stars are significantly less numerous than MS stars, the spatial resolution of the computed reddening map ($60\arcsec \times 60\arcsec$)
is coarser than that ($8\arcsec \times 8\arcsec$)  published in \cite{massari} for the {\it HST} ACS field of view. 
However, despite this difference in resolution, the WFI reddening map agrees quite well with that for ACS in the overlapping region.
Indeed for the stars in common between the two catalogs, the average difference between the differential reddening 
estimates is $\langle\Delta E(B-V)\rangle=0.01$ mag 
with a dispersion $\sigma=0.1$ mag. This latter  value is also the uncertainty that
we conservatively adopt for our color excess estimates.

\begin{figure}
\includegraphics[height=11cm,width=14cm]{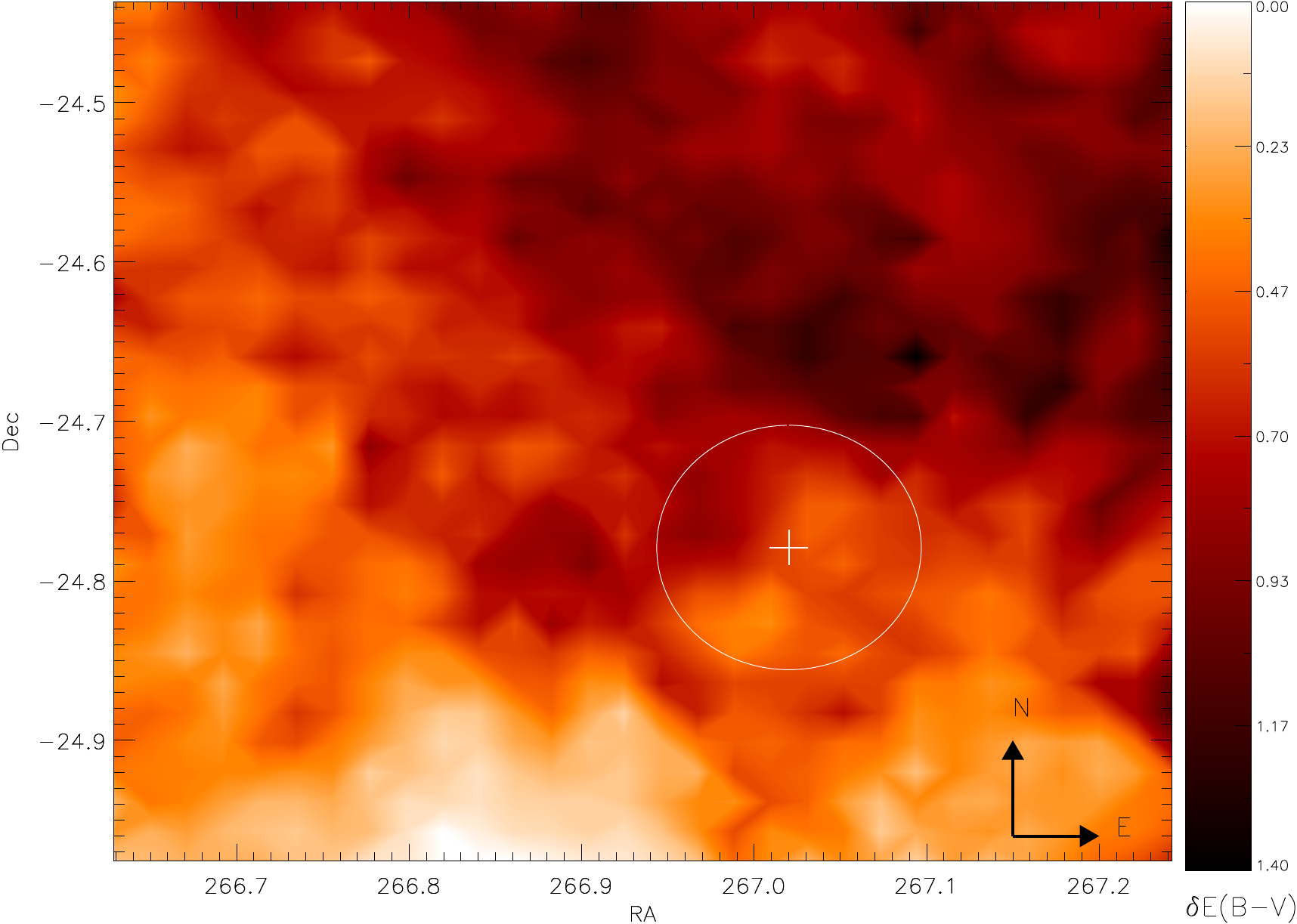}
\caption{\small Reddening map in the direction of Terzan 5 covering
  the entire $\sim 25$\arcmin$\times 25$\arcmin ~FoV.  Dark colors
  correspond to regions of large extinction (see the
  color bar on the right).  The center and tidal radius of Terzan 5
  are marked for sake of comparison.}
\label{redmap}
\end{figure} 

Figure \ref{cmd_f} shows the IR CMD after the internal reddening
correction, with the positions of the spectroscopic targets highlighted. 
To determine T$_{{\rm eff}}$, we adopted the (J-K)$_{0}-$T$_{{\rm eff}}$ empirical relation quoted by \cite{m98}.
Since the relation is calibrated onto the SAAO photometric system, we previously converted our 2MASS magnitudes
following the prescriptions in \cite{carpenter01}.
To estimate photometric gravities, we used the relation:
\begin{equation}
 \log g=\log g_\odot +4\log(T_{{\rm eff}}/T_{\odot})+\log(M/M_{\odot})+0.4(M_{{\rm bol}}-M_{{\rm bol},\odot})
\end{equation}
adopting $\log$ $g_{\odot}$=4.44 dex, T$_{\odot}$=$5770$ K, M$_{{\rm bol},\odot}=4.75$, M$=0.8$ M$_\odot$ and a distance of 8 kpc.
Such a distance is the average value predicted by the Besan\c{c}on model for a simulated field with the size of the FoV covered 
by our observations and centered around Terzan 5.  This value is also normally adopted when bulge stars are analyzed
(\citealt{zoccali08, alvesbrito10, hill11}, see also Sect. \ref{results_f} for a discussion on the impact of distance on our results). 
Bolometric corrections were taken from \cite{m98}.
The small number (about 10) of Fe~I lines available in the spectra prevents us to derive reliable values of
microturbulent velocity ($v_{turb}$; see \citealt{m11} for a review of the different methods to infer this parameter).
We therefore referred to the works of \cite{zoccali08} and \cite{johnson13} on large samples of bulge giant stars 
characterized by metallicities and atmospheric parameters similar to those of our targets.
Since no specific trend between $v_{turb}$ and the atmospheric parameters is found in these samples, we adopted their 
median velocity $v_{turb}=$1.5 km$\,$s$^{-1}$ ($\sigma=0.16$ km$\,$s$^{-1}$) for all the targets.

\begin{figure}
\plotone{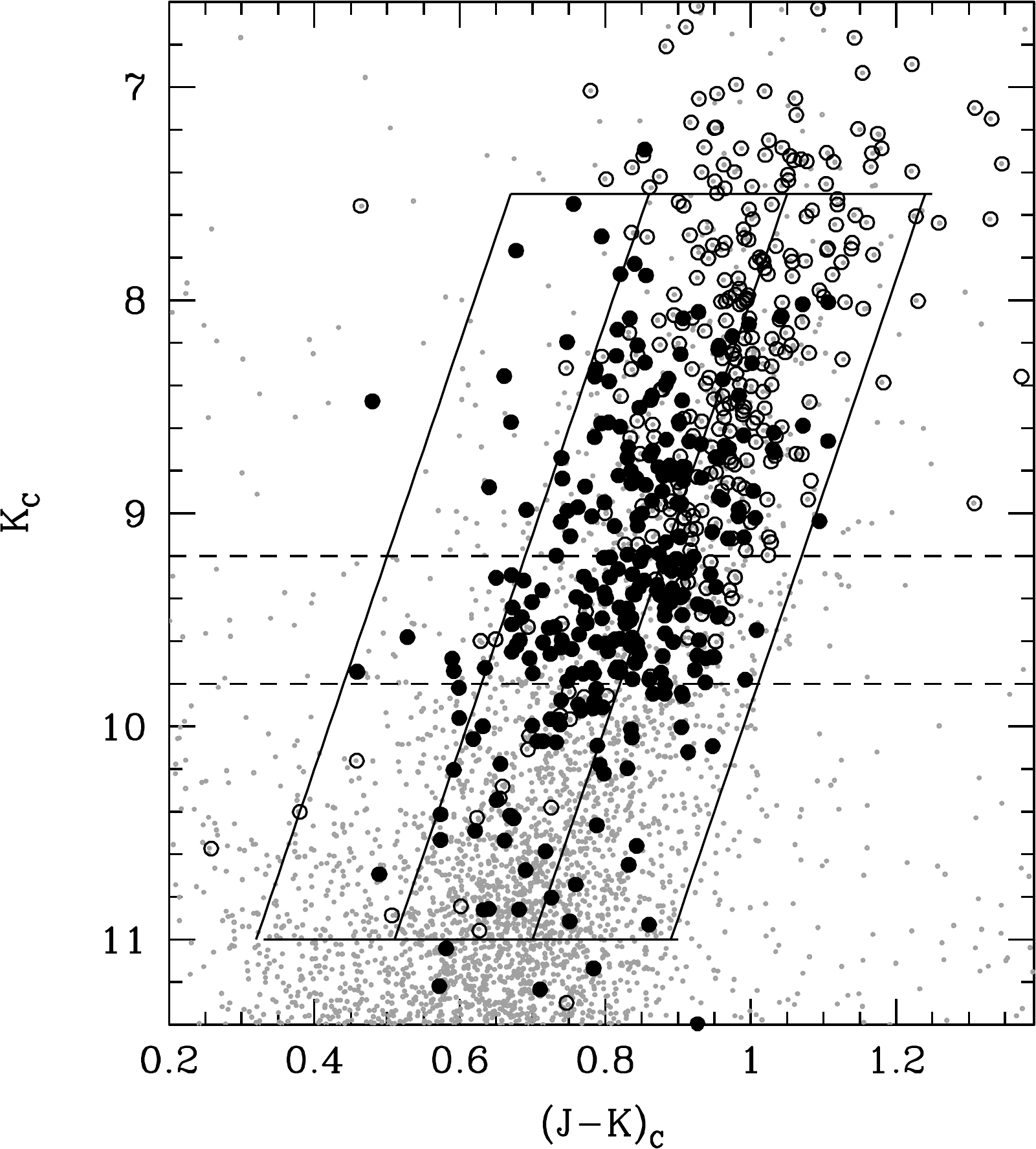}
\caption{\small(K, J-K) CMD corrected for internal differential
  extinction, for all the stars located at $400\arcsec<r<800\arcsec$
  from the center of Terzan 5, in our ESO-WFI photometric sample
  (small dots).  The spectroscopic targets are shown with large
  symbols. The targets for which iron abundance could be estimated are
  highlighted as large filled circles. The three strips adopted to
  evaluate the impact of the selection bias discussed in the text
  (Section \ref{results_f}) are shown. The horizontal dashed lines
  delimit the bias-free sample adopted to derive the metallicity
  distribution shown as a grey histogram  in Figure \ref{fedist}.}
\label{cmd_f}
\end{figure} 

\subsection{Chemical analysis}\label{analysis}

The Fe~I lines used for the chemical analysis were selected from the latest version of the Kurucz/Castelli dataset of atomic 
data\footnote{http://wwwuser.oat.ts.astro.it}. 
We included only Fe I transitions found to be unblended in synthetic spectra calculated 
with the typical atmospheric parameters of our targets and 
at the resolutions provided by the GIRAFFE and DEIMOS spectrographs. 
These synthetic spectra were calculated with the SYNTHE code, 
including the entire Kurucz/Castelli line-list (both for atomic and molecular lines) convolved with a 
Gaussian profile at the resolution of the observed spectra. 
Due to the different spectral resolution of the two datasets, we used two different techniques 
to analyze the spectral lines and determine the chemical abundances.

(1)~{\it FLAMES spectra}---
The chemical analysis was performed using the package GALA \citep{gala}\footnote{GALA 
is freely distributed at the Cosmic-Lab project website, 
http://www.cosmic-lab.eu/gala/gala.php}, an automatic tool to derive the chemical abundances 
of single, unblended lines by using their measured equivalent widths (EWs). 
The adopted model atmospheres were calculated with the ATLAS9 code \citep{atlas}.
In our analysis, we run GALA fixing all the atmospheric parameters estimated as described above and 
leaving only the metallicity of the model atmosphere free to vary iteratively in order to match the iron abundance
derived from EWs.
EWs were obtained with the code 4DAO (\citealt{4dao})\footnote{Also this code is freely distributed at the 
Cosmic-lab website: http://www.cosmic-lab.eu/4dao/4dao.php.}, 
aimed at running DAOSPEC \citep{daospec} for a large set of spectra, tuning automatically 
the main input parameters used by DAOSPEC and providing graphical outputs to visually 
inspect the quality of the fit for each individual spectral line.
The EWs were measured adopting a Gaussian function that is a reliable approximation for the line profile 
at the resolution of our spectra. EW errors were estimated by DAOSPEC 
from the standard deviation of the local flux residuals \citep[see][]{daospec} and lines with EW errors 
larger than 10\% were rejected.

(2)~{\it DEIMOS spectra}---
Due to the low resolution of the DEIMOS spectra,  
the high degree of line blending and blanketing makes the derivation of the abundances through the method of 
the EWs  quite complex and uncertain. Thus, the iron abundances were measured by 
comparing the observed spectra with a grid of synthetic spectra, 
following the procedure described in \citet{m12_2419}. Each Fe~I line was 
analyzed independently by performing a $\chi^2$-minimization between the normalized observed spectrum and 
a grid of synthetic spectra (computed with the code SYNTHE, convolved at DEIMOS resolution and 
resampled at the pixel-size of the observed spectra). Then, the normalization is readjusted locally 
in a region of $\sim$50-60 \AA{} in order to improve the quality of the fit\footnote{No systematic differences
in the iron abundances obtained from FLAMES and DEIMOS spectra have been found for the targets in common between the two datasets}. 
Uncertainties in the fitting procedure for each spectral line were estimated by using 
Monte Carlo simulations: for each line, Poissonian noise was added to the best-fit synthetic spectrum
in order to match the observed SNR, and then the fit was re-computed with the same procedure described 
above. A total of 1000 Monte Carlo realizations has computed for each line, and
the dispersion of the derived abundance distribution was adopted as the abundance uncertainty. 
Typical values are of about $\pm$0.2 dex. 

\subsection{Calibration stars}

Because of its prominence, the Ca~II triplet is commonly used as a
proxy of the metallicity. However, several Fe~I lines fall in the
spectral range of the adopted FLAMES and DEIMOS setups and we
therefore decided to measure the iron abundance directly from these
lines. To demonstrate the full reliability of the atomic data adopted
to derive the metallicity we performed the same analysis on a set of
high-resolution, high-SNR spectra of the Sun and of Arcturus.  For the
Sun we adopted the solar flux spectrum quoted by \citet{neckel}, while
for Arcturus we used the high-resolution spectrum of \cite{hinkle}.
Both spectra were analyzed by adopting the same linelist used for the
targets of this study. For the Sun, the solar model atmosphere
computed by
F. Castelli\footnote{http://wwwuser.oat.ts.astro.it/castelli/sun/ap00t5777g44377k1asp.dat}
was used (T$_{{\rm eff}}$=5777 K, $\log$~$g$=4.44 dex), and v$_{{\rm
    turb}}$=1.2 km$\,$s$^{-1}$ (\citealt{andersen99}) was adopted. For
Arcturus we calculated a suitable ATLAS9 model atmosphere with the
atmospheric parameters (T$_{{\rm eff}}$=5286 K, $\log$~$g$=1.66 dex,
v$_{{\rm turb}}$=1.7 km$\,$s$^{-1}$) listed by \citet{ramirez11}.  The
resulting iron abundance for the Sun is A(Fe~I)$_{Sun}$=7.49$\pm$0.03
dex, in very good agreement with that listed by
\citet[][A(Fe)$=7.50$]{gs98}.  For Arcturus we obtained
A(Fe~I)$_{Arcturus}$=7.00$\pm$0.02, corresponding to
[Fe/H]$=-0.50\pm0.02$ dex, in excellent agreement with the measure of
\citet{ramirez11} who quote [Fe/H]$=-0.52\pm0.02$.  Thus, we conclude
that the adopted atomic lines provide a reliable estimate of the iron
abundance.

As discussed in Section \ref{iron}, for the 158 stars with $0.6<q<0.8$
we limited the metallicity analysis to the iron lines in a restricted
wavelength range ($8680-8850$ \AA) poorly affected by the TiO bands.
In order to properly check for any possible systematic effect due to
the different line list adopted, we re-performed the metallicity
analysis of the 126 stars with $q>0.8$ using only the reduced line
list. We found a very small off-set in the derived abundance
($\delta[Fe/H]=$[Fe/H]$_{full}$-[Fe/H]$_{reduced}=-0.06\pm0.01$ dex), that
was finally applied to the iron abundance obtained for the 158 low-$q$
targets.

\subsection{Uncertainties}

The global uncertainty of our iron abundance estimates (typically
$\sim0.2$ dex) is computed as the sum in quadrature of two different
sources of error. The first one is the error arising from the
uncertainties on the atmospheric parameters. Since they have been
derived from photometry, the formal uncertainty on these quantities
depends on all those parameters which can affect the location of the
targets in the CMD, such as photometric errors, uncertainty on the
absolute and differential reddening and errors on the distance modulus
(DM). In order to evaluate the uncertainties on $T_{eff}$ and
$\log$~$g$ we therefore repeated their estimates for every single
target assuming $\sigma_{K}=0.04$, $\sigma_{J-K}=0.05$,
$\sigma_{\delta[E(B-V)]}=0.1$ (see Section \ref{atmpar}),
$\sigma_{[E(B-V)]}=0.05$ (\citealt{massari}) and a conservative value
$\sigma_{\rm DM}=0.3$ (corresponding to $\pm1$ kpc) for the DM.
Following this procedure we found uncertainties of $\pm160$ K in
T$_{{\rm eff}}$ and $\pm0.2$ dex in $\log$~$g$. For $v_{turb}$ we
adopted a conservative uncertainty of 0.2 km$\,$s$^{-1}$ (see Section
\ref{atmpar}).  To estimate the impact of these uncertainties on the
iron abundance, we repeated the chemical analysis assuming, each time, a
variation by 1$\sigma$ of any given parameter (keeping the other
ones fixed).

The second source of error comes from the internal abundance estimate
uncertainty. For each target this was estimated as the dispersion
around the mean of the abundances derived from the used lines, divided
by the root square of the number of lines. It is worth noticing that for
any given star the dispersion is calculated by weighting the abundance
of each line by its own uncertainty (as estimated by DAOSPEC for the
FLAMES targets, and from Monte Carlo simulations for the DEIMOS
targets).

\subsection{Results}
\label{results_f}
The iron abundances and their total uncertainties for each of the 284
targets analyzed are listed in Table \ref{tab1}, together with the adopted
atmospheric parameters. The [Fe/H] distribution for the entire
sample is shown as a dashed-line histogram in Figure \ref{fedist}. The
distribution is quite broad, extending from [Fe/H]$\simeq-1.2$ dex, up
to [Fe/H]$\simeq0.8$ dex, with a pronounced peak at [Fe/H]$=-0.25$
dex. However, the exclusion of a significant fraction of stars with
spectra seriously contaminated by TiO bands (see Setc. \ref{iron}),
possibly introduced a selection bias on the derived metallicity
distribution.  In fact prominent TiO bands are preferentially expected
in the coolest and reddest stars.  This is indeed confirmed by Figure
\ref{cmd_f}, showing that the targets for which no abundance measure was
feasible (open circles) preferentially populate the brightest and
coolest portion of the RGB.  Since this is also the region were the
most metal-rich stars are expected to be found, in order to provide a
meaningful metallicity distribution, representative of the bulge
population around Terzan 5, we restricted our analysis to a sub-sample
of targets likely not affected by such a bias.  To this purpose, in
the CMD corrected for internal reddening we selected only stars in the
magnitude range $9.2<K_c<9.8$ (see dashed lines in Figure \ref{cmd_f}),
where metallicity measurements have been possible for 82\% of the
surveyed stars (i.e., 112 objects over a total of 136).  The
metallicity distribution for this sub-sample is shown as a grey
histogram in the top panel of Figure \ref{fedist}. The
distribution is still quite broad, extending from [Fe/H]$\simeq-1.2$
dex up to [Fe/H]$\simeq0.7$ dex, but the sub-solar component (with
$-0.5<$[Fe/H]$<0$ dex) seems to be comparable in size to the
super-solar component (with $0<$[Fe/H]$<0.5$ dex).

\begin{figure}
\plotone{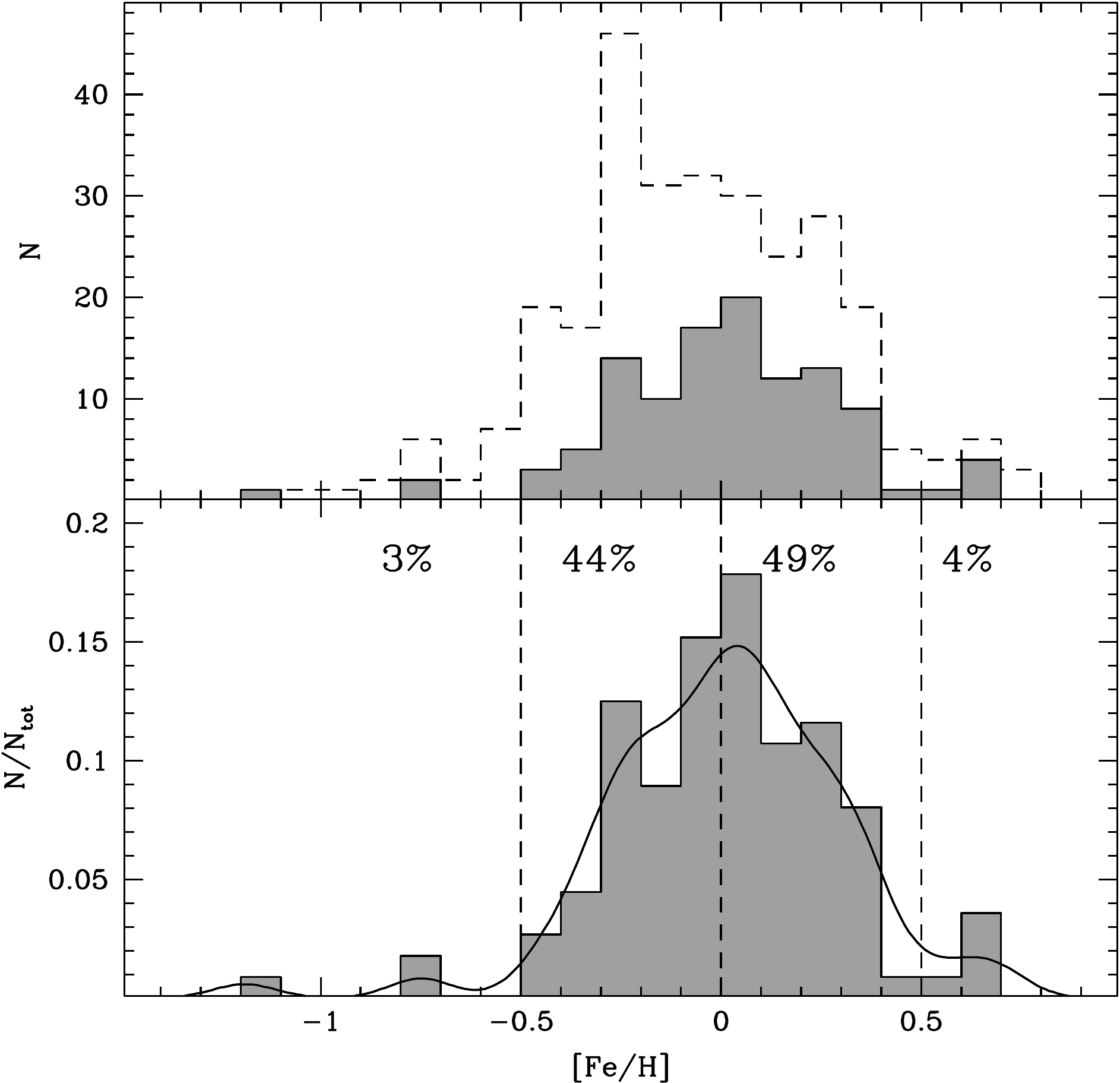}
\caption{\small {\it Top panel:} Metallicity distribution of the bulge
  field around Terzan 5 for the entire sample of 284 stars (dashed
  histogram) and for the sub-set of 112 targets selected at
  $9.2<K_c<9.8$ (grey histogram), free from the bias introduced by the
  TiO bands (which preferentially affects the spectra of the most
  metal-rich objects).  {\it Bottom panel:} Metallicity distribution
  observed in the unbiased sub-set of stars (the same grey histogram
  as above), compared to the generalized distribution (solid line)
  obtained from 1000 realization described in the text (Section
  \ref{results_f}).  The dashed lines delimit the metallicity ranges
  adopted to define the R$_{l/h}$ parameter (see text). The percentage
  of stars in each metallicity range is also marked.}
\label{fedist}
\end{figure} 
 
In order to properly evaluate the existence of any residual bias, we
followed the method described in \cite{zoccali08}.  We considered
three strips in the CMD roughly parallel to the slope of the bulge RGB
(Figure \ref{cmd_f}). In each strip, we computed the fraction $f$
defined as the ratio between the number of stars with measured
metallicity and the number of targets observed in the spectroscopic
surveys. In the selected sub-sample, the $f$ parameter ranges from
0.75 up to 0.90, with the peak in the central bin and with the reddest
bin being the less sampled. In order to evaluate the impact of this
residual inhomogeneity on the derived metallicity distribution we
randomly subtracted from the bluest and central bins a number of stars
(2 and 18, respectively) suitable to make the $f$ ratio constant in
all the strips. We repeated such a procedure $1000$ times, and for
each iteration a new metallicity distribution has been computed. The
bottom panel of Figure \ref{fedist} shows the generalized distribution
obtained from the entire procedure, overplotted to the observed one.
As can be seen, the two distributions are fully compatible, thus
providing the final confirmation that the observed distribution is not
affected by any substantial residual bias. Hence it has been adopted
as representative of the metallicity distribution of the bulge
population around Terzan 5.  

\begin{deluxetable}{rccccccc}
\tablewidth{0pc}
\tablecolumns{8}
\tiny
\tablecaption{Identification number, coordinates, atmospheric parameters, iron abundances and
their uncertainties for the Terzan 5 field stars in our sample.}
\tablehead{\colhead{ID} & \colhead{RA} & \colhead{Dec} & \colhead{T$_{{\rm eff}}$} & \colhead{log~$g$} & \colhead{[Fe/H]} 
& \colhead{$\sigma_{[Fe/H]}$} & Dataset \\
& & & \colhead{(K)} & \colhead{(dex)} & \colhead{(dex)} & \colhead{(dex)} & }
\startdata
 & & \\
  1030711  &  267.1829348  &  -24.6923402  &  3922   &   0.6   &  -0.22    &   0.22   &  FLAMES \\
  1052484  &  267.1470436  &  -24.6830821  &  4220   &   0.8   &  -0.19    &   0.19   &  FLAMES \\
  1071029  &  267.1182806  &  -24.6825530  &  3971   &   1.9   &   0.39    &   0.18   &  FLAMES \\
  1071950  &  267.1169145  &  -24.6630493  &  3832   &   0.9   &   0.24    &   0.19   &  FLAMES \\
  1072160  &  267.1165783  &  -24.6588856  &  4111   &   0.9   &   0.63    &   0.20   &  FLAMES \\
  2009060  &  267.0686838  &  -24.6638585  &  4434   &   0.9   &  -0.34    &   0.17   &  FLAMES \\
  2029939  &  267.0366131  &  -24.6637204  &  4366   &   1.4   &   0.69    &   0.15   &  FLAMES \\
  2065353  &  266.9816844  &  -24.6414183  &  4013   &   0.9   &  -0.04    &   0.26   &  FLAMES \\
  2066891  &  266.9791962  &  -24.6560448  &  4371   &   1.1   &   0.22    &   0.19   &  FLAMES \\
  2068105  &  266.9771994  &  -24.6119254  &  3888   &   0.9   &   0.04    &   0.22   &  FLAMES \\
\enddata
\tablecomments{\small The entire table is available in the online version of the journal.}
\label{tab1}
\end{deluxetable}

In order to compare our results with previous studies, in Figure
\ref{distribs} we show the metallicity distribution obtained in the
present work and those derived in different regions of the Galactic
bulge: $-6$\textdegree$<$b$<-2$\textdegree \citep[for a sub-sample of
  micro-lensed dwarfs][]{bensby13}, the Baade's window (for a
sub-sample of giants; \citealp{hill11} and \citealp{zoccali08}),
l$=-5.5$\textdegree, b$=-7$\textdegree (for a sample of giants;
\citealp{johnson13}), and the \citet{ness13} field closest to the Galactic disk, at
b$=-5$\textdegree. The distributions appear quite
different. However a few common characteristics can be noted and
deserve a brief discussion. Apart from the presence of more or less
pronounced peaks, all the distributions show: ($i$) a major sub-solar
([Fe/H]$\simeq-0.2$ dex) component; ($ii$) a super-solar component
([Fe/H]$\simeq0.2$ dex); ($iii$) a quite extended tail towards low
metallicities (reaching [Fe/H]$\simeq-1.5$ dex). However, the relative
percentage in the two prominent components appears to be different
from one field to another. In order to properly quantify this feature,
we defined the ratio R$_{l/h}=$N$_{l}/$N$_{h}$, where N$_{l}$ is the
number of stars in the sub-solar component (with $-0.5<$[Fe/H]$<0$
dex) and N$_{h}$ is the number of stars in the super-solar component
(with $0<$[Fe/H]$<0.5$ dex). The value of R$_{l/h}$ is labelled in
each panel of Figure \ref{distribs}.

\begin{figure}
\plotone{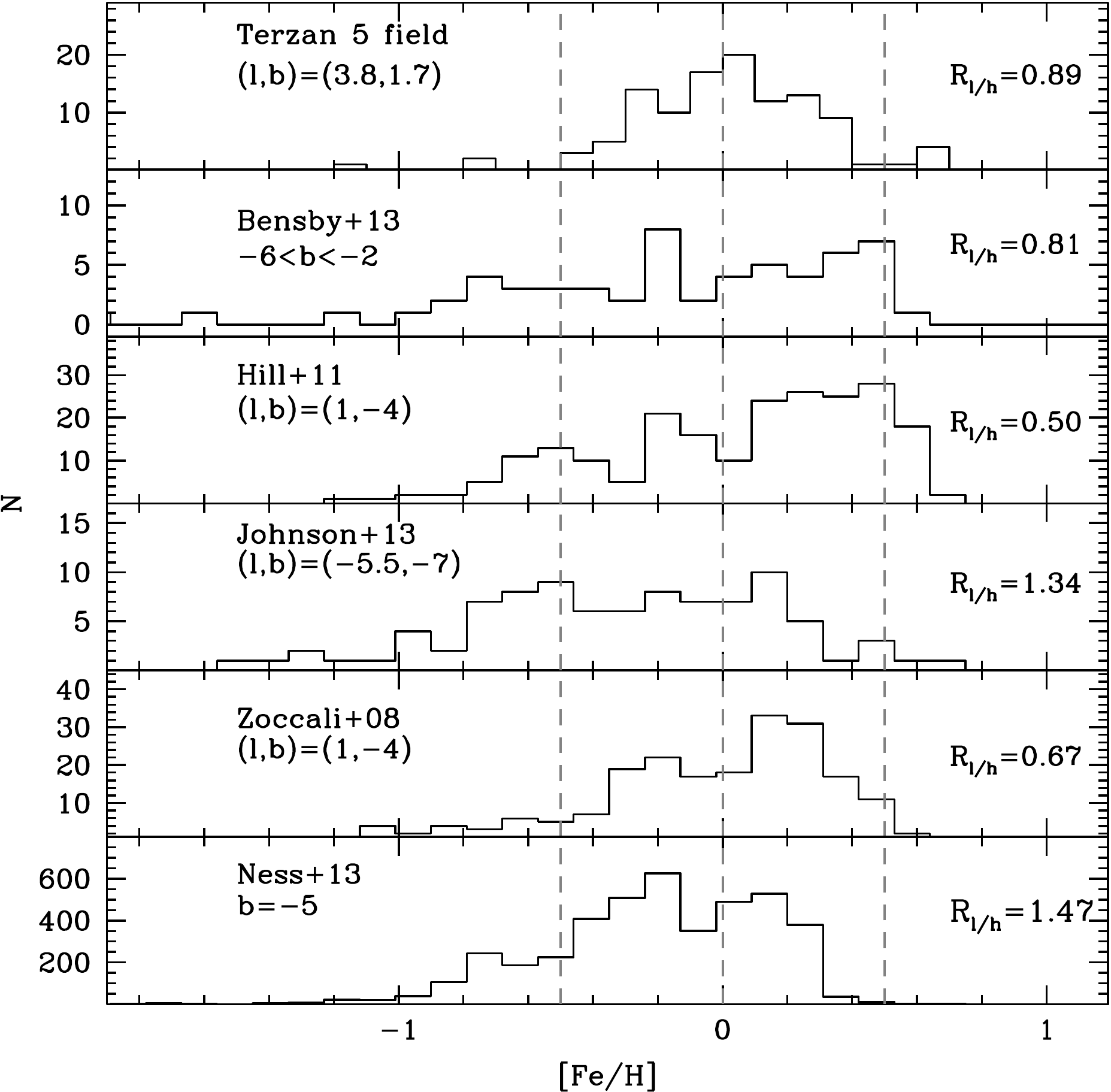}
\caption{\small Comparison of the iron distribution in a few bulge
  fields at different Galactocentric locations.  The corresponding
  references and Galactic coordinates are indicated in each
  panel. Vertical dashed lines delimitate the metallicity ranges
  defining the sub- and super-solar metallicity components.  The value
  of the R$_{l/h}$ parameter defined in the text (Section
  \ref{results_f}) is also reported in each panel. }
\label{distribs}
\end{figure} 

In Figure \ref{grad} we plot the value of R$_{l/h}$ for 13 bulge
regions at different latitudes published in the literature (see
e.g. \citealt{bensby13}, \citealt{hill11}, \citealt{zoccali08},
\citealt{johnson11, johnson13}, \citealt{gonzalez11},
\citealt{ness13}). The value obtained for the bulge field around
Terzan 5 is shown as a large filled circle. It is interesting to note
how the populations observed in the 13 reference fields define a clear
trend, suggesting that the super-solar component tends to be dominant
at latitudes below $|$b$|<5$\textdegree.  The field around Terzan 5
is located at the lowest latitude observed so far. It
nicely fits into this trend, and it suggests the presence of a {\it
  plateau} at R$_{l/h}\sim 0.8$ for $|$b$|<4$\textdegree (see also
\citealt{rich12}). 
On the other hand, the Galactic location of the field can possibly be
the reason for the small amount of stars detected with [Fe/H]$<-0.5$.
Figure \ref{gradpoor} shows the fraction (f$_{{\rm MP}}$) of metal-poor
objects (with [Fe/H]$<-0.5$) with respect to the total number of stars
for each of the samples described in Figure \ref{grad}, as a function of
$|$b$|$.  Also in this case a clear trend is defined: the percentage of
metal-poor stars  drops from $\sim40$\% at $|$b$|\sim10$\textdegree~ to
a few percent at the latitude of Terzan 5, the only exception being
the most external field of \cite{zoccali08}.
Our findings are in good agreement with several recent results 
about the general properties of the Galactic bulge. Indeed, metal-rich 
stars are dominant at low Galactic latitudes, that is closer to the Galactic plane
\citep[see e.g.][and references therein]{ness14}.
Also, we checked the impact on these findings of the assumption of a $8$ kpc distance. 
We repeated the chemical analysis by adopting distances of $6$ and $10$ kpc. 
The change of the surface gravity (on average $+0.25$ dex and $-0.2$ dex, respectively) 
leads to only small differences in the measured 
[Fe/H], with the stellar metallicities differing on average by $0.06$ dex 
($\sigma=0.04$ dex) and $-0.05$ dex ($\sigma=0.02$ dex), in the two cases.
Such a tiny difference moves the value of R$_{l/h}$ from $0.89$ to $0.58$ and $1.26$, 
respectively, leaving it fully compatible with a flat behavior for 
$|$b$|<4$\textdegree~in both cases, while it does not change significantly 
(less than 1\%) the value of f$_{{\rm MP}}$.
We underline that this is the first determination of the 
{\it spectroscopic} metallicity distribution for a significant sample of stars 
at these low and positive Galactic latitudes.  Other spectroscopic surveys at 
low latitudes are needed to confirm the existence of these features.

\begin{figure}
\plotone{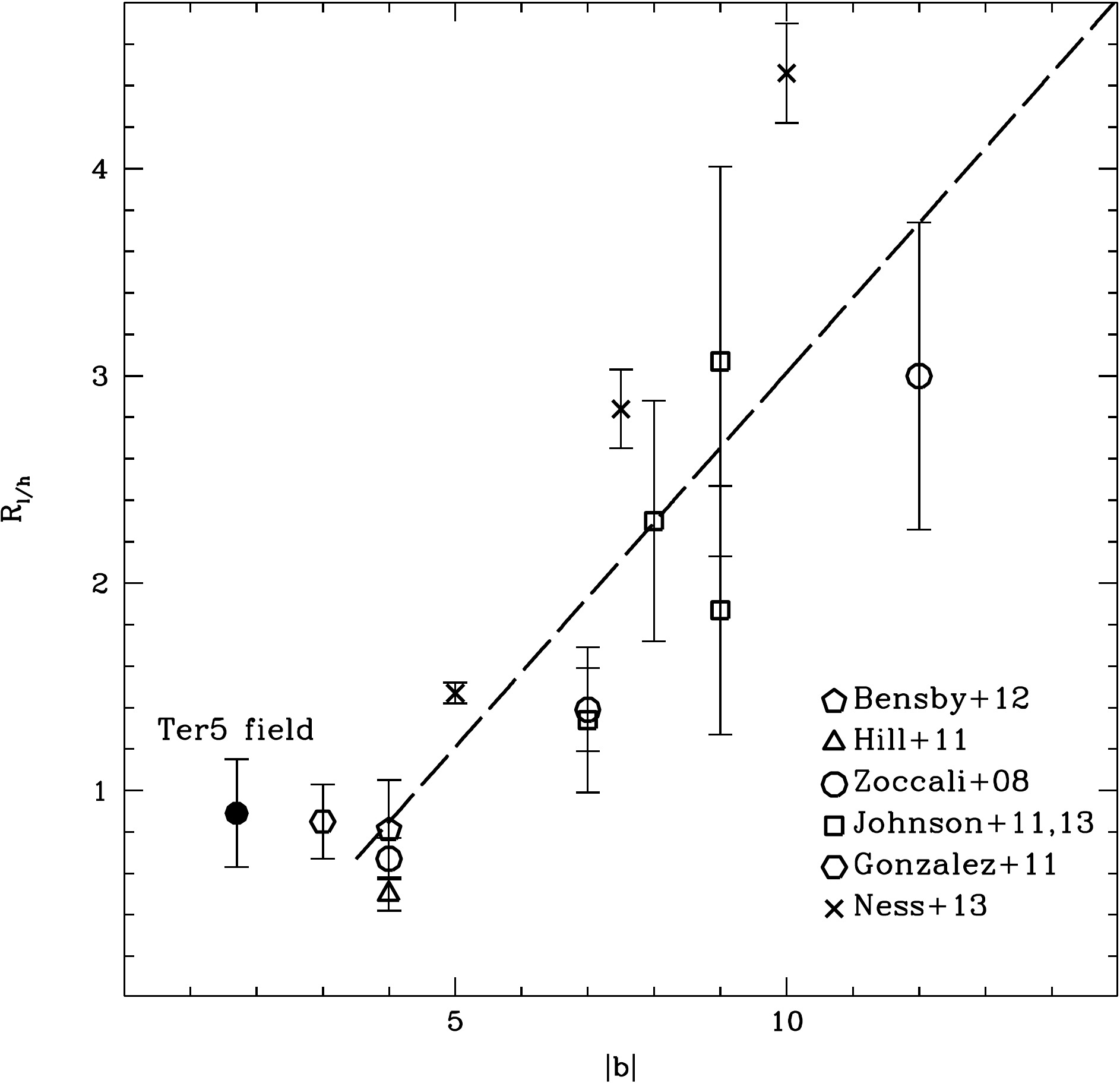}
\caption{\small R$_{l/h}$ parameter as a function of the Galactic
  latitude (absolute value).  The 13 fields taken from the literature
  nicely describe a metallicity gradient, suggesting that the
  super-solar component increases with decreasing Galactic
  latitude. The field measured around Terzan 5 is highlighted with a
  large filled circle and possibly suggests the presence of a
  ``plateau'' at R$_{l/h}\simeq 0.8$ for $|$b$|<4$\textdegree.  }
\label{grad}
\end{figure}

Finally it is worth commenting on the velocity dispersion obtained for
the two main metallicity components in the field surrounding Terzan
5. We found two similar values, $\sigma_v=108\pm8$ km s$^{-1}$ and
$\sigma_v=111\pm11$ km s$^{-1}$ for the sub-solar and the super-solar
component, respectively.  The two measured values are in agreement
with those observed in the fields at b$=-5$\textdegree ~and at low
longitudes (l$<5$\textdegree) by \citet[][see the red diamonds in the
  lower panels of their Figure 7]{ness13kin} in the same metallicity
range (-0.5$<$[Fe/H]$<0.5$ dex).

\begin{figure}
\plotone{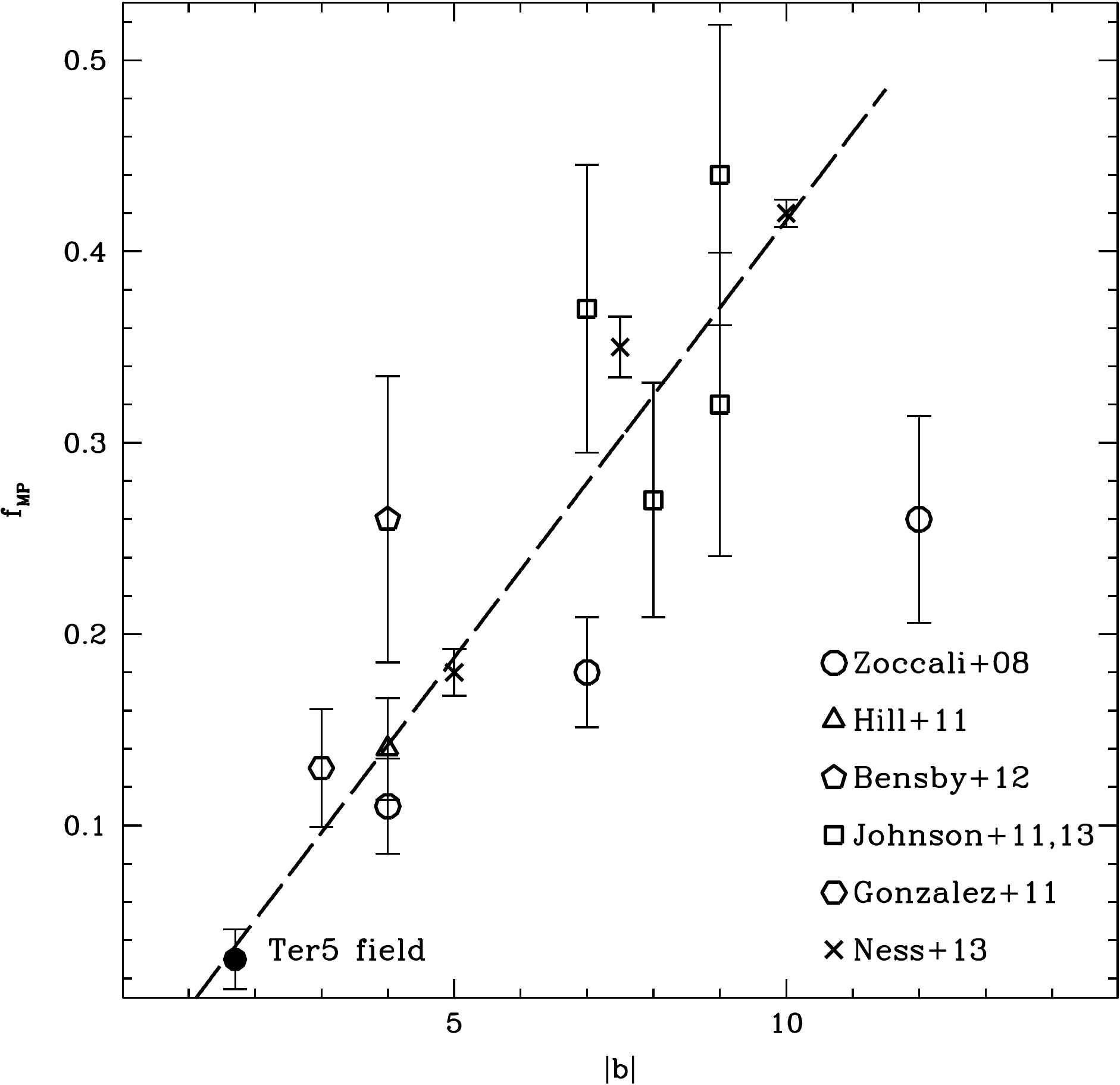}
\caption{\small Fraction of metal-poor stars with [Fe/H]$<-0.5$ dex (f$_{{\rm MP}}$) as a function of the
  absolute value of the Galactic latitude.  The considered bulge fields are the same as those
  in Fig. \ref{grad} and also in this case they describe a clear trend, with the only exception of
  the survey at the largest latitude (from \citealt{zoccali08}). The field measured around Terzan 5 
  is highlighted with a large filled circle and fits very well into the correlation.}
\label{gradpoor}
\end{figure}

\section{Summary}
\label{summary} 
We determined the radial velocity distribution for a sample of 615
stars at (l,b)=(3.7\textdegree,1.8\textdegree), representative of the
bulge field population surrounding the peculiar stellar system Terzan
5. We found that the distribution is well fitted by a Gaussian
function with $<v_{\rm rad}>=21.0\pm4.6$ km s$^{-1}$ and
$\sigma_{v}=113.0\pm2.7$ km s$^{-1}$. Once converted to Galactocentric
velocities, these values are in agreement with the determinations
obtained in other bulge fields previously investigated.  We did
not find evidence for the high-velocity sub-component recently
identified in \cite{nidever}.

Because of the strong contamination of TiO bands, we were able to
measure the iron abundance only for a sample of 284 stars
(corresponding to $\sim 46\%$ of the entire sample) and we could
derive an unbiased metallicity distribution only from a sub-sample
of 112 stars with $9.2<K_c<9.8$. 
Statistical checks have been used to demonstrate that
this is a bias-free sample representative of the bulge population
around Terzan 5.  The metallicity distribution turns out to be quite
broad with a peak at [Fe/H]$\simeq+0.05$ dex and 
it follows the general metallicity-latitude trend
found in previous studies, with the number of super-solar bulge
stars systematically increasing with respect to the number of sub-solar ones
for decreasing latitude. Indeed the population ratio between the sub-solar
and super solar components (quantified here by the parameter R$_{l/h}$)
measured around Terzan 5 nicely agrees with that observed in other low
latitude bulge fields, possibly suggesting the presence of a plateau
for $|$b$|<4$\textdegree. Moreover, also the fraction of stars with
[Fe/H]$<-0.5$ measured around Terzan 5 fits well into the correlation 
with $|$b$|$ found from previous studies.

\clearpage{\pagestyle{empty}\cleardoublepage}

\chapter{The Terzan~5 puzzle: discovery of a third, metal-poor component}\label{chaporiglia}

In this Chapter we present the discovery of 3 red giant stars belonging to
Terzan~5, with metallicity [Fe/H]$\simeq-0.8$ dex, significantly smaller than that of
the sub-Solar component previously detected (F09, O11).

For all the details regarding such a discovery see \cite{o13}.

\section{Observations and chemical abundance analysis}
In the context of an ongoing
spectroscopic survey 
with VLT-FLAMES and Keck-DEIMOS 
of the Terzan 5 stellar populations, 
aimed at constructing a massive 
database of radial velocities and metallicities (\citealt{massari14a}; Ferraro et al.; 2014 in preparation),
we found some indications of the presence of a minor
($\sim 3\%$) component significantly more metal-poor than the
sub-Solar population of Terzan~5.  
We acquired high
resolution spectra of 3 radial velocity candidate metal-poor giants 
members of Terzan~5.
Observations using NIRSPEC \citep{nirspec} at Keck II  
were undertaken on 17 June 2013.   
We used the
NIRSPEC-5 setting to enable observations
in the $H$-band and a
$0.43\arcsec$ slit width that provides an overall spectral resolution
R=25,000. 

Data reduction has been performed by using 
the REDSPEC IDL-based package developed at the UCLA IR Laboratory.
Each spectrum has been sky subtracted by using nod pairs, corrected for flat-field  
and calibrated in wavelength using arc lamps.
An  O-star spectrum observed during the same night has been used to remove
to check and remove telluric features. 
The SNR per resolution element of the final spectra is always $>$30. 
Figure~\ref{spec} shows portions of the observed spectra and the comparison with 
a Terzan~5 giant with similar stellar parameters and higher iron content from the sub-Solar population 
studied by O11.

\begin{figure*}[!hp]
\begin{center}
\includegraphics[scale=0.7]{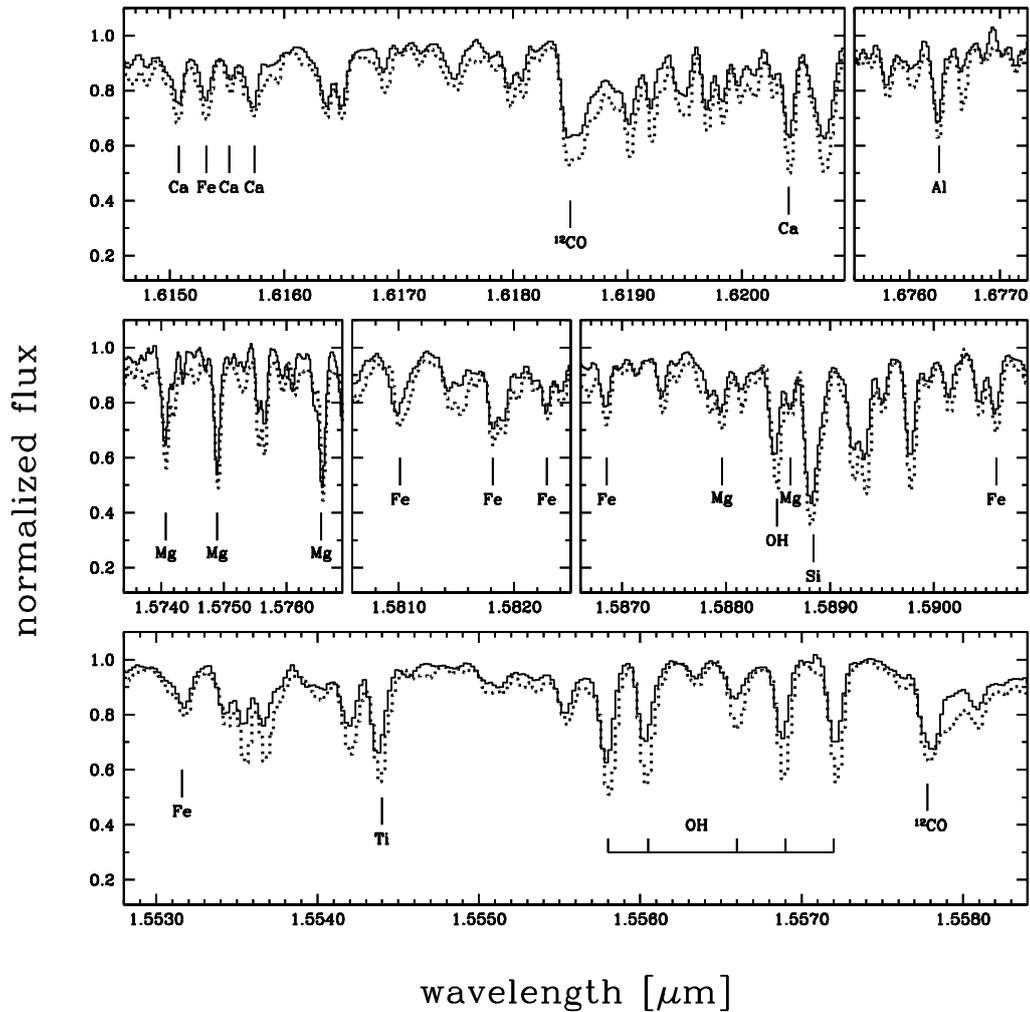}
\caption{Portion of the NIRSPEC $H$-band
  spectra of two red giants of Terzan~5 with similar temperature ($\rm
  T_{eff}\approx 3800$ K), but different chemical abundance patterns
  (solid line for the metal-poor star \#243, dotted line for a
  sub-Solar star at [Fe/H]$\approx$-0.22 from O11).  
  The metal poor giant \#243 has significantly shallower features. 
  A few atomic lines and molecular bands of interest are marked.}
\label{spec}
\end{center}
\end{figure*}  

We compare the observed spectra with synthetic ones 
and we obtain accurate chemical abundances of C and O using molecular lines and 
of Fe, Ca, Si, Mg, Ti and Al using neutral atomic lines, as also 
described in O11 and references therein.

We made use of both spectral synthesis analysis and equivalent width measurements 
of isolated lines.
Synthetic spectra covering a wide range of stellar parameters and elemental abundances  
have been computed by using 
the same code as in O11 and described in detail
in \citet{ori02} and \citet{ori04}. 
The code uses the LTE approximation, the 
molecular blanketed model atmospheres of
\citet{jbk80} at temperatures $\le $4000~K, and 
the \citet{gs98} abundances for the Solar reference.

Stellar temperatures have been first estimated from colors, by using
the reddening estimates by \citet{massari} and the color-temperature
scale by \citet{m98}, calibrated on globular
cluster giants.  Gravity has been estimated from theoretical
isochrones \citep{pie04,pie06}, according to the position of the stars on the
RGB. An average microturbulence velocity of 2 km/s has been adopted 
\citep[see e.g.][ for a detailed discussion]{origlia97}.  
The simultaneous spectral fitting of the CO and OH molecular
lines 
that are especially sensitive to temperature, gravity and
microturbulence variations \citep[see also][]{ori02},
allow us to fine-tune our best-fit adopted stellar parameters.

\section{Results}
Our provisional estimate
for the systemic velocity of Terzan~5, as inferred from
our VLT-FLAMES and Keck-DEIMOS survey,  
is --82 km/s  with a velocity dispersion of $\approx$15~km/s.

From the NIRSPEC spectra we first measured 
the radial velocity of the 3 stars under study and confirm values within $\approx 1\sigma$ 
from the systemic velocity of Terzan~5 (see Table~\ref{tab1_o}).
These stars are located in the central region of Terzan~5, at distances between 13 and 71 arcsec 
from the center (see Table~\ref{tab1_o}). 
Our VLT-FLAMES and Keck-DEIMOS survey shows that in 
this central region the contamination by field stars with similar radial velocities 
and metallicity is negligible (well below 1\%).
Preliminary analysis of proper motions also indicates that these stars are likely members of Terzan~5.

\begin{deluxetable}{lclcrlllllllc}
\tabletypesize{\scriptsize} \tablecaption{Stellar parameters and
abundances for the 3 observed giants in Terzan~5.}
\tablewidth{0pt} \tablehead{ 
\colhead{\#} &  
\colhead{$\rm T_{eff}$}& 
\colhead{log~g}&
\colhead{$\rm v_r^a$}&
\colhead{r$^b$}&
\colhead{$\rm [Fe/H]$}& 
\colhead{$\rm [O/Fe]$}& 
\colhead{$\rm [Si/Fe]$}& 
\colhead{$\rm [Mg/Fe]$}&
\colhead{$\rm [Ca/Fe]$}& 
\colhead{$\rm [Ti/Fe]$}& 
\colhead{$\rm [Al/Fe]$}& 
\colhead{$\rm [C/Fe]$} } \startdata
243 & 3800 & 1.0   & -74 &71  & -0.78  &+0.36  &+0.53  &+0.30  &+0.38  &+0.35 &+0.24  &-0.12  \\
    &&&&  &$\pm$0.02  &$\pm$0.05  &$\pm$0.10  &$\pm$0.03  &$\pm$0.04  &$\pm$0.10  &$\pm$0.10  &$\pm$0.07\\
262 & 4000 & 1.0   &-64  &13  &-0.83  &+0.26  &+0.22  &+0.46  &+0.39  &+0.31  &+0.39  &-0.47  \\
    &&&&  &$\pm$0.08  &$\pm$0.09  &$\pm$0.13  &$\pm$0.08  &$\pm$0.08  &$\pm$0.13  &$\pm$0.13  &$\pm$0.11\\
284 & 3800 & 0.5   &-92 &24    &-0.75  &+0.25  &+0.44  &+0.33  &+0.36  &+0.55  &+0.60  &-0.05  \\
    &&&&  &$\pm$0.05  &$\pm$0.08  &$\pm$0.11  &$\pm$0.13  &$\pm$0.08  &$\pm$0.11  &$\pm$0.11  &$\pm$0.09\\
\enddata                       
\tablenotetext{a}{Heliocentric radial velocity in $\rm km~s^{-1}$.}
\tablenotetext{b}{Radial distance from the center of Terzan~5 in arcsec.}
\label{tab1_o}
\end{deluxetable}

We then measured the chemical abundances of iron, 
alpha-elements, carbon and aluminum.  
Our best-fit estimates of the stellar temperature and gravity,  
radial velocity and chemical abundances with 
$\rm 1\sigma$ random errors are listed in Table~\ref{tab1_o}.  
In the evaluation of the overall error budget we also estimate
that systematics due to 
$\rm \Delta T_{eff}\pm$200~K, 
$\rm \Delta log~g\pm$0.5 dex, $\rm \Delta \xi\pm$0.5 km/s variations in the adopted stellar parameters can affect 
the inferred abundances by $\approx \pm 0.15$ dex. 
However, the derived abundance ratios 
are less dependent 
on the systematic error, since 
most of the spectral features used to measure abundance ratios have similar trends with varying 
the stellar parameters, and 
at least some degeneracy between abundance and the latter is canceled out.

We find the average iron abundance [Fe/H]=--0.79$\pm$0.04 r.m.s.
to be significantly lower (by a factor of $\sim 3$) than the
value of the sub-Solar population 
([Fe/H]$=-0.25$), pointing towards the presence of 
a distinct population in Terzan~5, rather than
to the low metallicity tail of the sub-Solar component.

As shown in Figure~\ref{alpha}, our newly discovered metal-poor 
population has an average $\alpha$-enhancement ([$\alpha$/Fe]$=
+0.36 \pm 0.04$ r.m.s.) similar to that of the sub-Solar one, 
indicating that both populations likely formed early
and on short timescales from a gas polluted by type II SNe.  

\begin{figure*}
\begin{center}
\includegraphics[scale=0.7]{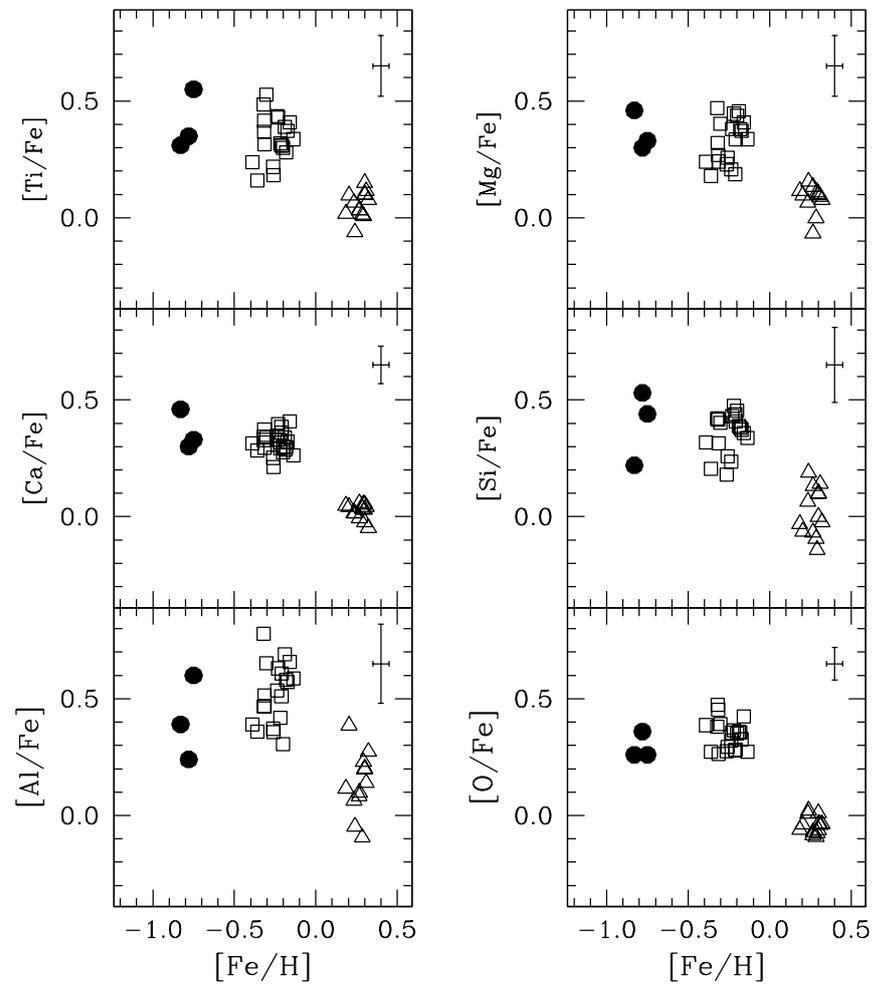}
\caption{Individual [$\alpha$/Fe] and [Al/Fe] abundance ratios as a function of
  [Fe/H] for the 3 observed metal-poor giants (solid dots), and the 20 sub-Solar (open squares) and 13 super-Solar 
  (open triangles) giants from O11, for comparison.
  Typical errorbars are plotted in the top-right corner of each panel.  
  }
\label{alpha}
\end{center}
\end{figure*}

As the stars belonging to the sub-Solar component,
also these other giants with low iron content show an enhanced
[Al/Fe] abundance ratio (average [Al/Fe]$=+0.41 \pm 0.18$ r.m.s.) and
no evidence of Al-Mg and Al-O anti-correlations, and/or large [O/Fe]
and [Al/Fe] scatters, although no firm conclusion can be drawn with
3 stars only.

We also measured some [C/Fe] depletion 
(at least in stars \#243 and \#262),
as commonly found in giant stars and explained with mixing
processes in the stellar interiors during the evolution along the RGB.

\section{Discussion and Conclusions}\label{discussoriglia}

New spectroscopic observations of 3 stars, members of
Terzan~5, have provided a further evidence of the complex nature of
this stellar system and of its likely connection with the bulge
formation and evolution history.

We find that Terzan~5 hosts a third, metal-poorer population with
average [Fe/H]=$-0.79 \pm 0.04$ r.m.s. and [$\alpha$/Fe] enhancement.
From our VLT-FLAMES/Keck-DEIMOS survey,
we estimate that this component represents a minor fraction (a few percent) of 
the stellar populations in Terzan~5. 

Notably, a similar fraction ($\approx 5\%$) of metal-poor stars ([Fe/H]$\approx
-1$) has been also detected in the bulge \citep[see e.g.][and
references therein]{ness13,ness13kin}. 
This metal-poor population shows a kinematics
typical of a slowly rotating spheroidal or a metal weak thick disk
component.  

Our discovery significantly enlarges the metallicity range
covered by Terzan~5, which amounts to $\Delta$[Fe/H]$\approx 1$ dex.
Such a value is completely unexpected
and unobserved in genuine globular clusters. Indeed, within the Galaxy
only another globular-like system, namely $\omega$Centauri, 
harbors stellar populations with a
large ($>$1~dex) spread in iron \citep{norris95, sollima05, johnson10, pancino11}. 
This evidence strongly sets Terzan 5 and
$\omega$Centauri apart from the class of genuine globular clusters,
and suggests a more complex formation and evolutionary history for these two
multi-iron systems. 

It is also interesting to note that detailed
spectroscopic screening recently performed in $\omega$Centauri
revealed an additional sub-component 
significantly more metal-poor (by $\Delta$[Fe/H]$\sim 0.3-0.4$ dex)
than the dominant population \citep{pancino11}. 
The authors suggest that this is best accounted for in 
a self-enrichment scenario, where
these stars could be the remnants of the fist stellar generation in
$\omega$Centauri.  

The three populations of Terzan~5 
may also be explained with some self-enrichment.
The narrow peaks in their metallicity distribution can be the result of  
a quite bursty star formation activity in the proto-Terzan~5, which should
have been much more massive in the past to retain the SN ejecta and
progressively enrich in metals its gas.  However, Terzan~5 might also
be the result of an early merging of fragments with sub-Solar metallicity
at the epoch of the bulge/bar formation, and with younger and more
metal-rich sub-structures following subsequent interactions with the
central disk.

However, apart from the similarity in terms of large iron range and possible 
self-enrichment, $\omega$Centauri and Terzan 5 likely had quite different origins 
and evolution.
It is now commonly accepted that $\omega$Centauri can be the remnant of a dwarf
galaxy accreted from outside the Milky Way \citep[e.g.][]{bekki03}.
At variance, the much higher metallicity of Terzan~5 and its chemical similarity 
to the bulge populations suggests some \emph{symbiotic} evolution
between these two stellar systems.

\clearpage{\pagestyle{empty}\cleardoublepage}

\chapter{The metallicity distribution of the stellar system Terzan~5}\label{chapmdf}

In order to accurately reconstruct the evolutionary history of Terzan 5, a first crucial 
step is to precisely determine the metallicity distribution of its stellar populations,
based on a statistically significant sample of stars. In this Chapter
we present and discuss the iron abundances measured for a sample of 220 giants distributed 
over the entire radial extent of Terzan 5, from the innermost regions, out to the tidal
radius.

All the details of this work can be found in \cite{massari14b}.

\section{Observations and data reduction}\label{obsmdf}

This work is part of a large spectroscopic survey of stars in the direction of
Terzan 5, aimed at characterizing the kinematical and chemical properties of the stellar populations
within the system and in the surrounding Galactic bulge field. While the overall
survey will be described in a forthcoming paper (Ferraro et al. 2014 in preparation) and the 
properties of the field around Terzan 5 have
been discussed in Chapter \ref{chapfield} and in \citet[][hereafter M14a]{massari14a}, here we focus on 
the metallicity distribution of Terzan 5.

This study is based on a sample of stars located within the tidal radius of Terzan 5 
(r$_{t}\simeq300\arcsec$; \citealt{l10, miocchi13}) observed with two different instruments:
FLAMES \citep{pasquini} at the ESO-VLT and DEIMOS (\citealt{faber}) at the Keck II
Telescope. The spectroscopic targets have been selected from the optical photometric
catalog of Terzan~5  described in \citet{l10} along the brightest portion ($I<17$) of the
RGB. In order to avoid contamination from other sources, in the selection
process of the spectroscopic targets we avoided stars with bright
neighbors (I$_{neighbor}<{\rm I_{star}}+1.0$) within a distance of 2\arcsec.    
The spatial distribution of the observed targets is shown in Fig.~\ref{mapmdf}.
 
\begin{figure}
\plotone{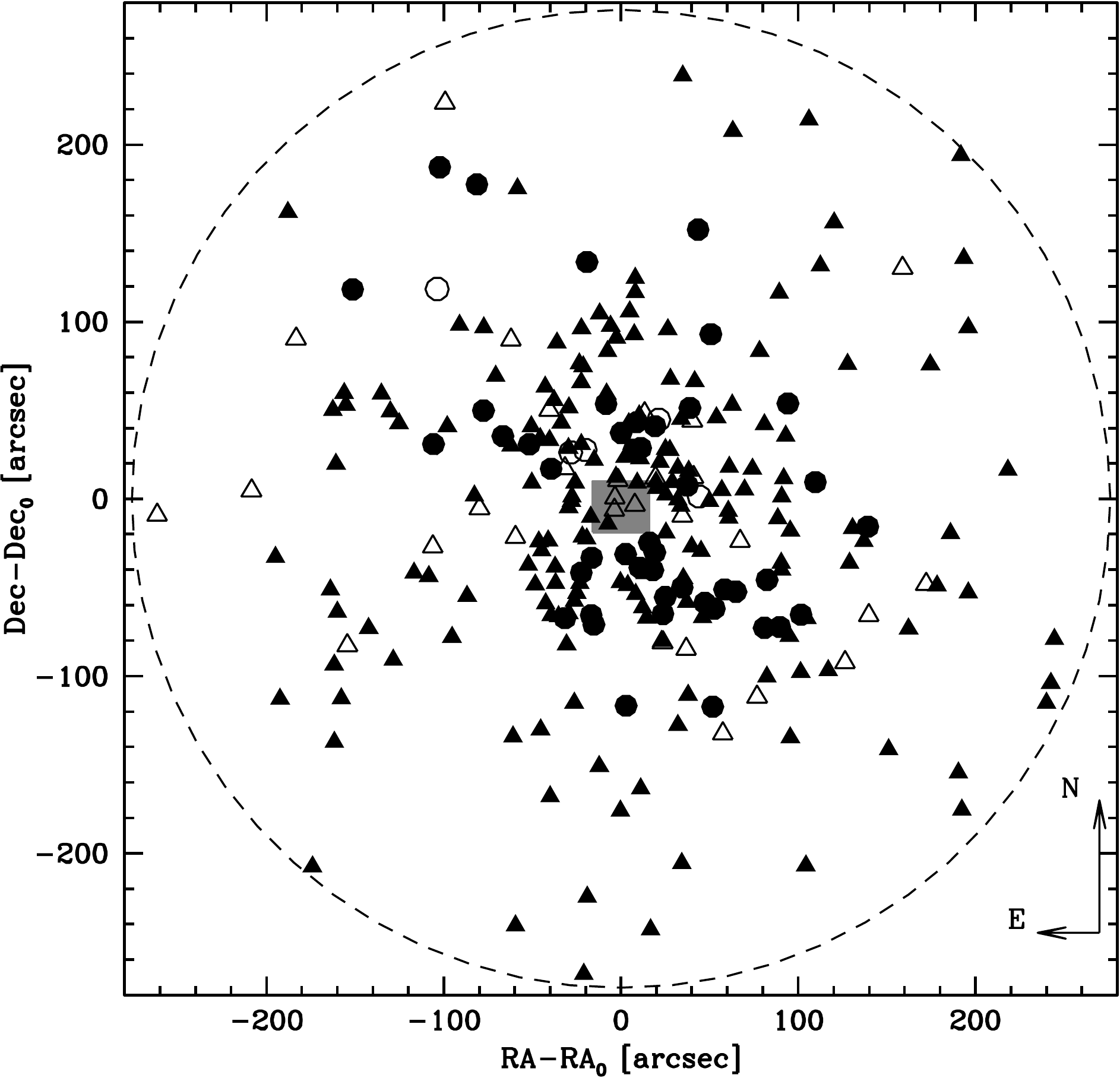}
\caption{\small Spatial distribution of the spectroscopic targets in Terzan 5. 
FLAMES and DEIMOS targets are shown as triangles and circles, respectively. 
The central gray square marks the region where the NIRSPEC targets are located
(see Section \ref{samplemdf} for the details about the membership). 
Filled symbols mark targets for which the iron abundance was measured while 
empty symbols are used to indicate targets affected by TiO contamination for which 
no abundance determination was possible. The dashed circle
marks the tidal radius of the system, $r_t=276\arcsec =7.9$ pc ($100\arcsec$ corresponding to 2.86 pc at the distance of
Terzan 5).}
\label{mapmdf}
\end{figure}

(1)~{\it FLAMES dataset}--- This dataset has been collected under three different programs (ID: 087.D-0716(B), PI: Ferraro, ID:
087.D-0748(A), PI: Lovisi and ID: 283.D-5027(A), PI: Ferraro). As already described in
M14a, all the spectra have been obtained using the HR21 setup in the GIRAFFE/MEDUSA mode,
providing a resolving power of R$\sim16200$ and a spectral coverage ranging from 8484 \AA{} to
9001 \AA{}.  This grating has been chosen because it includes the prominent Ca~II triplet
lines, which are widely used features for radial velocity estimates, even in low
SNR spectra.  Several metal lines (mainly of Fe~I) lie in this spectral range, thus
allowing a direct measurement of [Fe/H]. In order to reach SNR$\sim$40-50 even for the
faintest ($I\sim17$) targets, multiple exposures with integration  times ranging from 1500 s to
2400 s (depending on the magnitude of the targets) have been secured for the majority  of
the stars.  In order to reduce the acquired spectra we used the FLAMES-GIRAFFE ESO
pipeline\footnote{http://www.eso.org/sci/software/pipelines/}. This
includes bias-subtraction, flat-field correction, wavelength calibration with a
standard Th-Ar lamp, resampling at  a constant pixel-size and extraction of one-dimensional
spectra.  Because of the large number of O$_{2}$ and OH emission lines in this spectral range,
a correct sky subtraction is a primary requirement.  Thus, in each exposure $15$-$20$
fibers have been used to measure the sky. The master sky spectrum obtained as the median of
these spectra has been then subtracted from the stellar ones. Finally, all the spectra have
been reported to zero-velocity and in the case of multiple exposures they have been co-added
together. 

(2)~{\it DEIMOS dataset}--- This spectral dataset has been acquired 
by using the 1200 line/mm grating coupled  with the GG495 and GG550
order-blocking filters.  The spectra cover the $\sim$6500-9500 \AA{} wavelength range with  a
resolution of R$\sim7000$ at $\lambda\sim8500$ \AA{}.  An exposure time of $600$ s for
each pointing allowed to reach SNR$\sim50-60$ for the brightest stars and  SNR$\sim15-20$ for
the faintest ones ($I\sim17$ mag). We used the package  described in \cite{ibata11} for an
optimal reduction and extraction of the DEIMOS spectra. 

For sake of comparison, Fig.~\ref{specmdf} shows two spectra of the same star observed with 
FLAMES (top panels) and with DEIMOS (bottom panels).

\begin{figure}
\plotone{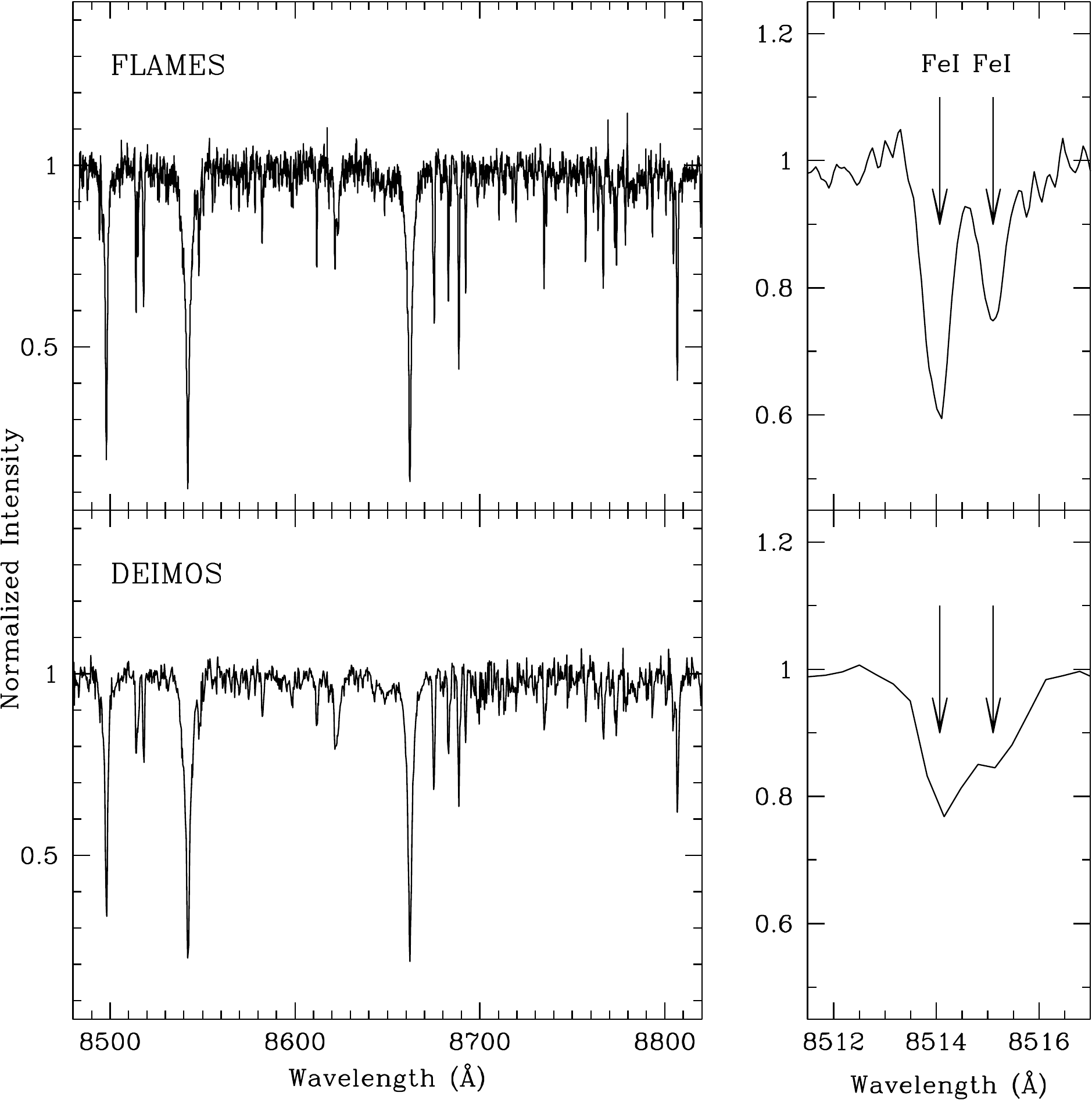}
\caption{\small The Ca II triplet spectral region for star 34, as obtained from  FLAMES (left-upper panel) and  
DEIMOS (left-lower panel) observations. The right panels show the zoomed spectra around two Fe~I lines 
used in the analysis.}
\label{specmdf}
\end{figure} 

\section{Analysis}\label{atmparmdf}

\subsection{Atmospheric parameters}\label{vturb}

T$_{{\rm eff}}$ and $\log$ $g$ for each target have been derived 
from near infrared photometry in order to minimize the effect of possible residuals  in the
differential reddening correction.  The $(K, J-K)$ CMD has been obtained by combining the SOFI
catalog of \citet{valenti07} for the central  $2.5$\arcmin$ \times 2.5$\arcmin~ and 2MASS
photometry in the outermost regions. 
Magnitudes and colors of each star 
have been corrected for differential extinction according to their spatial location with respect to the 
center of Terzan 5. For stars in the innermost regions, lying within the FoV of the 
ACS/HST observations (see \citealt{l10}), the reddening map 
published in \citet[][, see Chapter \ref{chapred}]{massari} has been adopted. Instead  
the correction for stars in the outer regions has been estimated from
the new differential reddening map described in M14a (see Chapter \ref{chapfield}). 
The target positions in the reddening-corrected CMD 
are shown in Fig.~\ref{cmdsmdf}. In
order to estimate T$_{{\rm eff}}$ and $\log$~$g$, the position of each target in the
reddening-corrected CMD has been projected onto a reference isochrone. 
Following F09,  we adopted a 12 Gyr-old isochrone 
extracted from the BaSTI database (\citealt{pie06}) 
with metallicity Z$=0.01$ (corresponding to [Fe/H]$=-0.25$),
$\alpha$-enhanced chemical mixture and helium content Y$=0.26$ dex  (well reproducing the dominant stellar population 
in Terzan 5, see O11). The isochrone is shown as dashed line in Fig.~\ref{cmdsmdf}.
Since Terzan~5 hosts at least two stellar populations, but they are photometrically
indistinguishable in the  near-infrared plane, in Section \ref{caveatsmdf}, we discuss the effect
of using isochrones with different metallicities and ages.

\begin{figure}
\plotone{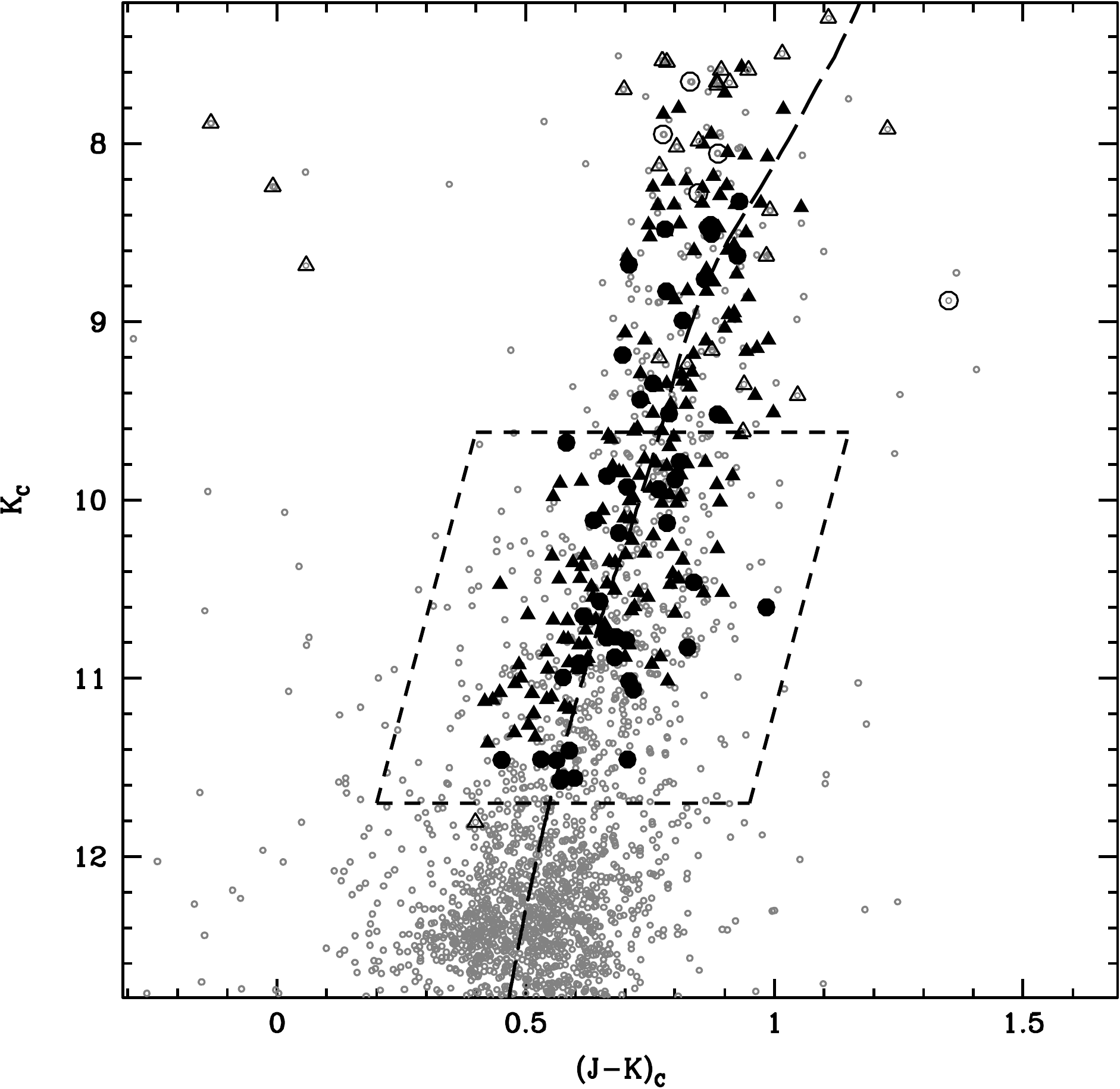}
\caption{\small Infrared CMD of Terzan~5 corrected for differential reddening. 
Symbols are as in Fig.~\ref{mapmdf}, with empty  symbols marking the targets affected
by TiO contamination.
The BaSTI isochrone with an age of 12 Gyr and metallicity Z=0.01 used to derive the atmospheric 
parameters is also shown as a long-dashed line.
The box delimited by the short-dashed line indicates the sample not affected by TiO contamination 
that was selected 
to compute  
the metallicity distribution.}
\label{cmdsmdf}
\end{figure}

As already explained in M14a, the small number (about 10) of Fe~I lines
observed in the FLAMES and DEIMOS spectra  (see Section \ref{analysismdf}) prevents us from deriving a reliable 
spectroscopic determination of $v_{turb}$ (see \citealt{m11} for a review of the different
methods to estimate this parameter). Therefore, for homogeneity with our previous work we adopted
the same value, $v_{turb}=$1.5 km$\,$s$^{-1}$, which is a reasonable assumption for cool giant stars
(see also \citealt{zoccali08, johnson13}).

\subsection{Chemical analysis}\label{analysismdf}

We adopted the same Fe~I linelist and the same techniques to analyze the spectra and to determine the chemical 
abundances as those used in M14a.

(1)~{\it FLAMES data-set}---
We performed the chemical analysis using the package GALA \citep{gala}, an
automatic tool to derive chemical abundances  of single, unblended lines by using their measured
EWs.  The adopted model atmospheres have been calculated with the ATLAS9 code
\citep{atlas}. Following the  prescriptions by M14a, we performed the analysis
running GALA with  all the model atmosphere parameters fixed and allowing 
only the metallicity to vary  iteratively in order to match the iron
abundance measured from EWs. The latter were measured by using
the code 4DAO (\citealt{4dao}).  This code runs DAOSPEC \citep{daospec} for
large sets of spectra, tuning automatically the main input parameters used by DAOSPEC. It also
provides graphical outputs  that are fundamental to visually check the quality of the fit for each
individual spectral line. EW errors are estimated by DAOSPEC as the standard deviation of the local
flux residuals \citep[see][]{daospec}.  All the lines with EW errors
larger than 10\% were excluded from the analysis.

(2)~{\it DEIMOS data-set}---
The lower resolution of DEIMOS causes a high degree of line blending and blanketing in the
observed spectra. The
derivation of the abundances through the method of the EWs is therefore
quite uncertain. 
Thus, the iron abundances for this dataset have been measured by  comparing the observed spectra
with a grid of synthetic spectra,  according to the procedure described in \citet{m12_2419}. Each
Fe~I line has been  analyzed individually by performing a $\chi^2$-minimization between the
normalized observed spectrum and  a grid of synthetic spectra. The synthetic spectra have been computed with the
code SYNTHE (\citealt{sbordone04}) assuming the proper atmospheric parameters for each
individual star, then convolved at the DEIMOS resolution and finally resampled at the pixel size of the observed spectra.
To improve the quality of the fit, the normalization is iteratively readjusted locally
in a region of $\sim$50-60 \AA{}.  We estimated the uncertainties in the fitting procedure for each
spectral line by using  Monte Carlo simulations: for each line, Poissonian noise is added to the
best-fit synthetic spectrum in order to reproduce the observed SNR and then the fit is re-computed
as described  above.  The dispersion of the abundance distribution derived from 1000 Monte Carlo
realizations has been  adopted as the abundance uncertainty (typically about $\pm$0.2
dex). 

\section{Error budget}\label{caveatsmdf}

In order to verify the robustness of our abundance analysis, in the following we discuss
the effect of each specific assumption we made and the global uncertainty on the
iron abundance estimates.

\subsection{Systematic effects}

\begin{enumerate}
\item {\it Choice of the isochrone.}
The atmospheric parameters of the selected targets have been determined from the projection
onto an isochrone corresponding to the old, sub-solar population 
(see Fig.\ref{cmdsmdf}). However, as discussed by F09 and O11, Terzan~5 hosts at least two stellar populations with different 
iron abundances and possibly ages. 
In order to quantify the effect of using isochrones with different metallicity/age, 
we re-derived the atmospheric parameters by using a BaSTI isochrone 
\citep{pie04} with an age of 6 Gyr, Z=+0.03 and a solar-scaled mixture (corresponding to [Fe/H]=+0.26 dex). 
The temperatures of the targets decrease by less than 200 K and the gravities 
increase by $\sim$0.2 (as a consequence of the larger evolutive mass). By re-analyzing the spectra of these stars 
with the new parameters, we obtained very similar iron abundances, the mean difference and rms scatter being
$\langle$[Fe/H]$_{{\rm 6~Gyr}}$-[Fe/H]$_{{\rm 12~Gyr}}\rangle=0.00$ dex and $\sigma=0.12$ dex, respectively.
We performed an additional check by adopting the metallicity of the extreme metal poor component  
([Fe/H]$\simeq-0.8$ dex), by using 
a BaSTI isochrone with an age of 12 Gyr, Z$=0.004$ and $\alpha$-enhanced (corresponding to [Fe/H]$\simeq-1$ dex), 
finding that iron abundances increase only by about 0.06 dex.

\item {\it Temperature scale.} To check the impact of different T$_{{\rm eff}}$ scales we derived the 
atmospheric parameters by adopting the Dartmouth \citep{dotter} and Padua \citep{marigo} isochrones, 
and we found negligible variations ($\delta T_{{\rm eff}}\le$50 K). 
Also the adoption of the ($J-K$)--T$_{eff}$ empirical scale by \citet{m98}
has a marginal impact (smaller than 100 K) on the derived temperatures. Such differences lead to iron 
variations smaller than 0.05 dex.

\item {\it Microturbulent velocities.} The assumption of a different 
value of $v_{turb}$ has the effect of shifting the metallicity distribution, without 
changing its shape. Typically, a variation of $\pm$0.1 km$\,$s$^{-1}$ 
leads to iron abundance variations of $\mp$0.07-0.1 dex. Given the typical dispersion of $v_{turb}$
for this kind of stars (see M14a), this effect would lead to a systematic shift of 
the distribution of a few tenths of dex. However the nice match between the abundances measured in these work and 
those obtained by O11 and \citet[][hereafter O13]{o13} from higher-resolution spectra for the targets in common 
(see Section \ref{samplemdf}) demonstrates that our choice of $v_{turb}$ is adequate.

\item {\it Model atmospheres.} We repeated the analysis of the targets by adopting MARCS \citep{gustaf} 
and ATLAS9-APOGEE \citep{mesz}
model atmospheres, instead of the ATLAS9 models by \citet{atlas}. The adoption of different model atmospheres 
calculated assuming different lists for opacity, atomic data and computation recipes leads to 
variations smaller than $\pm$0.1 dex in the [Fe/H] determination, 
and it does not change the shape of the metallicity distribution.

\end{enumerate}

\subsection{Abundance uncertainties}\label{uncert}

As discussed in M14a, the global uncertainty of the derived iron abundances (typically
$\sim0.2$ dex) has been computed as the sum in quadrature of two different
sources of error.

{\it (i)}  The first one is the error arising from the
uncertainties on the atmospheric parameters.
Since they have been derived from photometry, the formal uncertainty on these quantities
depends on all those parameters which can affect the location of the
targets in the CMD, such as photometric errors ($\sigma_{{\rm K}}$ and $\sigma_{{\rm J-K}}$
for the magnitude and the color, respectively), 
uncertainty on the absolute and differential reddening ($\sigma_{{\rm[E(B-V)]}}$ and 
$\sigma_{{\rm \delta[E(B-V)]}}$, respectively) and errors on the distance modulus
($\sigma_{{\rm DM}}$). In order to evaluate the uncertainties on $T_{{\rm eff}}$ and
$\log$~$g$ we therefore repeated the projection onto the isochrone for every single
target assuming $\sigma_{K}=0.04$, $\sigma_{J-K}=0.05$, $\sigma_{\delta[E(B-V)]}=0.05$
for the targets in the ACS sample (\citealt{massari}),
$\sigma_{\delta[E(B-V)]}=0.1$ for targets in the WFI FoV (M14a), and
$\sigma_{[E(B-V)]}=0.05$ and $\sigma_{DM}=0.05$ (\citealt{valenti10}). 
We found that uncertainties on T$_{{\rm eff}}$
range from $\sim$60 K up to $\sim$120 K, and those on $\log$~$g$ are of the order of 0.1-0.15 dex. 
For $v_{turb}$ we adopted a conservative uncertainty of 0.2 km$\,$s$^{-1}$.

{\it (ii)} The second source of error is the internal abundance 
uncertainty. For each target this was estimated as the dispersion
of the abundances derived from the lines used, divided
by the squared root of the number of lines. It is worth noticing that, for
any given star, the dispersion is calculated by weighting the abundance
of each line by its own uncertainty (as estimated by DAOSPEC for the
FLAMES targets, and from Monte Carlo simulations for the DEIMOS
targets).

\section{Results}\label{resultsmdf}

\subsection{Metallicity distribution}\label{samplemdf}

In order to build the metallicity distribution of Terzan 5, we selected bona fide
members according to the following criteria:

{\it (i)}~we considered only stars within the tidal radius of Terzan~5 ($\sim4.6$\arcmin, \citealt{l10}, see also \citealt{miocchi13});

{\it (ii)}~we considered stars with radial velocities within $\pm 2.5\sigma$
(between $-123$ km$\,$s$^{-1}$ and $-43$ km$\,$s$^{-1}$) 
around the systemic radial velocity of
Terzan 5 (v$_{{\rm rad}}\simeq-83$ km$\,$s$^{-1}$, Ferraro et al. 2014 in preparation);

{\it (iii)}~we discarded spectra affected by TiO molecular bands, which can make difficult the
evaluation of the continuum level and in the most extreme cases they completely hide the
spectral lines of interest. To evaluate the impact of TiO bands on the observed spectra we 
followed the strategy described in M14a, adopting the same q-parameter 
(defined as the ratio between the deepest feature of the TiO band at
$\sim8860$ \AA{} and the continuum level measured in the adjacent spectral range
$8850$\AA{}$<\lambda<8856$ \AA{}). 
Thus we analyzed the full set of absorption lines in all the targets
with q$>0.8$, while we adopted a reduced linelist (by selecting only iron absorption 
lines in the range $8680$ \AA{}$<\lambda<8850$ \AA{}, which are only marginally affected by
TiO contamination) for stars with $0.6<$q$<0.8$, and we
completely discarded all the targets with q$<0.6$ (see the empty symbols in Figure \ref{cmdsmdf}).

Following these criteria, we selected a sample of 224 stars (170 from the FLAMES dataset and 54
from the DEIMOS dataset). A few stars observed with different instruments were used to check the
internal consistency of the measures.   
In fact, three DEIMOS targets are in common with the FLAMES sample and the average difference between the
metallicity estimates is  $\langle$[Fe/H]$_{{\rm DEIMOS}}-$[Fe/H]$_{{\rm FLAMES}}\rangle$=+0.07$\pm$0.06 ($\sigma$=0.11
dex).  One DEIMOS target is in common with the NIRSPEC sample by O11 and we find 
[Fe/H]$_{{\rm DEIMOS}}$-[Fe/H]$_{{\rm NIRSPEC}}=+0.02$ dex. Finally, three metal-poor FLAMES stars have been
observed at higher spectral resolution with NIRSPEC by O13, and the average difference between
the iron abundance estimates is $0.01 \pm0.02$ dex ($\sigma=0.03$), only. Hence we can conclude that 
iron abundances obtained from different instruments are in good agreement (well within the
errors). For those stars with multiple measurements we adopted the iron abundance obtained from the dataset
observed at higher spectral resolution. Thus the selected sample numbers 220 stars.

As discussed in detail in M14a, the rejection of targets severely contaminated by
TiO bands introduces a bias that leads to the systematic exclusion of metal-rich stars.
To avoid such a bias, we will focus the analysis only on a sub-sample of stars selected
in a magnitude range ($9.6<{\rm K}_c<11.7$) where no targets have been discarded because of 
TiO contamination.
Thus, the final sample discussed in the following contains a total of 135 stars 
and their measured iron abundances and final uncertainties 
(computed as described in Section \ref{uncert}), together with the adopted atmospheric parameters, 
are listed in Table \ref{tab1mdf}. 
The [Fe/H] distribution for these 135 targets is shown in Figure \ref{mdf}.  

\begin{figure}
\plotone{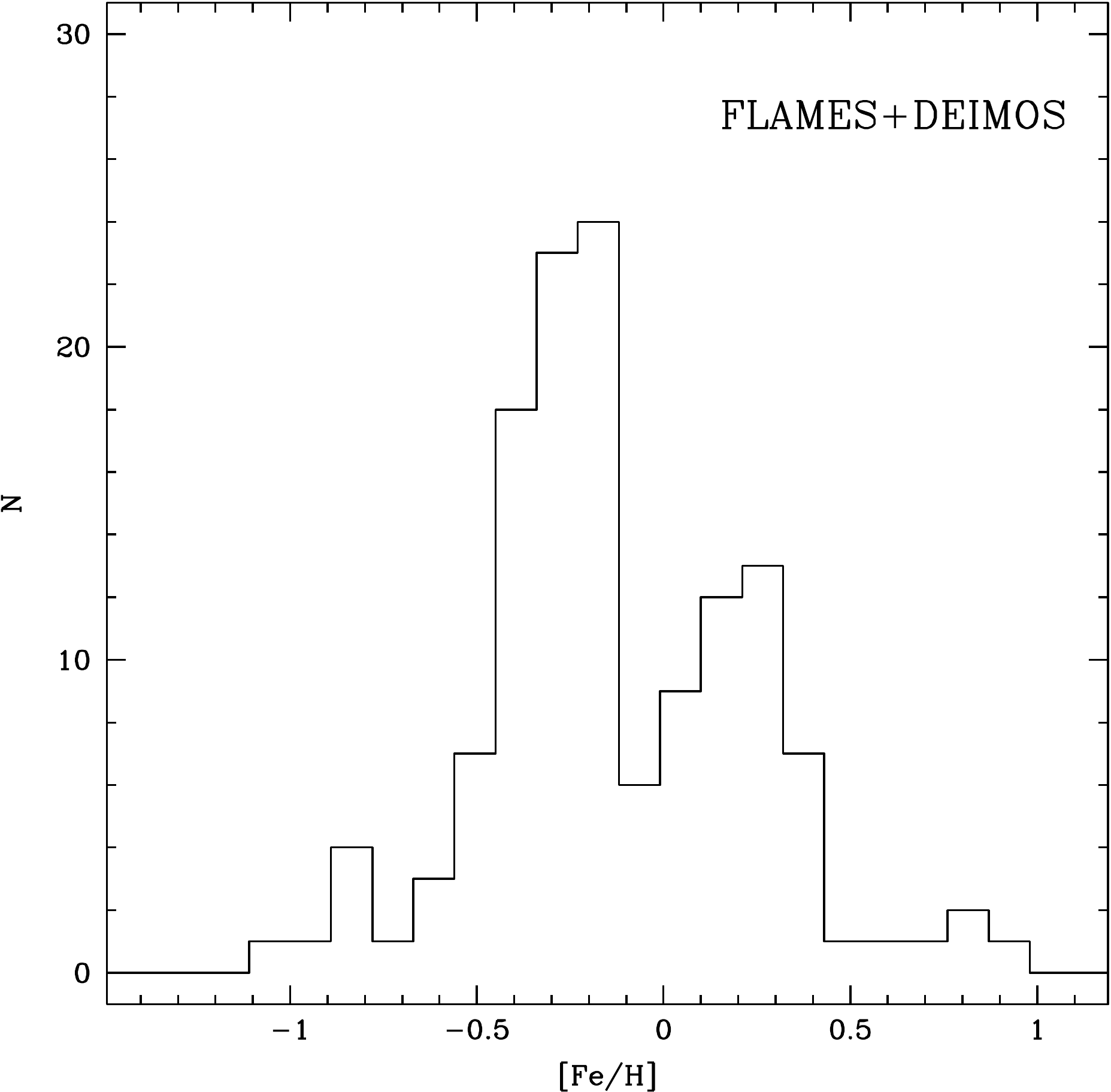}
\caption{\small Metallicity distribution obtained for the unbiased FLAMES+DEIMOS sample (135 targets selected
in the magnitude range $9.6<$K$_c<11.7$), before the statistical decontamination. }
\label{mdf}
\end{figure}

It is quite
broad, extending from [Fe/H]=--1.01 to +0.94 dex, with an average value of [Fe/H]$=-0.12$ and a dispersion
$\sigma=0.35$, much larger than the typical uncertainty on the abundance estimates. 
More in details, the observed distribution shows a main peak at [Fe/H]$\sim-0.30$ dex
and a secondary component at [Fe/H]$\sim+0.30$ dex, in very good agreement with the results of O11.
Also the third component discovered by O13 is clearly visible at [Fe/H]$\simeq-0.8$ dex.
The distribution also shows a very metal-rich tail, up to [Fe/H]$\sim+0.8$ dex. However,
only five stars have been measured with such an extreme metallicity value, with a somewhat 
larger uncertainty ($\sim 0.2$ dex).  
Figure \ref{supermr} shows the spectra of two such super metal-rich stars 
(7009197 and 7036045 with metallicity of [Fe/H]$=+0.77$ dex and [Fe/H]$=+0.74$ dex, respectively), 
and the spectrum of a star with [Fe/H]$=+0.26$ dex and very similar 
atmospheric parameters (T$_{{\rm eff}}=4325 K$, $\log$~$g=1.7$ dex for the two super metal-rich 
targets and T$_{{\rm eff}}=4269 K$ and $\log$~$g=1.6$ dex for the latter). As can be seen,
the super metal-rich stars have deeper iron absorption lines, thus
indicating a higher metal content with respect to the star at [Fe/H]$=+0.26$ dex. Note that
in order to fit these lines with an iron abundance of  0.3 dex, one needs to assume a
significantly warmer ($\sim500$ K)  temperature.
A spectroscopic follow-up at  higher spectral resolution
is needed to draw a more firm  conclusions about the metal content of these stars.
If their extremely high metallicity were confirmed,
they would be among the most metal-rich stars in the Galaxy.

\begin{deluxetable}{rrrrrcrrc}
\tablewidth{0pc}
\tablecolumns{8}
\tiny
\tablecaption{Iron abundance of Terzan 5 stars.}
\tablehead{\colhead{ID} & \colhead{RA} & \colhead{Dec} & \colhead{K$_{c}$} & \colhead{T$_{{\rm eff}}$} & \colhead{log~$g$} & \colhead{[Fe/H]} 
& \colhead{$\sigma_{{\rm [Fe/H]}}$} & Dataset \\
& & & \colhead{(mag)} & \colhead{(K)} & \colhead{(cm\,s$^{-2}$)} & \colhead{(dex)} & \colhead{(dex)} & }
\startdata
 & & \\
        109  &    266.9801977   &    -24.7835577  &  8.60  &   3741   &   0.7  &  --0.30   &    0.17   &     FLAMES \\
        126  &    267.0292394   &    -24.7803417  &  8.60  &   3736   &   0.7  &  --0.26   &    0.14   &     FLAMES \\
        134  &    267.0332227   &    -24.7953548  &  8.73  &   3771   &   0.7  &  --0.32   &    0.07   &     FLAMES \\
        146  &    267.0254477   &    -24.7817867  &  8.78  &   3786   &   0.8  &  --0.38   &    0.07   &     FLAMES \\
        148  &    267.0291700   &    -24.7969272  &  8.86  &   3804   &   0.8  &  --0.17   &    0.06   &     FLAMES \\
        155  &    267.0286940   &    -24.7786346  &  8.83  &   3799   &   0.8  &  --0.34   &    0.09   &     FLAMES \\
        158  &    267.0124475   &    -24.7843182  &  8.88  &   3814   &   0.8  &  --0.36   &    0.10   &     FLAMES \\
        159  &    267.0226507   &    -24.7624999  &  8.83  &   3800   &   0.8  &  --0.32   &    0.06   &     FLAMES \\
        164  &    267.0282685   &    -24.7949808  &  8.98  &   3838   &   0.8  &  --0.31   &    0.07   &     FLAMES \\
        165  &    267.0315361   &    -24.7896355  &  8.95  &   3831   &   0.8  &  --0.18   &    0.15   &     FLAMES \\
\enddata
\tablecomments{\small Identification number, coordinates, K$_c$ magnitude atmospheric parameters, iron abundances and
their uncertainties, and corresponding dataset for all the 220 stars members of Terzan 5 with iron 
abundance measured. All stars with K$_{c}<9.6$ or K$_{c}>11.7$ have been excluded from the analysis of the
MDF (see Section \ref{samplemdf}).}
\label{tab1mdf}
\end{deluxetable}

\begin{figure}
\plotone{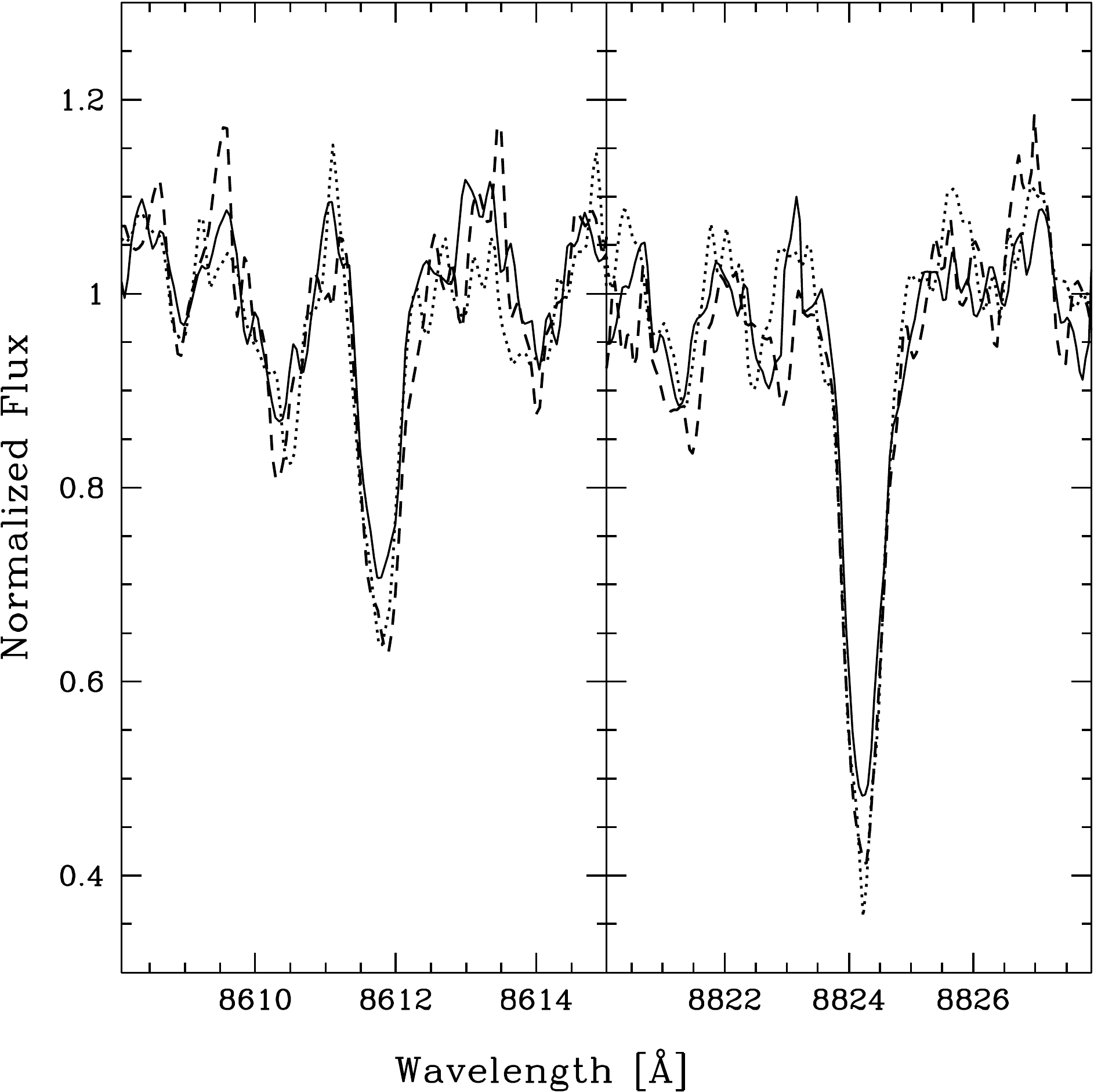}
\caption{\small Comparison of the spectra of two super metal-rich stars 
(namely 7009197 and 7036045, shown as dashed and dotted line, respectively)
and that of a star at [Fe/H]$=+0.26$ (solid line) with similar 
atmospheric parameters.  
The two super metal-rich stars show more pronounced absorption lines, 
thus indicating an actual, very high metallicity.}
\label{supermr}
\end{figure}

\subsection{Statistical decontamination}\label{statdeco}

Even though our sample has been selected within the narrow radial velocity range
around the systemic velocity of Terzan 5, we may expect some
contamination by a few bulge field stars. 
Hence, we performed a statistical decontamination by using the properties of the field
population surrounding Terzan 5 described in M14a.
As shown in detail in that paper, we found that the bulge field population has a very broad radial velocity 
distribution, peaking at v$_{{\rm rad,field}}\sim21$ km$\,$s$^{-1}$ and with a dispersion $\sigma\sim113$ 
km$\,$s$^{-1}$, thus overlapping the Terzan~5 distribution.
When considering different metallicity bins, the bulge population is distributed as follows:
($i$)~$3$\% with [Fe/H]$<-0.5$ dex; ($ii$)~$44$\% with $-0.5<$[Fe/H]$<0$ dex; 
($iii$)~$49$\% with $0<$[Fe/H]$<0.5$ dex; ($iv$)~$4$\% with [Fe/H]$>0.5$ dex.

To perform a meaningful statistical decontamination we first split our sample 
in three radially selected sub-samples (see Fig. \ref{decomdf}). 
The inner (r$<100$\arcsec) subsample is composed of 66 stars. 
The fractions of field stars expected (Ferraro et al. in prep.) to populate this inner region amounts to 2\%,
corresponding to a number of contaminating targets of about N$_{1,field}=2$.
The intermediate subsample ($100$\arcsec$<$r$<200$\arcsec) is composed of 48 stars. 
In this case, the number of expected field stars increases to N$_{2,field}$=16, 
i.e. the 32\% of the subsample.
Finally, in the outer sample ($200$\arcsec$<$r$<276$\arcsec), where we count 21 stars, the expected
contamination by non-member stars amounts to 73\% (corresponding 
to N$_{3,field}=16$).
Fig.~\ref{decomdf} summarizes the number of stars observed (in black) and the number of contaminants
expected (in grey, encircled) in each radial and metallicity bin considered.

For each radially selected sub-sample and metallicity bin, we then randomly subtracted the corresponding
number of expected contaminants, thus obtaining the decontaminated sample. 

\begin{figure}
\plotone{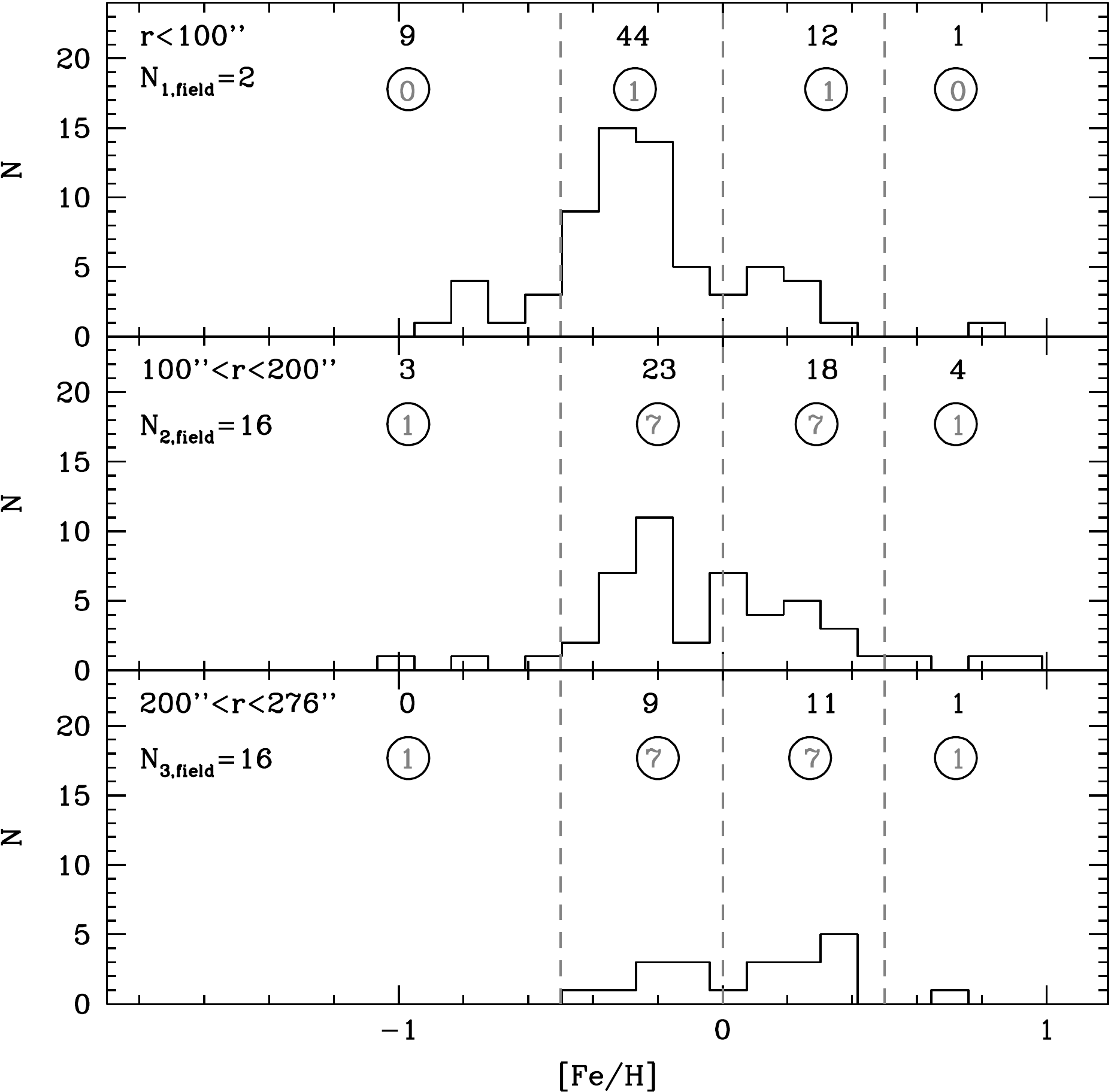}
\caption{\small Metallicity distributions of Terzan 5 stars in the   
inner r$<100$\arcsec ({\it upper panel}), intermediate $100$\arcsec$<$r$<200$\arcsec ({\it middle panel})  
and outer $170$\arcsec$<$r$<276$\arcsec ({\it lower panel}) annuli. The total
number of expected contaminants in each radial bin is reported in the upper-left corner of each panel.
The number of stars observed in each metallicity bin (delimited by vertical dashed lines)
is quoted, while the number of contaminants to be statistically subtracted is highlighted 
in grey and encircled in black. }
\label{decomdf}
\end{figure}

\subsection{Decontaminated distribution}\label{gmm}

The final decontaminated sample is composed of 101 stars and its metallicity
distribution is shown in the upper panel of Fig.~\ref{mdfdeco}. For
comparison, the lower panel shows the distribution of the 34 giant stars in the 
innermost region (r$< 22\arcsec$) of Terzan~5 analyzed in O11 and the three 
metal-poor stars studied in O13. 
The two main peaks at sub-solar and super-solar metallicity, as well as the peak of the minor (5\%) metal-poor component 
at [Fe/H]$\sim$-0.8 dex nicely match each other in the two distributions.

\begin{figure}
\plotone{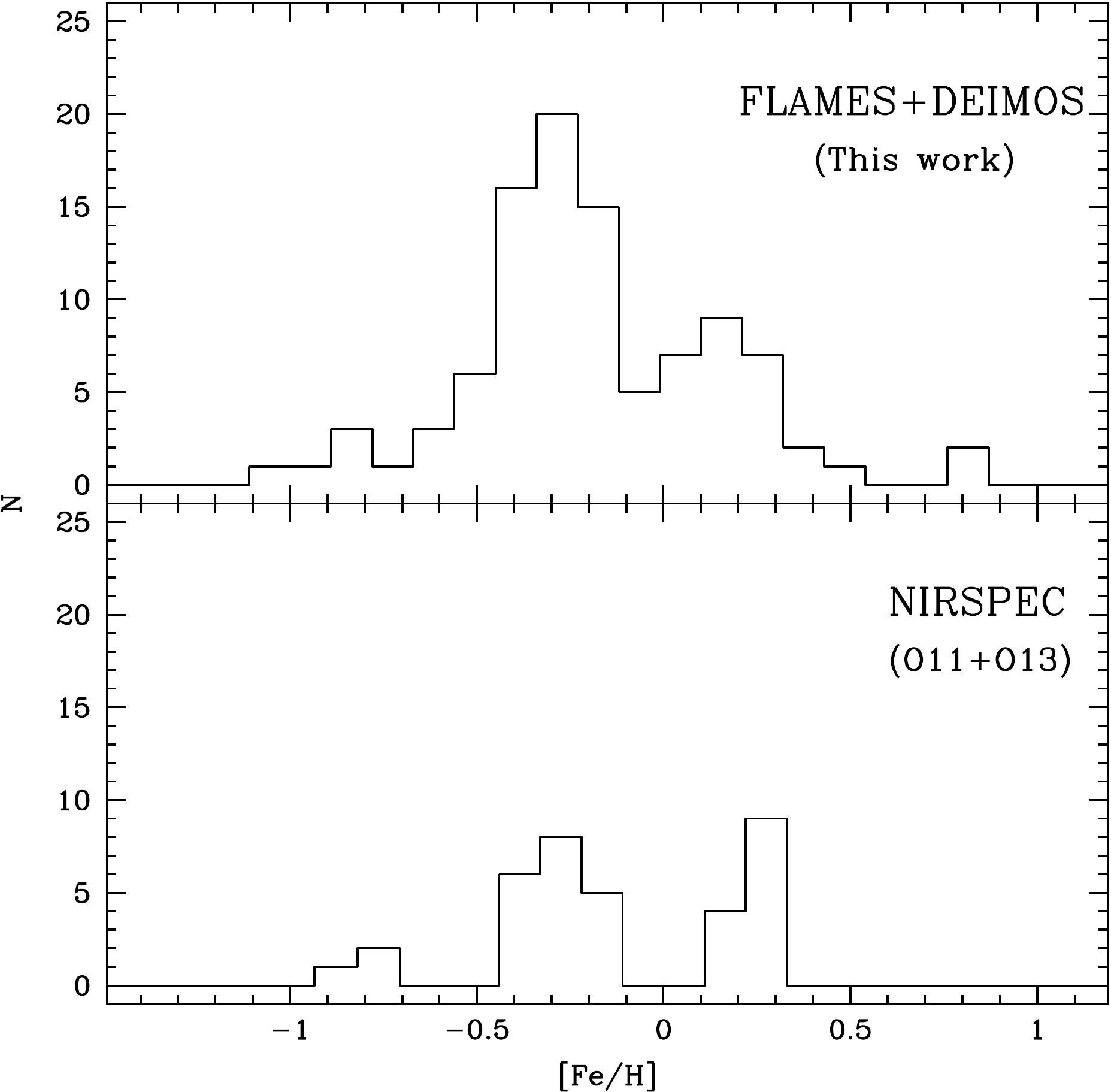}
\caption{\small Statistically decontaminated metallicity distribution for the FLAMES+DEIMOS sample (101 stars, upper panel), compared 
to that derived by O11 and O13 (34 and 3 stars respectively, lower panel). }
\label{mdfdeco}
\end{figure}

It is worth noticing that, while in the O11 sample the super-solar component is about as numerous as the sub-solar one 
(40\% and 60\%, respectively), in the FLAMES+DEIMOS 
distribution the component at $\sim-0.3$ dex is dominant. This essentially reflects the different radial 
distributions of the two stellar populations observed in Terzan~5, with the metal-rich
stars being more concentrated (at $r<20\arcsec$), 
and rapidly vanishing at $r\gtrsim50\arcsec$ (see F09 and \citealt{l10}). Note, in fact, that while the
34 RGB stars observed by O11 are located at $r<22\arcsec$, almost all the FLAMES+DEIMOS 
targets are at larger radial distances.
In the FLAMES+DEIMOS distribution there are also three stars with very high metallicities ([Fe/H]$>+0.7$ dex). 
Given the small number of objects, at the moment we conservatively do not consider it as an additional sub-population of Terzan 5. 

The overall metallicity distribution of Terzan~5, derived from a total of $135$ stars
(corresponding to $101$ targets from the decontaminated FLAMES+DEIMOS sample discussed here, plus 
$34$ NIRSPEC giants from O11) is shown in Fig.~\ref{mdf2}.  
In order to statistically verify the apparent multi-modal behavior of the
distribution, we used the Gaussian mixture modeling (GMM) algorithm proposed by 
\cite{gmm}. This algorithm determines whether a distribution is better
described by a unimodal or a bimodal Gaussian fit. In particular, three requirements
are needed to rule out the unimodality of a distribution:
\begin{enumerate}
 \item the separation D between the peaks, normalized to the widths of the 
 Gaussians, defined as in \cite{ashman94}, has to be strictly 
 larger than 2;
\item the kurtosis of the distribution has to be negative;
\item the likelihood ratio test (\citealt{wolfe}), which obeys
$\chi^{2}$ statistics, has to give sufficiently large values of $\chi^{2}$.
\end{enumerate}
The algorithm also performs a parametric bootstrap to determine the confidence 
level at which the unimodality hypothesis can be accepted or rejected.

\begin{figure}
\plotone{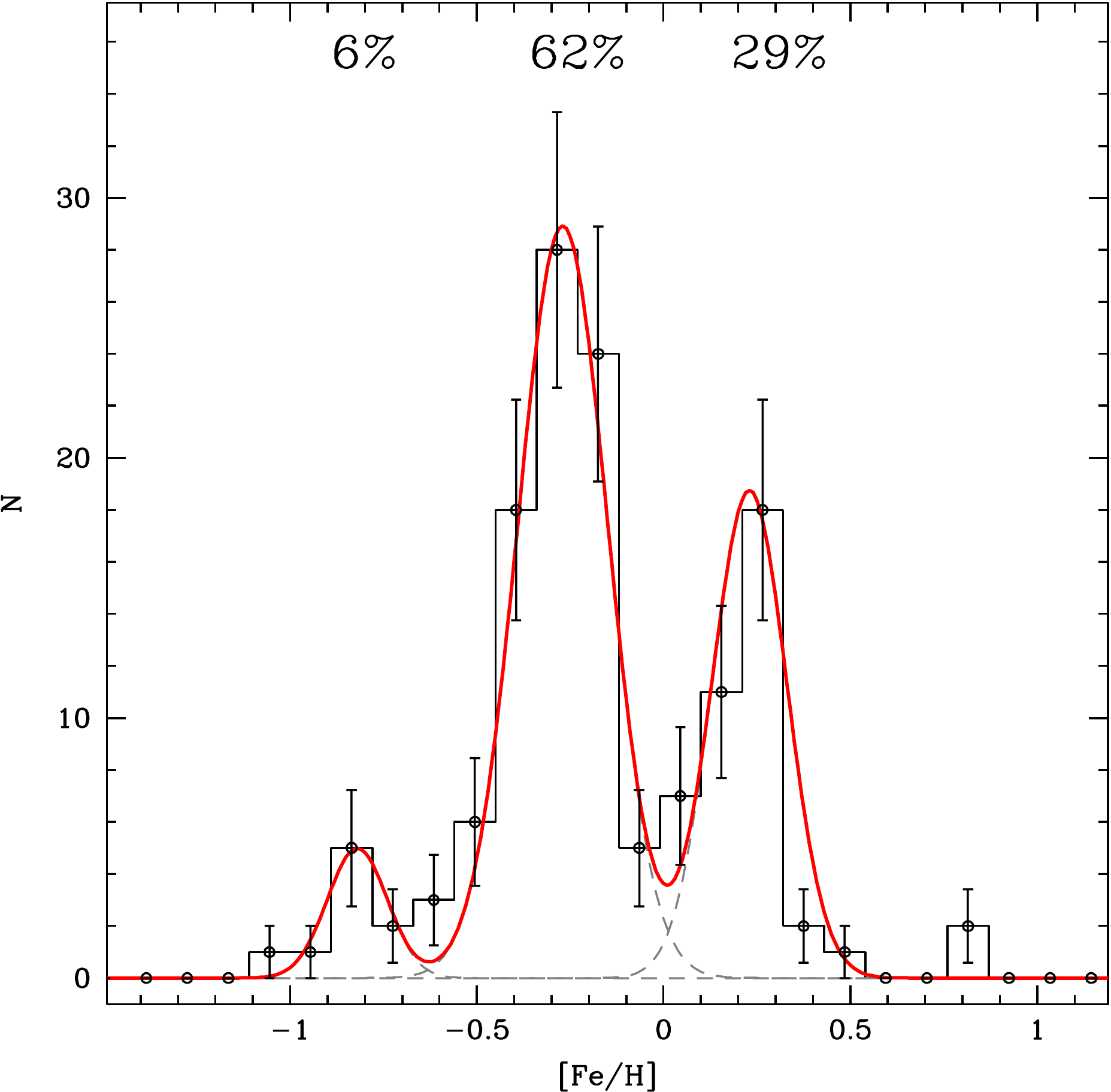}
\caption{\small Decontaminated metallicity distribution for the combined FLAMES+DEIMOS (101 stars, this work) and NIRSPEC (34 targets, O11) spectroscopic samples. 
The solid red line shows the fit that best reproduces the observed distribution using three Gaussian profiles. Individual Gaussian 
components are shown as grey dashed lines. The percentage of each individual component with respect to the total sample
of 135 stars is also reported.}
\label{mdf2}
\end{figure}

First of all, we computed the GMM test on the two main components.
In this case, all the three requirements are verified (D$=3.96$, kurtosis$=-0.89$
and $\chi^{2}=43.46$ with 4 degrees of freedom) and the unimodal fit is rejected 
with a probability P$>99.9$\%.
We then repeated the same procedure considering the most metal-poor component
at [Fe/H]$\simeq-0.8$ and the sub-solar one. 
Also in this case the unimodal fit is rejected with a probability P$>99.9$\%\footnote{
Note that because of the large difference in size between the two components, the computed
kurtosis turns out to be positive. However we checked that by reducing the size of the sample 
belonging to the sub-solar component, the kurtosis turns negative, as required by the GMM test.}.
We can therefore conclude that the metallicity distribution of Terzan 5 
is clearly \emph{multi-peaked}. 
We are able to reproduce its shape using three Gaussian profiles (red line in Fig. \ref{mdf2}).
Adopting the mean values and dispersions obtained from the GMM test, the two main 
peaks are located at [Fe/H]$\simeq -0.27$ dex (with $\sigma=0.12$) and [Fe/H]$\simeq +0.25$
dex (with $\sigma=0.12$), the sub-solar component being largely dominant 
(62\% of the total). A minor component (6\% of the total) is
located at [Fe/H]$\simeq-0.77$ dex (with $\sigma=0.11$).

Finally in  Fig. \ref{distvsfe}  we show the radial distribution of the 135 stars (101 from this study
and 34 from O11) adopted 
to construct the Terzan 5 metallicity distribution shown in Fig. \ref{mdf2}:
the multi-modal metallicity distribution is clearly evident also in this plot. It is worth 
of noticing that the most metal poor component is essentially located in the innermost 
$80$\arcsec~ from the cluster center, further supporting the membership of this minor component. 

\begin{figure}
\plotone{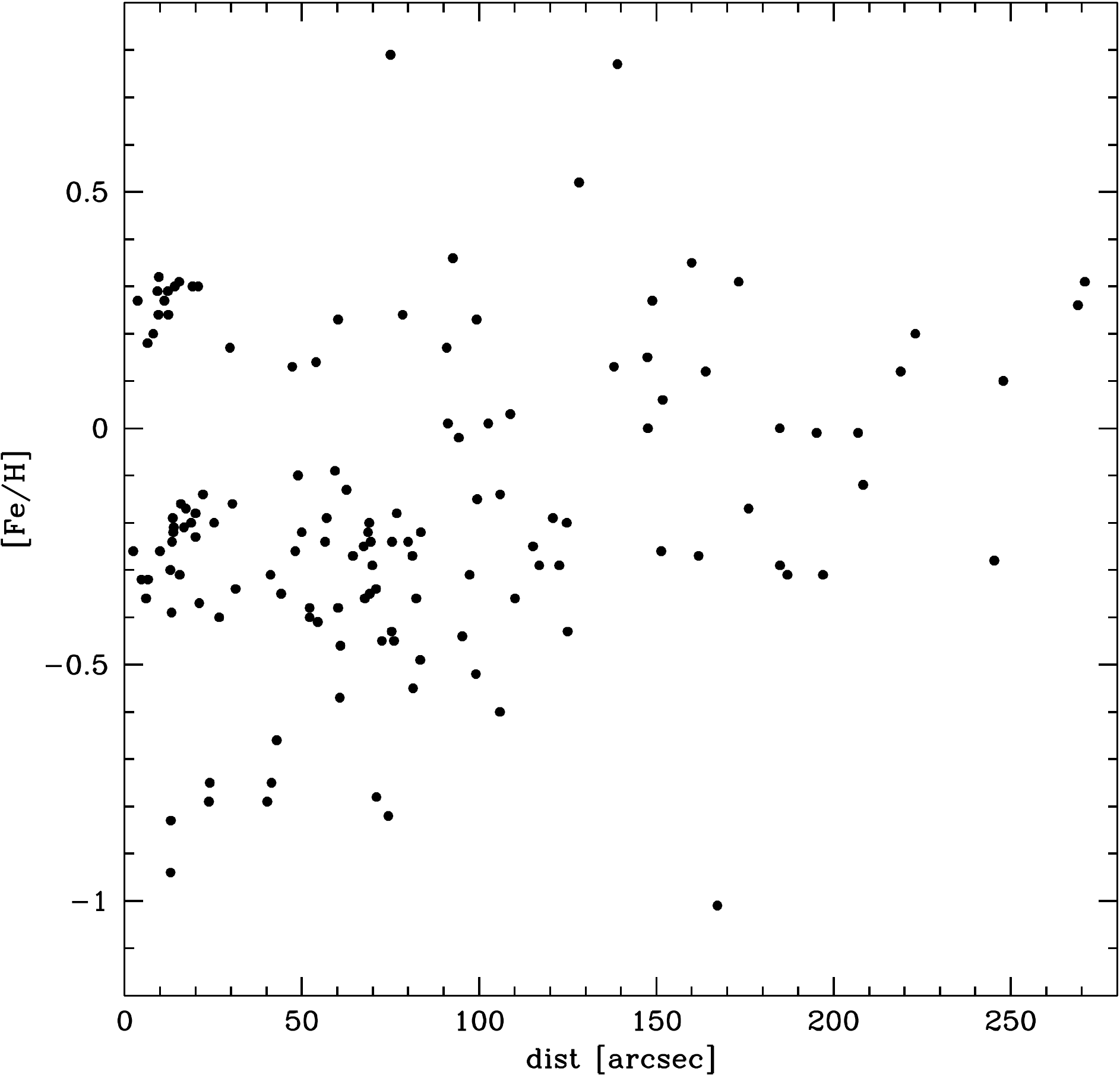}
\caption{\small  The [Fe/H] distribution as a function of the distance from the cluster center 
for the 135 stars composing the final decontaminated iron distribution shown in Fig \ref{mdf2}: 
the multi-modality of the metallicity distribution is clearly evident. The bulk of 
each of the three components
is located in the innermost $80$\arcsec~  from the cluster center,
 thus further confirming the actual membership of all the three populations.}
\label{distvsfe}
\end{figure}

\section{Discussion and Conclusions}\label{discussmdf}

The results presented in this work are based on a statistically significant sample of stars 
distributed over the entire radial extent of Terzan 5, thus solidly
sampling the metallicity distribution of this stellar system.
We confirm the previous claims by F09, O11 and O13 
that Terzan~5 hosts multiple stellar populations characterized by significantly different
iron contents.

The multi-modal iron distribution of Terzan~5 puts this stellar system
in a completely different framework with respect to that of genuine GCs.  
In fact, the latter systems, although showing significant spreads in 
the abundance of light elements (as sodium, oxygen,
aluminum etc.; see, e.g., \citealt{carretta10})\footnote{This suggests that
GC formation has been more complex than previously thought, having
re-processed the low-energy ejecta from
asymptotic giant branch stars \citep{ventura01} and/or fast rotating
massive stars \citep{decressin07}, with enrichment timescales of
$\sim10^8$ years or shorter 
\citep[e.g.,][]{dercole,valcarce}.},
still maintain a striking homogeneity in terms of iron content, thus
indicating that their stellar populations formed within a potential
well which was unable to retain the high-velocity gas ejected by violent SN
explosions.  Indeed, the iron content of stellar populations can be
considered the main feature to distinguish between genuine GCs and
more complex stellar systems (\citealt{willman}). Following this view,
Terzan 5 certainly belongs to the latter class of objects.  

Recent high-precision spectroscopic studies have shown some iron spread 
(but still with a range largely smaller than  1 dex) in a few GCs, namely 
M22 \citep{marino09,marino11a,marino12},
M2 \citep{yongm2}, and M54 \citep{m54}\footnote {Other two GCs
  have been proposed to harbor intrinsic iron dispersion, namely NGC
  5824 \citep{saviane12,dacosta14} and NGC~3201 \citep{simmerer13}. We exclude
  these two clusters from our discussion because their intrinsic iron
  scatter has been not firmly confirmed.  The analysis of NGC 5824 is
  based on the Calcium II triplet as a proxy of metallicity and direct
  measurements of iron lines from high-resolution spectra are not
  available yet. Moreover, based on HST photometry, \cite{sanna} have
  recently found that the color distribution of RGB stars is consistent
  with no metallicity spread. Concerning NGC 3201, the analysis of
  \citet{simmerer13} leads to an appreciable iron spread among the stars of
  this cluster, but the analysis of \citet{munoz} contradicts this
  result.}.  
However, the iron distributions observed in these systems are unimodal, 
with no evidence of multiple peaks, as we also verified 
by means of the GMM test described above.  Only M54 shows a tail towards the 
metal-rich side of its metallicity distribution, but this population can 
be severely contaminated by the Sagittarius field stars (see
\citealt{bellazzini99,bellazz08}). 

Only another GC-like system in the Galaxy ($\omega$ Centauri) is known
to host a large variety of stellar sub-populations \citep{lee99,
pancino2000, ferraro04, ferraro06, bellini09, bellini10, bellini13} 
with a large range of iron abundance ($\Delta$[Fe/H]$>1$ dex; \citealp{norris95,
origlia03, sollima04, sollima07, johnson10, villanova14}), similar to what is observed in
Terzan 5.  
As shown in Figure \ref{confrmdf}, a few similarities between
Terzan 5 and $\omega$ Centauri can be indeed recognized: {\it (i)} a
broad extension of the iron distribution ($\sim 1.8$ dex in Terzan 5
and $\sim 2$ dex in $\omega$ Centauri; see \citealt{johnson10} for the
latter); {\it (ii)} a multi-modal distribution; 
{\it (iii)} the presence of a numerically small
stellar population ($\sim5-10$\% of the total in both cases) which is more
metal-poor than the main peak, possibly corresponding to the
first generation of stars in the system (see \citealp{pancino11} for
$\omega$ Centauri).
The intrinsic large dispersion in [Fe/H] indicates that in the past
these systems were massive enough to retain the high-energy,
high-velocity ejecta of SNe, allowing for multiple bursts of star
formation from increasingly iron-enriched gas over timescales 
of the order of a few $10^9$ years.

\begin{figure}
\plotone{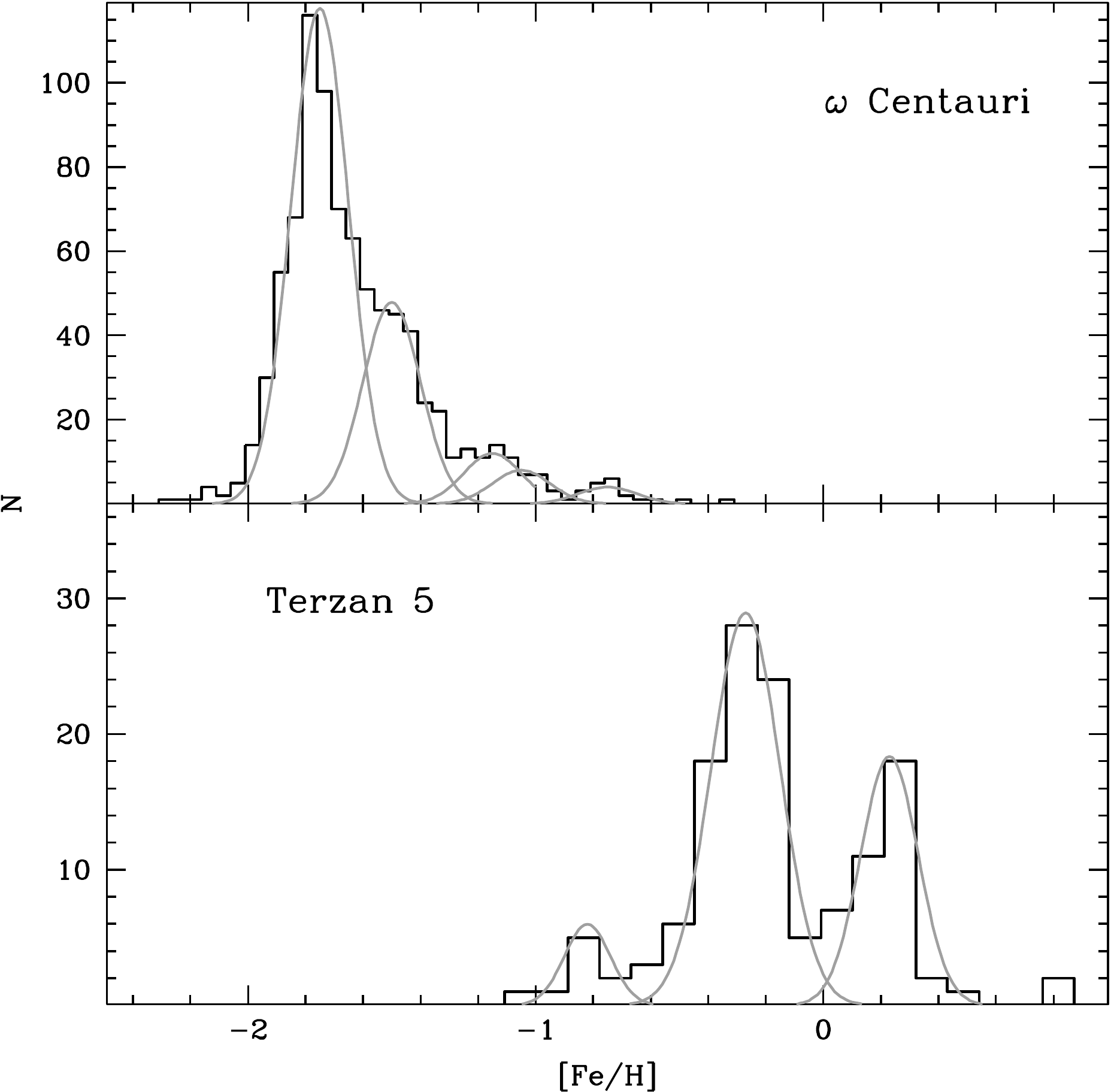}
\caption{\small The metallicity distribution of $\omega$ Centauri ({\it upper panel}) and Terzan 5 ({\it lower panel}).
The distribution of $\omega$ Centauri, together 
with the five Gaussians reproducing its multi-modality, have been taken from \cite{johnson10}.}
\label{confrmdf}
\end{figure}

$\omega$ Centauri is now believed
to be the remnant of a dwarf galaxy accreted by the Milky Way
\citep[e.g.,][]{bekki03}. In contrast, the high metallicity regime of Terzan 5 
(not observed in the known satellites of our Galaxy) and its
tight chemical link with the Galactic bulge (O11, O13, M14a) make very unlikely
that it has been accreted from outside the Milky Way, and 
favor an in-situ formation.
Terzan 5 could be the remnant of an early
giant structure which may plausibly have contributed to form the Galactic bulge. 
In principle, the low dispersion of the iron content within each sub-population of Terzan 5 
could be consistent with both a bursty star formation and chemical self-enrichment, and
the dry merging of individual sub-structures with different metallicity \citep[e.g.][]
{immeli04, elme08,fs11}.  
However, the fact that among the three distinct sub-populations, the
metal-rich one is more centrally concentrated than
the more metal-poor ones seems to favor 
a self-enrichment scenario, at least for the formation of the metal-rich component
\citep[e.g.][]{dercole}.

Certainly Terzan 5 is very peculiar, if not unique, system within the Galactic bulge.
In order to solve the puzzle of its true nature, some pieces of information are still missing, such as
the accurate estimate of the absolute ages of its populations, and a
proper characterization of the global kinematical properties of the
system.
\clearpage{\pagestyle{empty}\cleardoublepage}

\addcontentsline{toc}{chapter}{Conclusions}
\newpage
\markboth{CONCLUSIONS}{CONCLUSIONS}
\chapter*{Conclusions}

The core project of this Thesis is the detailed study of the physical, chemical and kinematic 
properties of the puzzling stellar system Terzan 5 in the Galactic bulge, with the ultimate goal 
of unveiling its true nature and verifying whether it may be the remnant of one of the pristine structures 
that merged together contributing to form the Galactic bulge at early epochs.

\subsection*{Towards the age determination of Terzan 5 stellar populations}

The accurate determination of the absolute ages of Terzan 5 stellar populations
requires to measure the magnitude of their respective MSTO points in the CMD. 
However, such a task is made quite complex
by the strong contamination from the underlying bulge stars and by the severe 
differential reddening in the direction of the system, which
stretches and mixes together the evolutionary sequences.
To solve these problems, we first measured accurate
relative PMs for more than 120\,000 stars in the direction of Terzan 5,
reaching $\sim4$ magnitudes below the TO in the CMD (see Chapter \ref{chaptermoti}).
These PMs allowed us to properly decontaminate both the optical and the IR photometric
catalogs of Terzan 5 from non member stars, further demonstrating that both the RCs identified 
in the Terzan 5 CMDs belong to the system.
We then built a high resolution reddening map for the inner $200\arcsec\times
200\arcsec$ of Terzan 5, finding a range of spatial variation for the
color excess as large as $\delta E(B-V)\sim 0.67$ mag
(see Chapter \ref{chapred}). A free tool providing the color excess values at any
coordinate within the ACS-WFC FoV can be found at the web site {\tt
http://www.cosmic-lab.eu/Cosmic-Lab/products}. 
This study has provided the necessary preparatory work in order to construct
a field-decontaminated and differential reddening-corrected CMD where the measure
of the absolute ages of the distinct stellar populations in Terzan 5 can be performed.

The measure of relative PMs is also a necessary requirement for the determination of 
absolute PMs, from which to derive Terzan 5 3-D orbit within the Galactic potential. Unfortunately,
up to now no extragalactic sources (the ideal objects to be used as an absolute motion
reference frame - see the Appendix of the Thesis) have been found because of the strong
extinction in the bulge direction. This lack makes
it necessary to adopt other PM zero-points, such as those coming from stars 
in the Terzan 5 FoV. In this sense, GAIA will secure remarkably useful data
to fix an absolute reference frame and properly address this issue.

\subsection*{Discovery of a third population}

From the spectroscopic point of view, F09 and O11 isolated two components
with different iron abundance based on a sample of 40 RC and RGB stars.
The two components appeared to be discrete and with a small intrinsic
dispersion ($\sim0.1$ dex), thus suggesting that Terzan 5 experienced two
separated bursts of star formation. However these features were derived from
a statistically limited sample of stars.
In order to determine the actual extent of the iron distribution in Terzan 5 stars,
the spectra of more than 1600 stars obtained
with several ground-based instruments have been analyzed. After the rejection of spectra severely
contaminated by TiO bands, we determined both the 
radial velocity and the metallicity distributions of the bulge stars surrounding Terzan 5
(see Chapter \ref{chapfield}),
finding that they present features similar to those typically observed in other bulge fields
at the same Galactic latitude. Then, by using this information to statistically
decontaminate the sample of likely member stars of Terzan 5 (as selected from radial velocities),
we discovered the presence of a third, metal-poor
([Fe/H]$\simeq-0.8$ dex) and $\alpha$-enhanced component 
(see Chapter \ref{chaporiglia}).

We finally built a bias-free metallicity distribution counting $135$ stars
(see Chapter \ref{chapmdf}).
Such a distribution demonstrates that the iron spread in Terzan 5 is quite large (larger
than 1 dex, ranging from [Fe/H]$\sim-0.9$ dex to [Fe/H]$\sim+0.5$ dex) and it shows three distinct components that are
consistent with three separated populations.
The stellar populations of Terzan 5 have also different [$\alpha$/Fe] abundance ratios:
the two sub-solar ones are enhanced with respect to the solar ratio, while the most metal rich 
has about solar [$\alpha$/Fe]. 
Both the metallicity and the [$\alpha$/Fe] distributions of Terzan 5
are very similar to the distributions observed in the bulge fields (see Chapter \ref{discussoriglia}).
This seems to favor a scenario where   
Terzan 5 formed and evolved in tight connection with the bulge.

\subsection*{Comparison with other resolved stellar systems}

Among the other Galactic stellar systems, only a few GCs, namely M22 \citep{marino09,marino11a,marino12},
M2 \citep{yongm2}, M54 \citep{m54} and possibly NGC 1851 \citep{carretta1851}, 
NGC 5824 \citep{saviane12,dacosta14} and NGC~3201 (\citealt{simmerer13}), 
show some intrinsic iron spread. However, such a spread is much smaller that that observed in
Terzan 5, being at the level of a few tenths dex only (see the discussion in Chapter \ref{discussmdf}) 
and the iron distributions of these GCs are broad (compared to vast majority of GCs) but unimodal.
Moreover, while all the Galactic GCs show distinctive anti-correlations in the abundances of light elements like Na and O
(\citealt{carretta09, muccia09}), the two main populations of Terzan 5 do not.
Therefore all these features demonstrate that Terzan 5 is not a genuine globular cluster. 
Also, the possibility that it is the merger-product of two or three independent stellar aggregates
appears to be quite unlikely. Such a possibility has been found to be implausible for GCs in the Galactic halo.
Although the chance of capturing a completely independent stellar system could be larger if the orbits were
confined within the Galactic bulge, it should be noted that ($i$) such metal-rich (and possibly young)
GCs are quite rare in the Galaxy and ($ii$) as shown in O11 all the sub-populations in Terzan 5 do not show
any evidence of the typical light-elements anti-correlations routinely found in genuine GCs.

There is only one other GC-like stellar system in the Galaxy showing an intrinsic metallicity spread as large
as that of Terzan 5, that is $\omega$ Centauri.
Its peculiar retrograde orbit (\citealt{bekki03}), its chemistry (e.g. \citealt{johnson10}) 
and the finding of a possible tidal debris in the solar neighborhood (e.g. \citealt{maj12}) 
suggest that it is not a genuine GC but rather the nucleus of  
an accreted dwarf galaxy. 
As already discussed in Chapter \ref{discussmdf}, Terzan 5 and $\omega$ Centauri
show several features in common: a similar extent in their iron spread, the
multi-modality of their metallicity distributions and the presence of a small,
metal-poor component that in both cases could be the first-born population.
However, there are also striking differences. Terzan 5
is much more metal-rich than $\omega$ Centauri.
In fact, the iron distribution of Terzan 5 ranges between about 
$-0.9<$[Fe/H]$<+0.6$ dex, while that of $\omega$ Centauri ranges between $-2.0<$[Fe/H]$<-0.5$. 

Such a high metallicity regime is very different from that typical of dwarf galaxies, with the 
only exception of dwarf ellipticals, (see e.g. \citealt{tolstoy12, carraro14}).
However, another chemical feature places Terzan 5 apart from the class of dwarf galaxies: the [$\alpha$/Fe] vs
[Fe/H] trend. 
In fact, in dwarf galaxies the knee point in the [$\alpha$/Fe] vs [Fe/H] diagram, indicating the metallicity reached by
the stellar system when SNIa start to contribute (and dominate) the iron chemical enrichment, 
is typically observed at low metallicity (for example the Sgr dSph,
in spite of being a rare example of metal-rich dwarf, displays its knee point at [Fe/H]$\sim-1$ dex, see \citealt{monaco07}).
Instead, in Terzan 5 the metallicity at which the $\alpha$-elements abundance stars to decrease
is around solar, thus indicating a completely different chemical enrichment history.


\subsection*{The emerging scenario}

The new observational picture of Terzan 5 arising from the results (both photometric and spectroscopic) 
described in this Thesis indicates that this stellar system: 

1) is not a genuine globular cluster, primarily because of its huge spread in metallicity;

2) has a striking chemical similarity with the bulge stars (in terms of iron content and distribution,
as well as alpha-element patterns).

On the other hand, it is quite unlike that it is:

3) the merger product of distinct GCs (in fact, no light-element anti-correlations
are observed in the two main components, see O11); 

4) the remnant of a satellite galaxy accreted by the Milky Way, since 
the metallicity regime and the [$\alpha$/Fe] abundance ratios are very different from those typical of these objects;

A bulge {\it in situ} origin therefore seems the most natural interpretation for Terzan 5. 
In this scenario the oldest populations of the system would trace the early stages of bulge formation, while
the most metal-rich component would contain crucial information on the bulge more recent chemical evolution.

Additional support to this interpretation would come from the finding of other 
stellar systems in the bulge with the same features of Terzan 5. To this aim,
we are currently collecting all the necessary data to properly investigate
other candidate fragments of the pristine bulge, namely Liller 1, Terzan 6, NGC6440 and Djorgovski 2,
showing structural features similar to those observed in Terzan5.

\clearpage{\pagestyle{empty}\cleardoublepage}

\addcontentsline{toc}{chapter}{Appendix}
\newpage
\markboth{APPENDIX }{APPENDIX }
\setcounter{equation}{0}
\renewcommand{\theequation}{\Roman{equation}}
\appendix
\chapter{Appendix}

In this Appendix the method described in Chapter \ref{chaptermoti} to measure relative proper motions for Terzan 5 stars
has been successfully applied to other two Galactic GCs, namely NGC~6681 and NGC~362, in order to derive the absolute
proper motion in the first case and to analyze the BSS population in the second.
In the following, the main results obtained from these analyses are described.

\section{\textit{HST} absolute proper motions of NGC 6681 (M70) and the
  Sagittarius Dwarf Spheroidal Galaxy}

NGC~6681 is located in an extremely interesting 
region of the sky. In fact it overlaps the main body of the Sagittarius 
Dwarf Spheroidal galaxy (Sgr dSph).  Thanks to the extraordinarily
high photometric and astrometric accuracy of {\it HST} we have
been able to kinematically separate the stellar populations belonging to the
two systems and the measure the absolute PMs of the cluster and the dwarf spheroidal
separately by using distant galaxies as zero-motion reference frame.  
This is the first time that the absolute PM of NGC~6681 has been estimated.
All the details regarding this study,
which uses techniques developed in the context of the HSTPROMO collaboration
(see Chapter \ref{chaptermoti}) are described in \cite{massari13}.

\subsection{General context}

Galactic GCs provide a powerful tool to
investigate the structure and the formation history of the Milky Way.
Indeed, they are fundamental probes of the Galactic gravitational
potential shape, from the outer region of the Galaxy (see
\citealt{casetti07}) to the inner Bulge (\citealt{casetti10}).  The
currently and most widely accepted picture for the formation of the
Galactic GC system (\citealt{zinn93}, \citealt{fb10}) points toward an
accreted origin for the outer ($r>10$ kpc) young halo (YH) GCs, while
a large number of the inner, old halo (OH) clusters probably formed
via dissipationless collapse, coevally with the collapse of the
protogalaxy.  The finding of several OH, metal-poor GCs with a thick
disc-like kinematics (\citealt{dinescu99}), sets a tight constraint on
the epoch of the formation of the Galactic disc.  Moreover, the
demonstration that several YH GCs are kinematically associated with
satellites of the Milky Way, such as the Sgr dSph (see for instance \citealt{bellazzini03}),
gives important clues as to how the Galaxy was built up through merger
episodes.  Also the existence of other peculiar systems like $\omega$~Centauri
(\citealt{norris95}) and Terzan~5 (F09) harboring stellar
populations with significant iron-abundance differences
($\Delta$[Fe/H]$>0.5$ dex) supports a complex formation scenario for
the Galactic halo and the Bulge.  Therefore, a detailed description of
the kinematical properties of the Galactic GC system is a crucial
requirement to obtain new and stronger constraints on the formation
history of our Galaxy.  

One of the best opportunities to study the shape, orientation and mass of the
Milky Way dark matter halo is provided by the Sgr dSph (\citealt{ibata94}, 
\citealt{bellazzini99}) through the investigation of its luminous
tidal streams. Recent studies have highlighted a so-called halo
conundrum (\citealt{law05}), showing that the available models were
not able to reproduce simultaneously the angular position, distance
and radial-velocity trends of leading tidal debris.  \cite{lm10a}
claim to have solved this conundrum by introducing a non-axisymmetric
component to the Galactic gravitational potential that can be
described as a triaxial halo perpendicular to the Milky Way disc. Even
if poorly motivated within the current Cold Dark Matter paradigm,
these findings have subsequently been confirmed by \cite{dw13}. 
However, \cite{debattista13} fail to reproduce
plausible models of disc galaxies using such a scenario.  In order to
make substantial progress towards a solution of this debate, new
observational data are needed, starting from accurate proper motions
(PMs).

In this sense, publicly-available catalogs of
absolute PMs for several Galactic GCs are of great importance.  A
notable example is the ground-based Yale/San Juan Southern Proper Motion catalog
(\citealt{platais}, \citealt{dinescu97} and the following papers of
the series).
These kinds of studies are extremely difficult in regions of the sky where
different stellar populations overlap (such as towards the Bulge) and
the associated uncertainties are typically large, ranging between
$0.4$ \masyr and $0.9$ \masyr (\citealt{casetti07},
\citealt{casetti10}). In this sense the {\it Hubble Space Telescope}
({\it HST}) provides a unique opportunity to measure
high-accuracy stellar PMs even in the most crowded and complex regions
of the Galaxy, as seen in \cite{clark08}
or \cite{jay10}, for example.

\subsection{Observations and data reduction}\label{append_rel}

In order to measure the PMs in the direction of NGC~6681 we used two
{\it HST} data sets.  The one used as first epoch was acquired under
GO-10775 (PI:\ Sarajedini). It consists of a set of high-resolution
images obtained with the WFC of the ACS.
For our investigation we used four deep exposures in both the F606W
and the F814W filters (with exposure times of $140$ sec and $150$ sec,
respectively), taken on May 20, 2006.  We work here exclusively with
the \_FLC images, which have been corrected with the pixel-based
correction in the pipeline (\citealt{jaybedin10}, and \citealt{ubedajay})
as described in Chapter \ref{chaptermoti}.

The second-epoch data set is composed of proprietary data obtained
through GO-12516 (PI:\ Ferraro). This program consists of several
deep, high-resolution images taken with the UVIS channel of the
WFC3 in the F390W, F555W and F814W filters. 
The sample analyzed in this work consists of $9\times150\,$s
images in F555W and $13\times348\,$s images in
F814W. These images have not been corrected for CTE losses, since 
no pixel-based correction was available at the time of this reduction.  
These images were taken relatively soon after installation and background 
in these images is greater than 12 electrons, so any CTE losses should 
be small, particularly for the bright stars we are focusing on here 
(see \citealt{jayisr12}).
Since these observations were taken on November 5, 2011, the two data
sets provide a temporal baseline of $\sim 5.464$ yrs.

The data reduction procedure and the relative PMs determination follow 
the prescriptions already discussed 
in Chapter \ref{chaptermoti}. For the WFC3 UVIS dataset we used the program 
\texttt{img2xym$\_$wfc3uv}, which is similar to that developed for the
analysis of ACS data. The output of this analysis is
summarized in Figure \ref{relpm}, where in the upper panels we show
the VPDs and in the lower panels the
corresponding CMDs. Close inspections of the VPDs suggest that at
least three populations with distinct kinematics can be identified in
the direction of NGC~6681.

\begin{figure*}[!htb]
\includegraphics[scale=0.62]{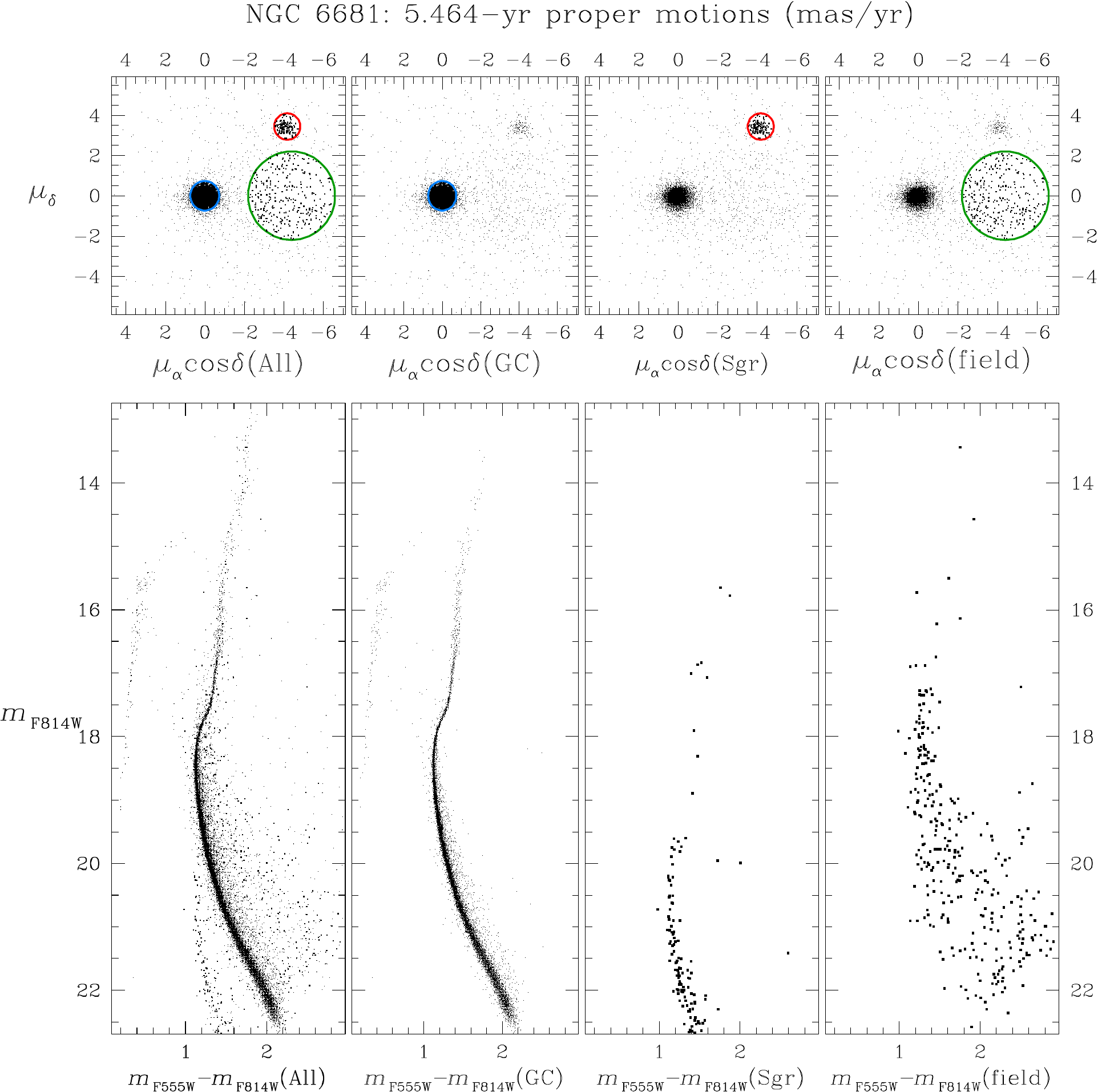}
\caption{\small The upper panels show the VPDs
  of the relative PMs. In the lower panels the CMDs corresponding to
  the selections applied in the VPDs are displayed. {\it First
    column}:\ in the VPD the different populations are indicated with
  different colors (a sample of cluster members in blue, of Sgr dSph
  stars in red, of the field in green), but no selection is
  applied. The corresponding CMD shows the entire PM catalog. {\it
    Second column}:\ in the VPD cluster members are selected within
  the blue circle and the corresponding CMD displays only well-defined
  cluster evolutionary sequences. {\it Third column}:\ Sgr dSph
  selection within the red circle and corresponding CMD.  {\it Fourth
    column}:\ the selection in the VPD (in green) of the bulk-motion
  of field stars and their location on the CMD.}
\label{relpm}
\end{figure*}

\begin{enumerate}
\item The cluster population is identified by the clump of stars at
  (0,0) \masyr. By selecting stars within the blue circle in the
  second upper panel, a clean CMD of the cluster is obtained (second
  lower panel of Fig.~\ref{relpm}).
\item A secondary clump of stars is located at roughly $(-4,3.5)$
  \masyr.  Stars selected within the red circle in the third upper
  panel of Fig.~\ref{relpm} define in the CMD (the third lower panel)
  a sequence significantly fainter than that defined by cluster stars.
  Therefore, these
  stars belong to a population that is both kinematically and
  photometrically different from that of the cluster.  This population
  appears uniformly distributed across the FoV of our observations and
  thus it can be associated to the Sgr dSph, whose main body is
  located in the background of NGC~6681.
\item A much sparser population of stars is centered around $(-4.5,0)$
  \masyr.  The bulk of this population is highlighted with the green
  circle in the last upper panel of Fig.~\ref{relpm}.  The
  corresponding CMD suggests that this is essentially due to
  fore/background sources.
\end{enumerate}

\subsection{Absolute reference frame}

In order to measure absolute PMs, an absolute zero point is required.
The best option to define this zero point is to use extragalactic
sources, since they are essentially stationary on account of their enormous
distances.
This method has already been adopted in several previous
works, such as \cite{dinescu99}, \cite{bellini10b}, and \cite{sohn12}.
In order to find extragalactic sources we first tried to use the Nasa
Extragalactic Database (NED) but no sources in our FoV were reported. This is
due to the strong incompleteness of of the NED catalog in the innermost
regions of dense stellar systems.  We then performed a careful visual inspection of
our images. Thirty-one galaxies were identified by eye, but only 11 of
them have point-like nuclei and thus are successfully fitted by the
adopted PSF.  Out of these, we selected only the 5 galaxies with an
associated QFIT value (a parameter which describes how well the adopted PSF fits
the brightness profile of the source, see \citealt{jayking06} for details) smaller
than $0.6$:\ this was necessary to guarantee a measurement of the
source centroid accurate enough to provide a precise determination of
the zero point for the absolute PMs.  Figure \ref{gal} shows how these
galaxies appear in the F814W band.

\begin{figure}[!htb]
\includegraphics[width=\columnwidth]{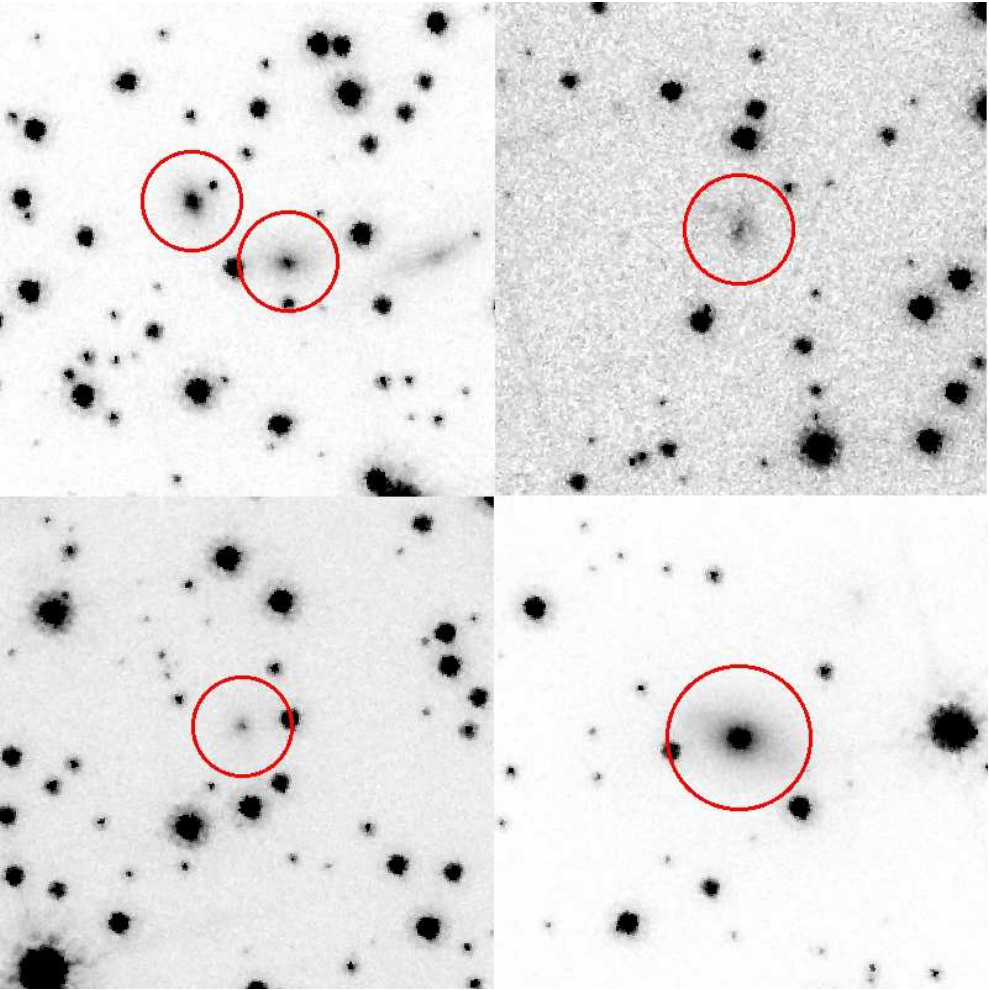}
\caption{\small The five selected background galaxies as they appear
  in the F814W images. They differ from the stellar sources since
  their light is more diffuse across the surrounding pixels. Their
  point-like nuclei allow us to accurately determine their centroid
  and thus to obtain a precise measure of their relative proper
  motions.}
\label{gal}
\end{figure}

The selected galaxies are located very close to each other in the
relative-PM VPD (Fig.~\ref{absolute}), as expected for distant
sources.  Therefore, we defined the zero-point of the absolute
reference frame as the weighted mean of their relative PMs (see the
blue dot in Figure \ref{absolute}):
\begin{equation}
  (\mu_{\alpha}\cos\delta, \mu_{\delta})_{{\rm gals}}=(-1.58\pm0.18, 4.57\pm0.16)~{\rm mas\, yr^{-1}},
\label{eqabsolute}
\end{equation}
as measured with respect to the mean NGC~6681 motion derived in
Section \ref{append_rel}. The uncertainties correspond to the error on the
calculated weighted means. 

\subsection{NGC~6681}

In order to measure the absolute PM of NGC~6681 we selected only stars
within $1.0$ \masyr from the cluster mean motion and in the magnitude
interval $17.5<${\it m}$_{{\rm F555W}}<22.5$ mag. We iteratively
refined the selection by applying a $3\sigma$ rejection and
re-calculating the barycenter of the PMs as the weighted mean value of
the PMs of the selected stars, until the difference between two
subsequent steps was smaller than $0.01$ \masyr. After the last iterative 
step, a total of $N_{clu}=14\,030$ stars survived the selection
criteria.  We used the sum in quadrature between each single measurement
error and the velocity dispersion of the cluster $\sigma_{v}=0.12$ mas/yr
(based on the line-of-sight velocity dispersion and distance given by \citealt{harris}) 
as weights. To estimate the error $\Delta PM$ on the weighted mean PM in 
each coordinate we use the standard error-in-the-mean, i.e., the dispersion of 
the surviving stars around the weighted mean PM, divided by $\sqrt(N_{\rm clu}-1)$. 
This includes scatter from the internal dispersion of NGC~6681 stars, which 
therefore does not need to be estimated explicitly. We find that the resulting 
error $\Delta PM$ is negligible compared to the error on the absolute reference 
frame. Therefore, the latter dominates the uncertainty on the final absolute PM 
of NGC~6681, which is:\
\begin{equation}
 (\mu_{\alpha}\cos\delta, \mu_{\delta})=(1.58\pm0.18, -4.57\pm0.16)~
  {\rm mas\, yr^{-1}}.
\end{equation}

The PM derived here can be combined with the known distance and
line-of-sight velocity of NGC~6681 from \cite{harris}, to determine
the motion of the cluster in the Galactocentric rest frame. Using the
same formalism, conventions, and solar motion as in \cite{marel12},
this yields $(V_X,V_Y,V_Z) = (203 \pm 2, 111 \pm 9, -179 \pm 7)$
km$\,$s$^{-1}$. This corresponds to a total Galactocentric velocity $|{\vec V}|
= 292 \pm 5$ km$\,$s$^{-1}$. This significantly exceeds the central velocity
dispersion $\sigma \approx 120$ km$\,$s$^{-1}$ of the Milky Way's spheroidal
components (e.g., \citealt{deason12}). Hence, NGC~6681 must spend most
of the time along its orbit at significantly larger distances from the
Galactic Center than its current distance of $2.2$ kpc
(\citealt{harris}).

\subsection{Sagittarius Dwarf Galaxy}

In order to determine the absolute PM of the Sgr dSph we basically
followed the same procedure previously described for NGC~6681.  
In setting the weights for the PM averaging, we used the dispersion 
$\sigma_v \sim 0.3$ \masyr implied by Figure 6. This includes both 
contributions from the internal velocity dispersion of the Sgr dSph (see
e.g. \citealt{frinchaboy12}), and unquantified systematic errors.
We selected stars within $1.0$ \masyr from the Sgr dSph mean motion and
in the interval $17.5<${\it m}$_{{\rm F555W}}<23.5$, which is one
magnitude fainter with respect to the case of NGC~6681, since most of
the Sgr dSph stars belong to its faint MS. The resulting absolute PM
is:\
\begin{equation}
 (\mu_{\alpha}\cos\delta, \mu_{\delta})=(-2.54\pm0.18,
  -1.19\pm0.16)~{\rm mas\, yr^{-1}}.
\end{equation}

We compared this value with previous estimates.  With the aim of
reconstructing the kinematical history of this galaxy and to predict
its evolution in a triaxial Milky Way halo, \cite{lm10a} built a
N-body model able to reproduce most of the system's observed
properties. In the Law \& Majewski model, the Sgr dSph
has a Galactocentric motion $(V_X,V_Y,V_Z) = (230, -35, 195)$ km$\,$s$^{-1}$,
corresponding to a total velocity $|{\vec V}| = 304$ km$\,$s$^{-1}$.
The absolute PM predicted by the model is
($\mu_{\alpha}\cos\delta, \mu_{\delta})=(-2.45, -1.30$) \masyr (light
green dot in Figure \ref{absolute}).  An estimate of the
absolute PM of the Sgr dSph based on {\it HST} data has been recently presented by
\cite{pryor10}. The authors used foreground Galactic
stellar populations as reference frame and they determined an absolute
PM of ($\mu_{\alpha}\cos\delta, \mu_{\delta})=(-2.37\pm0.2,
-1.65\pm0.22$) \masyr, which is shown as a magenta ellipse in Figure
\ref{absolute}.  A ground-based estimate of the absolute PM of the Sgr
dSph was presented by \cite{dinescu05}. Using the Southern Proper
Motion Catalog 3 they determined that $(\mu_\alpha \cos\delta,
\mu_\delta) = (-2.83 \pm 0.20, -1.33 \pm 0.20)$ \masyr, which is shown
as the dark green ellipse in Figure \ref{absolute}.  These previous
estimates are in rough agreement with the value determined here.

It is worth noting, however, that these other determinations are not
directly comparable with ours, since they refer to different regions
of the Sgr dSph. Indeed, this has two possible effects. The first one
is that possible internal motions, such as rotation, could translate
into different mean motions, thus introducing a systematic effect.
This should not be a problem for the Sgr dSph, since this galaxy does
not show any evidence of rotation (\citealt{pena}).  The second effect
is that if the whole galaxy has a 3D velocity vector different from
zero, then the observed PMs for different pointings are not the same,
because of perspective effects due to the imperfect parallelism
between the lines of sight (\citealt{marel02}).  Since the Sgr dSph is
a nearby galaxy, this effect could be relevant and we calculated the
correction to apply (as in \citealt{marel08}) in order to obtain
comparable estimates at the center of mass of the Sgr dSph.

Under the hypothesis that the center of mass of the Sgr dSph is moving
as the \cite{lm10a} prediction, our perspective-corrected PM
measurement becomes ($\mu_{\alpha}\cos\delta,
\mu_{\delta})=(-2.56\pm0.18, -1.29\pm0.16$) \masyr. The corrected
\cite{pryor10} estimate becomes ($\mu_{\alpha}\cos\delta,
\mu_{\delta})=(-2.37\pm0.20, -1.63\pm0.22$) \masyr, and the corrected
\cite{dinescu05} estimate becomes $(\mu_\alpha \cos\delta, \mu_\delta)
= -2.83 \pm 0.20, -1.56 \pm 0.20)$ \masyr. Thus our measurement is
consistent with the previous observations. The weighted average of all
observational estimates of the center-of-mass PM of the Sgr dSph is
$(\mu_\alpha \cos\delta, \mu_\delta) = (-2.59 \pm 0.11, -1.45 \pm
0.11)$ \masyr. This is consistent with the theoretical model of
\cite{lm10a}, once the uncertainties on transforming that into a PM
value (e.g., from uncertainties in the distance and solar motion) are
taken into account as well.  Therefore, our
measurement is consistent within about a $1\sigma$ uncertainty both with
theoretical predictions (\citealt{lm10a}) and the previous {\it HST}
observations (\citealt{pryor10}).

\subsection{Field}

We compared the absolute PMs of Field stars in our catalog with those
predicted in the same region of sky by the Besan\c{c}on Galactic model
(\citealt{robin}). We generated a simulation over a 0.01 square
degrees ($6\arcmin\times6\arcmin$) FoV around the center of NGC~6681
($l=2$\textdegree\!.$85$, $b=-12$\textdegree\!.$51$) and 50 kpc deep.
To minimize any possible bias, we have constructed a sample as similar
as possible to the observed stars, based on a comparison between the
observed and the simulated CMDs.  Simulated field stars were selected
within the magnitude range:\ $17.5<${\it m}$_{{\rm F555W}}<22.5$ mag
and ({\it m}$_{{\rm F555W}}-${\it m}$_{{\rm F814W}}$)$>1.5$ mag and
1378 stars survived these criteria. 
The average predicted motion is shown in Figure \ref{absolute} as a
cyan ellipse, which corresponds to ($\mu_{\alpha}\cos\delta,
\mu_{\delta})=(-0.91\pm0.08, -2.39\pm0.09$) \masyr.\footnote{It 
would be easy to reduce the random uncertainty on this model prediction 
by drawing a larger number of simulated stars. However, we have not 
pursued this since the accuracy of the prediction is dominated largely 
by systematic errors in the model assumptions anyway.}

Field stars in our observed catalog were selected following the same
color and magnitude cuts. We also required these stars to have PM
errors smaller than $0.2$ \masyr in each coordinate. Finally, we
excluded those stars within $1.8$ \masyr of the cluster mean motion
and within $1.0$ \masyr of the Sgr dSph mean motion.  We iteratively
removed field stars in symmetric locations with respect to the Sgr dSph
and NGC~6681 exclusions in order to better define the mean motion of
the Field population and adjusted the weighted mean motion after each
iteration (thus following the method described by \citealt{jay10} for
the determination of the center of $\omega$~Cen).  The 281 selected
field stars used for the final estimate are shown as black crosses in
Figure \ref{absolute}.  Since these stars display a large 
scatter in the VPD due to their velocity dispersion and not to their random errors, 
in this case we computed a statistically more appropriate $3\sigma$-clipped 
unweighted mean motion. It is
shown as a blue ellipse in Figure \ref{absolute} and its value is:\
\begin{equation}
 (\mu_{\alpha}\cos\delta, \mu_{\delta})=(-1.21\pm0.27, -4.39\pm0.26)
  {\rm mas\, yr^{-1}}.
\end{equation}

Our PM measurement is similar to the prediction of the Besan\c{c}on
model, in that it points in the same direction on the sky (see
Figure~\ref{absolute}). However, the sizes of the PM vectors are not
formally consistent to within the random errors.  Since our
measurements for Sgr dSph stars are entirely consistent with both
previous measurements and theoretical predictions, this cannot be due
to systematic errors in our measurements (which would affect all point
sources equally). Instead, the mismatch is most likely due to
shortcomings in the Besan\c{c}on models. In particular, for pointings
this close to the Galactic Plane, the model predicted PM distribution
is likely to depend sensitively on the adopted dust extinction model,
which is poorly constrained observationally. Also, the model predicted
PM distribution depends on the solar motion in the Milky Way, which
continues to be debated (e.g., \citealt{mcmillan11};
\citealt{bovy12}).

\begin{figure*}[!htb]
\includegraphics[scale=0.62]{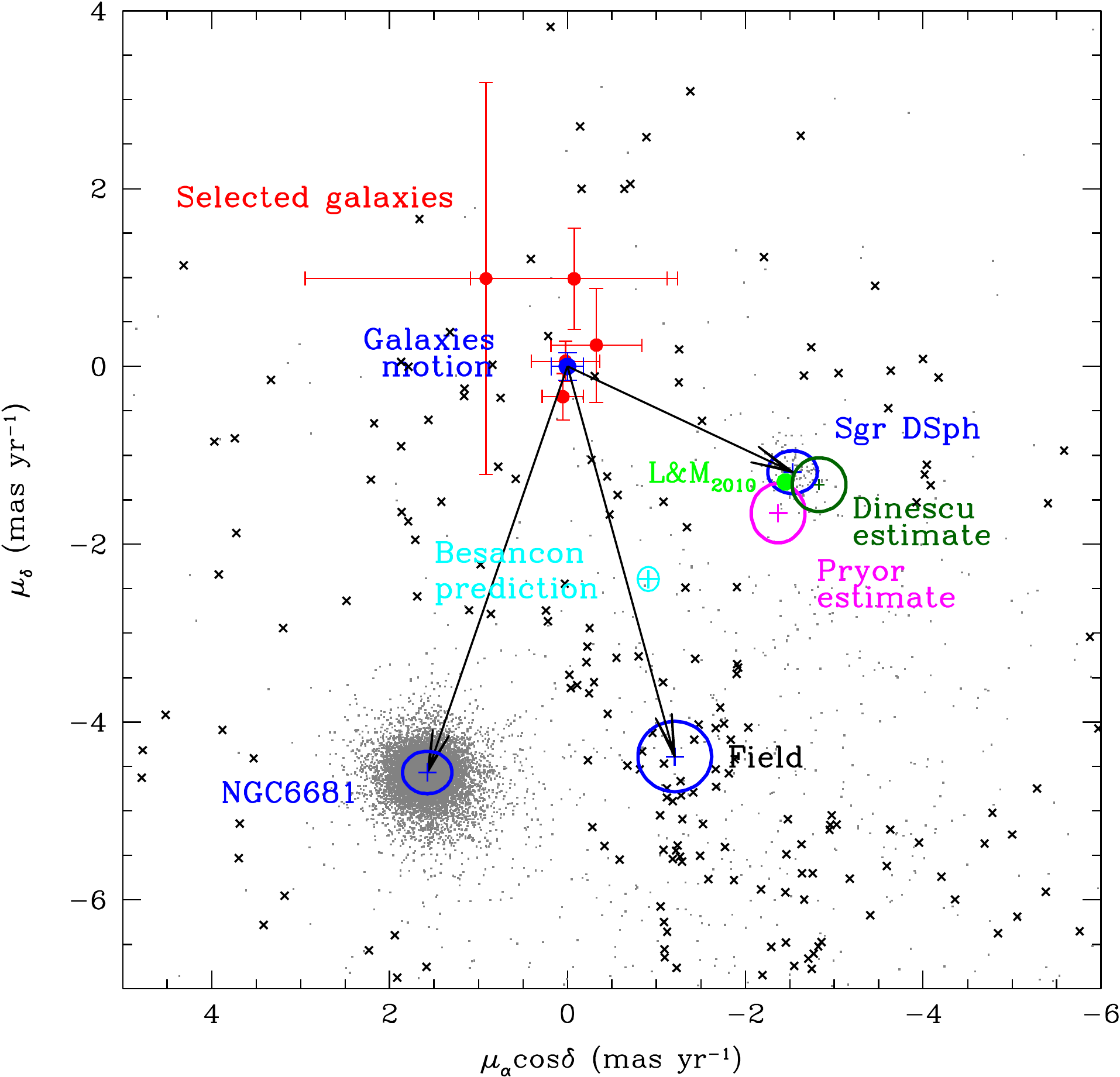}
\caption{\small VPD of the absolute PMs. The red dots indicate the
  selected background galaxies, whose
  mean motion corresponds to the zero point of the VPD. The blue
  ellipses are centered on the measured absolute PMs of the three
  populations (marked with a blue cross) and their size corresponds to
  the calculated 68.3\% confidence region. The black arrows indicate
  their absolute PM vectors.  In the proximity of the Sgr dSph
  estimate, the PM value predicted by \cite{lm10a} is shown as a light
  green dot, while the \cite{pryor10} and \cite{dinescu05}
  measurements and their 68.3\% confidence regions are shown as
  magenta and dark green ellipses, respectively. Finally, the cyan
  ellipse describes the prediction on the PM of the field population
  by the Besan\c{c}on model, which differs from our estimate obtained
  using the stars in the same magnitude and color range (marked with
  black crosses, see the text for the selection criteria).}
\label{absolute}
\end{figure*}

\section{Double Blue Straggler sequences in GCs: the case of NGC~362}

The second case under study concerns the BSS population
of the Galactic GC NGC 362. A series of suitably planned photometric observations
with the {\it HST} and
the analysis of the proper-motion-selected CMD allowed us to discover two distinct 
sequences of BSS, a rare feature observed only in one other GC in the Galaxy (namely M30,
Ferraro et al 2009b).
All the details regarding this study can be found in \cite{dalex13}.

\subsection{General context}

Among the large variety of exotic objects (like X-ray binaries,
MSPs, etc.) which populate the dense environment of
Galactic GCs (see Bailyn 1995, Paresce et
al. 1992, Bellazzini et al. 1995, Ferraro et al. 2001, 2003, 2006c, 
Ransom et al. 2005, Freire et al. 2008), BSS surely
represent the most numerous and ubiquitous population.  Indeed, BSS 
have been detected in any properly
observed stellar system (GCs, see Fusi Pecci et al. 1992, Ferraro et al. 1992 Piotto et al. 2004; open clusters 
-- Mathieu \& Geller 2009 -- dwarf galaxies --
Mapelli et al. 2009). BSS are brighter and bluer than the MSTO, 
thus mimicking a population significantly
younger than normal cluster stars. Indeed, observations demonstrated
that they have masses larger than that of MSTO stars ($m= 1.2 - 1.7~
M_\odot$; Shara et al. 1997; Gilliland et al. 1998; De Marco et
al. 2004, Fiorentino et al. 2014).  However, stellar evolution models predict that single
stars of comparable mass generated at the epoch of the cluster
formation should have already evolved away from the MS; thus some
mechanisms must have been at work to increase the mass of these objects in
their relatively recent past ($2-3$ Gyr ago; Sills et al. 2002).

Two main formation scenarios for BSS have been proposed over the
years: mass transfer (MT-BSS) and direct collision (COL-BSS).  The
collisional formation channel between two single stars was theorized
for the first time by Hills \& Day (1976). Following works (Lombardi
et al. 2002; Fregeau et al. 2004) showed that BSS may form also via
collision between binary-single and binary-binary systems.  In the
mass transfer scenario (McCrea 1964), the primary star transfers material to the secondary one
through the inner Lagrangian point when its Roche Lobe is filled. In
this picture the secondary star becomes a more massive MS star (with a
lifetime increased by a factor of 2 with respect to a normal star of
the same mass -- McCrea 1964) with an envelope rich of gas accreted
from the donor star.  Chemical anomalies are expected for MT-BSS
(Sarna \& De Greve 1996), since the accreted material (currently
settled at the BSS surface) could come from the inner region of the
donor star, where nuclear processing occurred.
Spectroscopic results  supporting the occurrence of the MT formation channel in a few BSS have been recently obtained 
(Ferraro et al. 2006a; Lovisi et al. 2013). Conversely, surface
chemical anomalies are not expected for COL-BSS (Lombardi et al. 1995), since no 
significant mixing should occur
between the inner core and the outer envelope.

In GCs, where the stellar density significantly varies from the center
to the external regions, BSS can be generated by both processes (Fusi
Pecci et al. 1992, Ferraro et al 1995).  Recent works
suggested that MT is the dominant formation mechanism in low density
clusters (Sollima et al. 2008) and possibly also in high-density
clusters (Knigge et al. 2009).  However the discovery of two distinct
sequences of BSS in M~30 (Ferraro et al. 2009b) clearly
separated in color further supports the possibility that both
formation channels can coexist within the cluster core. In fact the
blue BSS sequence is nicely reproduced by collisional models (Sills et
al 2009), while the red one is compatible with binary systems
undergoing MT (see Tian et al. 2006).  The origin of the double
sequence might be possibly related to the core-collapse process that
can trigger the formation of both red and blue BSS, enhancing the
probability of collisions and boosting the mass-transfer process in
relatively close binaries.  Given the evolutionary time-scales for
stars in the BSS mass range, the fact that the two sequences are still
well distinguishable is a clear indication that core collapse occurred
no more than $1-2$ Gyr ago.

In order to further explore the link between the presence of a double sequence of BSS and the occurrence of core collapse, 
we acquired {\it HST}
images with the WFC3 for a sample of suspected post-core collapse GGCs.  
Here we report on the discovery of the second case of a BSS double sequence
 in NGC~362 .

\subsection{Observations and Data analysis}

In the present work we used data acquired with the UVIS channel of the 
WFC3 on 2012 April 13 (Proposal ID: 12516; PI: Ferraro).
The dataset is composed of fourteen exposures obtained through the F390W (hereafter U\footnote{Although
the F390W filter almost corresponds to the broad filter C  
of the Washington photometric system (Canterna 1976),
we prefer to label it ``U filter" since it is more popular.}) filter, each one with an exposure 
time $t_{\rm exp}=348$ sec, ten F555W (V) images with  $t_{\rm exp}=150$ sec and fifteen F814W (I) frames with 
$t_{\rm exp}=348$ sec.
Each pointing is dithered by a few pixels in order to allow a 
better subtraction of CCD defects, artifacts and false detections.
The data reduction follows the same procedure described in Chapter \ref{chaptermoti}.

As evident in Figure \ref{bss_vpd},
the CMD of NGC~362 is contaminated by fore and background Galaxy stars and even 
more strongly by the Small Magellanic Cloud (SMC) populations. In particular the MS of the SMC defines a quite clear sequence
at $V>20$ and $0<(U-I)<2$. 
In order to evaluate the level of contamination in the direction of the cluster, we performed the
already described relative PM analysis. For this purpose we complemented our HST WFC3 images with 
a first epoch data-set consisting of WFC/ACS images 
obtained in June 2006 (Prop ID 10775; PI: Sarajedini). Also in this case, we repeated the procedure
already followed in Chapter \ref{chaptermoti}.

\begin{figure}
\includegraphics[scale=0.7]{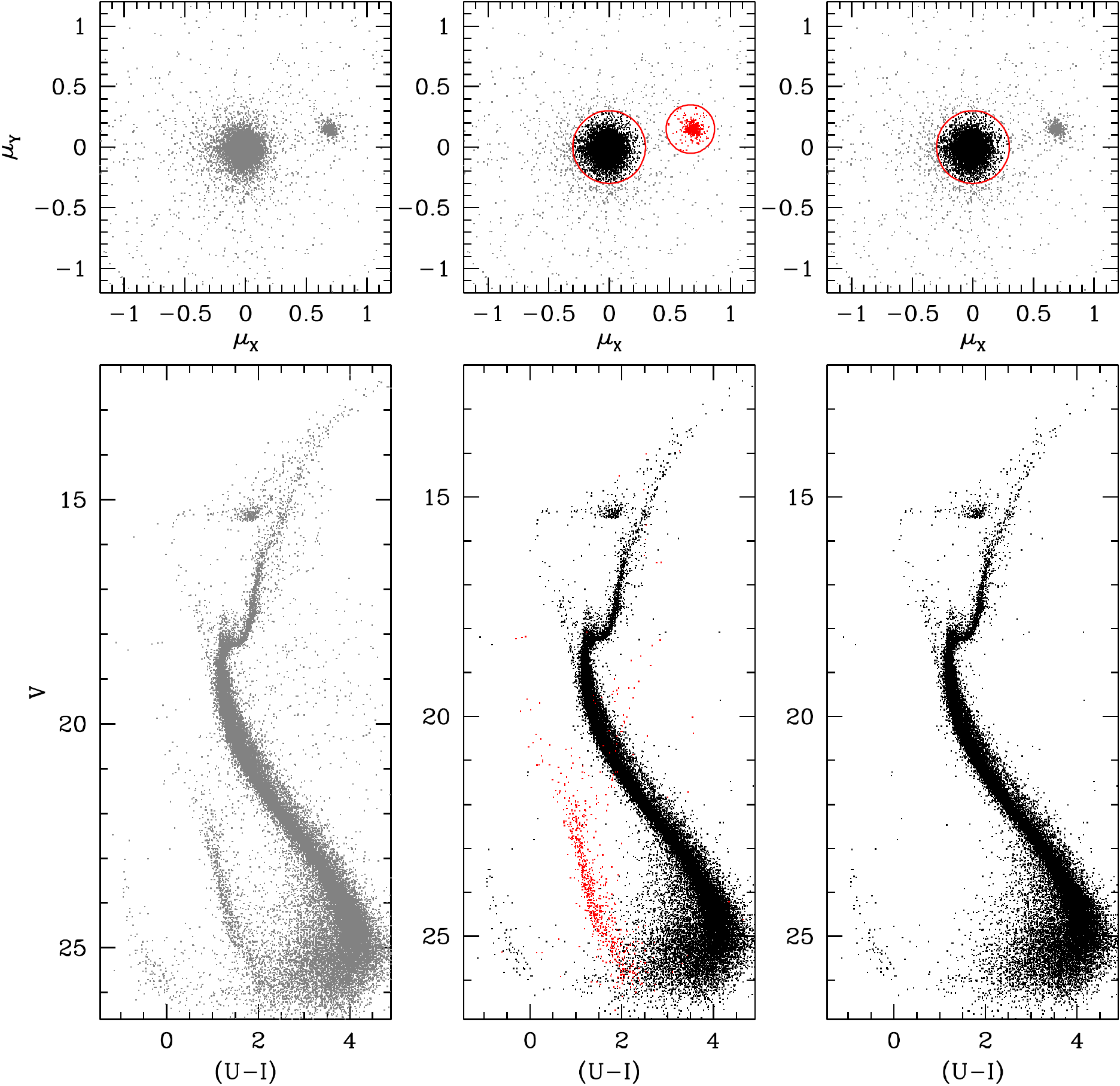}
\caption{{\it Upper panels}. In the leftmost panel VPDs in pixel/yr of all the stars identified in common between the first 
and second epoch data sets. In the middle panel the cluster and the SMC populations are selected
in black and red respectively. In the rightmost panel only cluster member are selected. 
{\it Lower panels}. From left to right, (V,U-I) CMDs for all the detected stars, for stars with a  high cluster membership probability and SMC selected stars (red dots), 
and for cluster members  only.}\label{bss_vpd}
\end{figure}

The upper panels of Figure \ref{bss_vpd} present the VPD obtained, where we can distinguish 
two main sub-populations.
The first and dominant one, centered at ($\mu_X=0$ pixel yr$^{-1}$, $\mu_Y=0$ pixel yr$^{-1}$) is, 
by construction, the cluster population. 
The second, at ($\mu_X=0.7$ pixel yr$^{-1}$, $\mu_Y=0.2$ pixel yr$^{-1}$) is instead populated by the SMC stars. The separation between the two 
components appears clearly in the (V,U-I) CMDs as shown in the lower panels of Figure \ref{bss_vpd}.
In Figure \ref{bss_clean}, we show the VPD at different magnitude levels. As expected, 
the population belonging to the SMC starts to appear in the VPD diagram for $V>20$. In addition the distribution of NGC~362
and SMC gets broader as a function of increasing magnitudes, because of the increasing uncertainties on 
the centroid positions of faint stars.
The same behavior is visible at bright magnitudes ($V<17$) because of non-linearity and saturation problems.

To build a clean sample of stars with a high membership probability, we defined in the VPD and for each magnitude bin 
a different fiducial region centered on ($\mu_X=0$ pixel yr$^{-1}$, $\mu_Y=0$ pixel yr$^{-1}$). The fiducial regions have 
radii of $2\times \sigma$, where $\sigma$ is
the dispersion of fiducial member stars, i.e. those with a distance $r<0.3$ pixel yr$^{-1}$  
from ($\mu_X=0$ pixel yr$^{-1}$, $\mu_Y=0$ pixel yr$^{-1}$) (see the member selection in the upper panels of Figure \ref{bss_vpd}).
It is worth noting that 
slightly different criteria for the membership selection do not appreciably affect the results about the BSS population.

\subsection{The BSS double sequence}

Given its observed density profile that shows a moderate
cusp in the central region of the cluster (see details in \citealt{dalex13}), NGC~362 is
suspected to have started the core collapse process.  This makes it particularly
interesting in the context of the working hypothesis proposed by Ferraro et al. (2009b)
that clusters undergoing the core-collapse process could develop a
double BSS sequence.  For this reason, the BSS population of NGC~362
has been analyzed first in the (V,V-I) CMD, in order to perform a
direct comparison with the observations of M~30 (Ferraro et al. 2009b).  Following the
approach adopted in previous papers (Lanzoni et al. 2007a,b; Dalessandro
et al. 2008a,b) we selected BSS candidates by defining a box which
roughly selects stars brighter and bluer than the TO point,
corresponding to $V_{TO}\sim19$ and $(V-I)_{\rm TO}\sim 0.7$. As
usual, in our selection we tried to avoid possible contamination of
blends from the SGB, TO region, and the saturation limit of the deep
exposures (dashed line in Figure \ref{bss_clean}).  With these limits 65 candidates
BSS have been identified.  We emphasize that the selection criteria
are not a critical issue here, since the inclusion or exclusion of a
few stars does not affect in any way the results of this work.  The
selected BSS are shown in the zoomed region of the CMD in Figure~\ref{bss_box}. At
a close look it is possible to distinguish two almost parallel and
similarly populated sequences, separated by about $\Delta V\sim0.4$
and $\Delta (V-I)\sim0.15$.  Such a feature resembles the one observed
in M~30.

\begin{figure}
\includegraphics[scale=0.7]{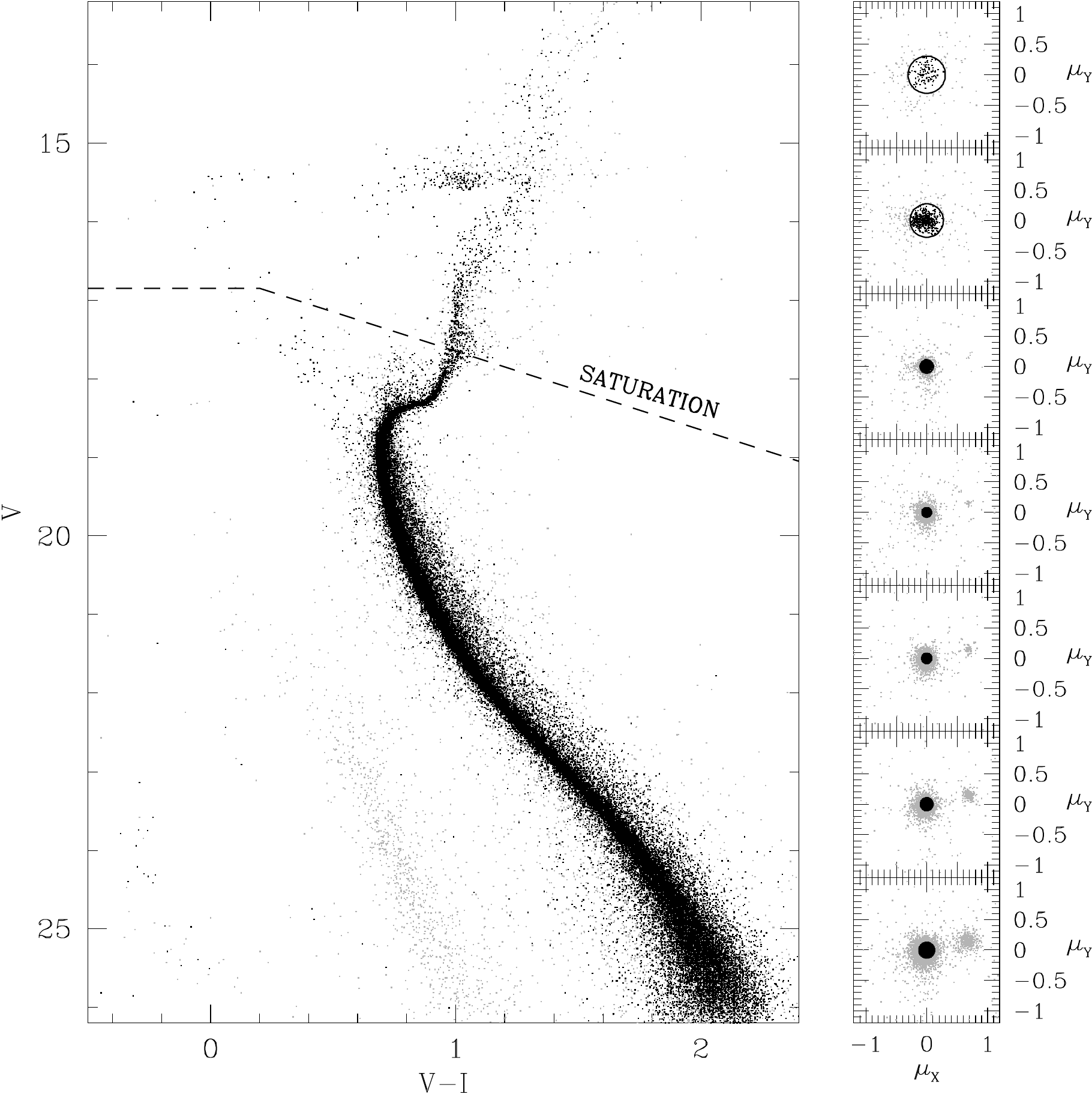}
\caption{On the left (V,V-I) CMD of stars in the WFC3 FoV. In black stars selected 
according to their membership probability, in grey stars excluded on the basis of the criteria 
highlighted in the VPDs in the right panels. On the right, VPDs at different magnitude levels. 
The distribution of stars gets broader moving to very faint and bright magnitudes because of the uncertainties on the centroid
determination. The black circle represent the $2\sigma$ fiducial region used to clean our sample from non-member stars.}\label{bss_clean}
\end{figure}

In order to perform a more direct comparison with  M~30, we over-plotted 
two fiducial areas (grey regions in Figure~\ref{bss_box}) representative of the color 
and magnitude distribution of the red
and blue BSS populations  in that cluster.
Differences in distance moduli and reddening have been properly taken into account.
Figure \ref{bss_box} shows that the BSS of NGC~362 nicely fall within the same fiducial regions used 
to separate the BSS sequences in the case of M~30 (Ferraro et al., 2009b) and only sparsely populate the region between the two. 
Moreover the BSS population of NGC~362 show  luminosity and color extensions  similar to the 
ones observed in M~30.  The red sequence appears slightly more scattered than that observed in M~30. 
This is mainly due to the fact that in the case of NGC~362 a few candidate BSS at 
$(V-I)\sim 0.6$ have been included
into the sample. However, we can safely conclude that, within the photometric uncertainties,
the two BSS sequences of NGC~362 well resemble the red and blue BSS sequences of M~30.
The blue BSS sample counts 30 stars which are distributed along a narrow 
and well defined sequence in the (V, V-I) CMD, while the red sequence counts 35 stars.
The relative sizes of the two populations is very similar to what found in M~30.

\begin{figure}
\includegraphics[scale=0.7]{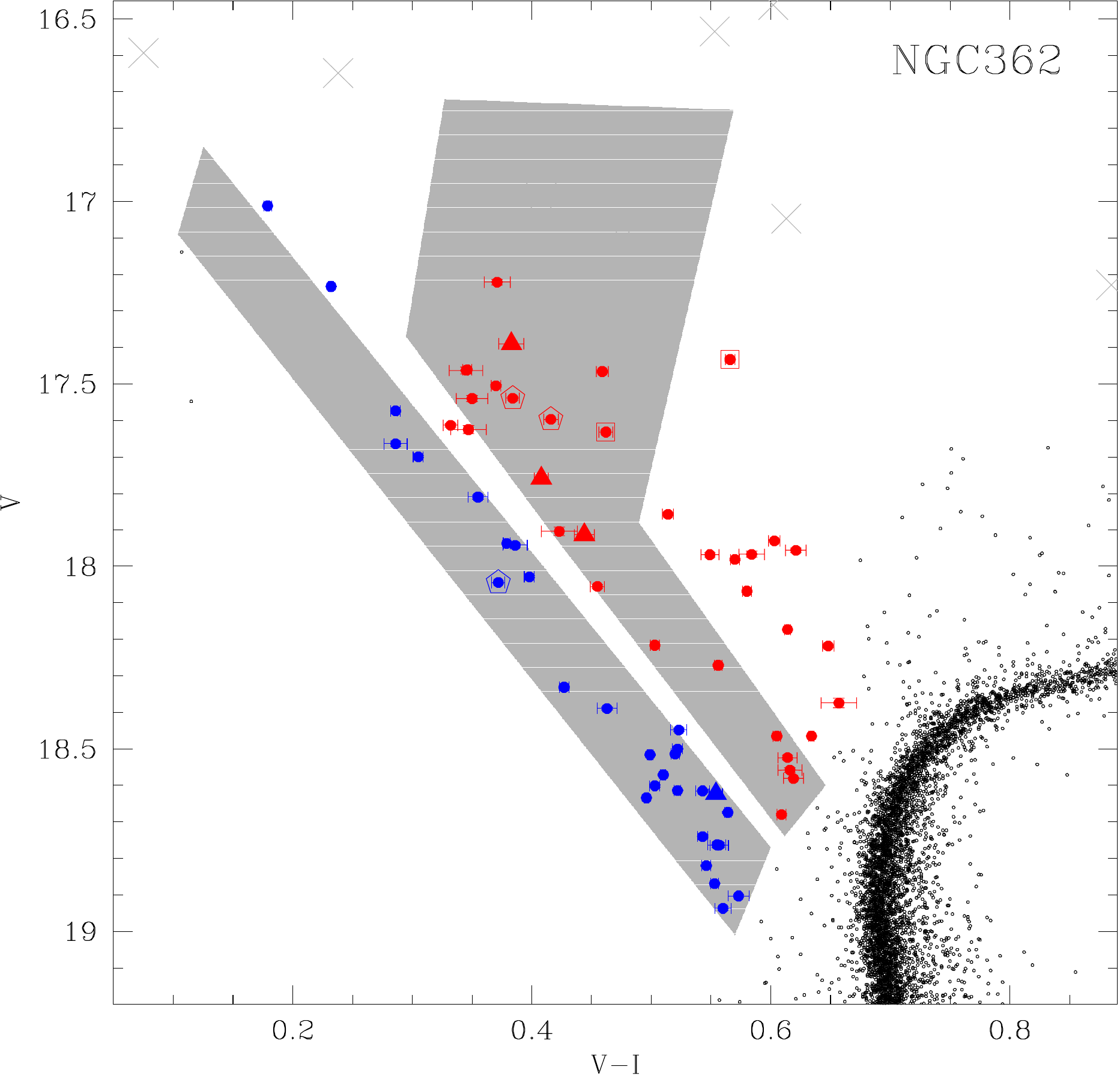}
\caption{A zoomed view of the (V,V-I) CMD of NGC~362 on the BSS region.
BSS are highlighted as red and blue symbols, and the photometric errors are shown as error bars.
The grey areas represent the fiducial loci of the red BSS  and blue BSS 
of M~30 (from Ferraro et al., 2009b). Open squares are SX-Phoenicis found by Szekely et al. (2007).
Open pentagons and filled triangles are respectively SX-Phoenicis and WUMa stars identified in this work.
Grey crosses are stars excluded for saturation or non linearity problems.
 }\label{bss_box}
\end{figure}

In order to quantify the significance of the bi-modality of the distribution shown in Figure~\ref{bss_box}, we 
used the Gaussian mixture modeling algorithm presented by Muratov \& Gnedin (2010) and already described in Chapter \ref{chapmdf}. 
We obtain that a unimodal distribution is rejected with a $99.6\%$ probability; 
hence it is bimodal with a confidence level of $3\sigma$.
    
We analyzed the spatial distribution of the red and blue BSS populations by looking at their cumulative 
radial distribution (see Figure \ref{bss_ks}).
We used  SGB stars in the magnitude interval $18<$V$<18.5$ as reference population. 
Both the red and the blue BSS samples are more centrally concentrated than the reference population. 
A Kolmogorov-Smirnov test  gives a probability 
$P\sim10^{-5}$ that they are extracted from the same parent population. Moreover, red BSS  are
more centrally segregated than the blue ones, with a high ($3\sigma$) confidence level that 
they are extracted from different populations. 
In addition, in striking agreement with what found by Ferraro et al. (2009b) in M~30, 
we do not observe any blue BSS within $5\arcsec-6\arcsec$ from $C_{grav}$ and both the blue and
red samples completely disappear (within the WFC3 FoV) at distances $r>75\arcsec$. 
The observational evidence collected so far leads us to conclude that NGC~362 is the second cluster,
after M~30, showing a clear double sequence of BSS. 

\begin{figure}
\includegraphics[scale=0.7]{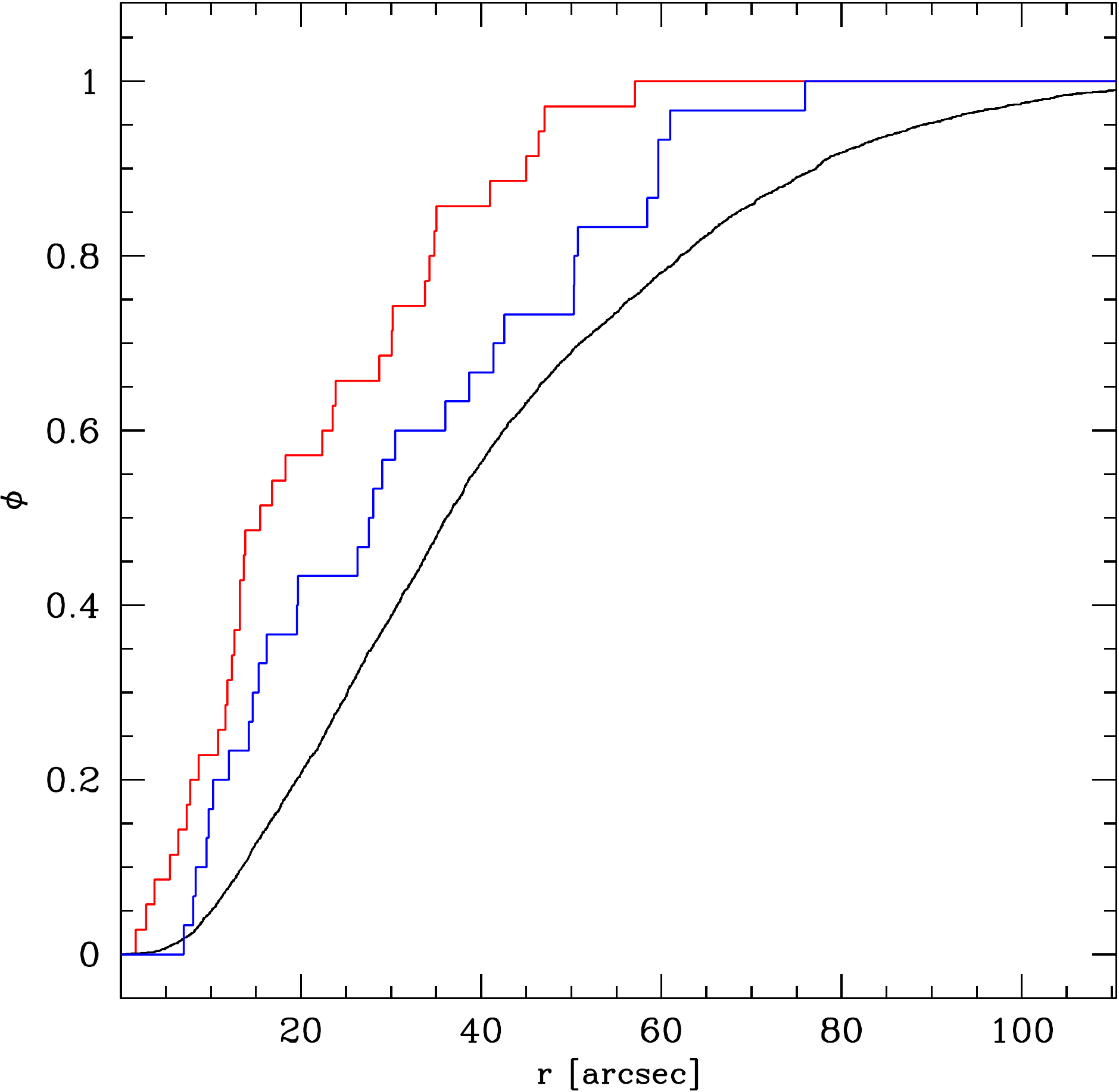}
\caption{Cumulative radial distribution of the red and blue BSS samples. 
In black  the distribution of 
SGB stars, taken as reference population. This analysis is limited to the WFC3 FoV which extends 
to a distance from $C_{grav}$ $r\sim100\arcsec$, corresponding to about $8-9 r_c $.}\label{bss_ks}
\end{figure}

\subsection{Discussion}  

The accurate HST WFC3 photometry presented in these Sections has revealed
the presence of two almost parallel BSS sequences in the core of
NGC~362.  This represents the second case, after M~30 (Ferraro et al. 2009b), for which
a double BSS sequence has been observed. The red and blue BSS
populations are well separated in the (V, V-I) CMD by
$\Delta$V$\sim0.4$ and $\Delta$(V-I)$\sim0.15$, and they nicely
overlap with the distribution of BSS shown in Ferraro et al (2009b, see Figures \ref{bss_box}).
As in the case of M~30, the red population is significantly more
centrally concentrated than the blue one (Figure \ref{bss_ks}) and their sizes
are very similar. Also the total number of BSS normalized to the total
cluster luminosity is basically the same in these two systems. In fact
we count, within $r_t$ and after field stars subtraction, 77 BSS in NGC~362 
which has $L_V\sim2\times10^5 L_{\odot}$,
and 51 in M~30 (Ferraro et al 2009b) which has instead $L_V\sim 10^5 L_{\odot}$. 
Ferraro et al. (2009b)
argued that blue BSS are likely the result of collisions while red BSS
are binary systems in an active phase of mass-transfer. Observational
hints supporting this interpretative scenario have been recently shown
by Lovisi et al. (2013).  A similar approach can be followed to
interpret the two sequences in NGC~362. The results are shown in
Figure \ref{bss_models}.  Indeed the position of the blue sequence in the CMD can be
nicely reproduced by a collisional isochrone (Sills et al. 2009) of
proper metallicity ([Fe/H]=-1.31) and age $t=0.2$~Gyr and by assuming
a distance modulus $(m-M)_0=14.68$ and reddening $E(B-V)=0.05$
(\citealt{ferr99}). 

\begin{figure}
\includegraphics[scale=0.7]{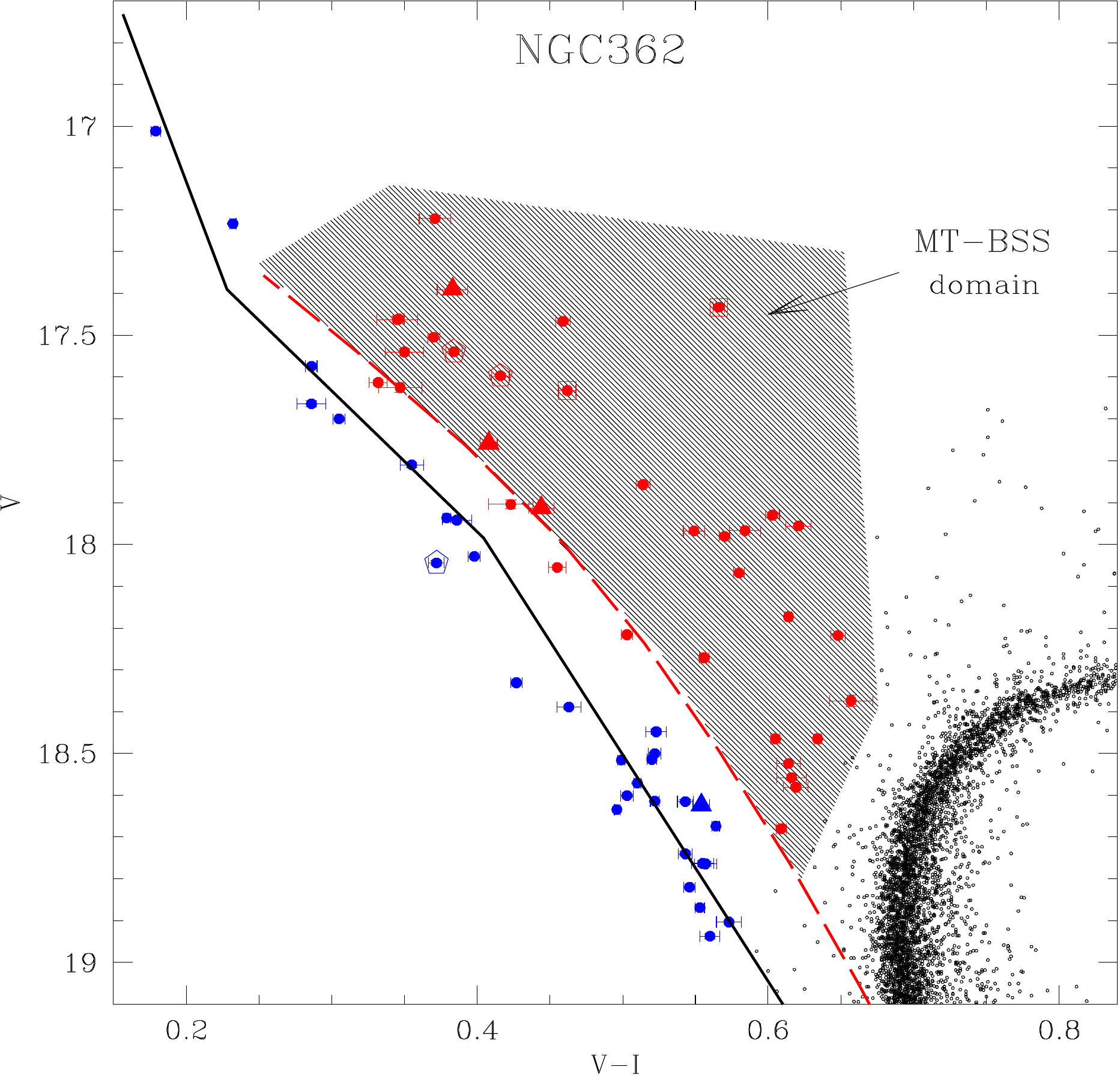}
\caption{As in Figure~7, the grey shaded area (here defined  as ``MT-BSS domain'') approximately indicates 
the region populated by mass-transfer binaries in M~67 (Tian et al. 2006),
"translated" into the CMD of NGC~362. The solid black line is a 0.2 Gyr collisional isochrone (Sills et al. 2009).}\label{bss_models}
\end{figure}

In Ferraro et al. (2009b) the position of the red BSS sequence in the CMD has been found
to be well reproduced by the lower luminosity boundary defined by the
distribution of binary stars with ongoing mass-transfer, as found in
Monte Carlo simulations by Tian et al. (2006).  This boundary
approximately corresponds to the locus defined by the Zero-Age-MS
(ZAMS) shifted to brighter magnitudes by $0.75$~mag. The locus
obtained for NGC~362 is shown as a red dashed line in Figure \ref{bss_models}. As
can be seen, red BSS lie in a sparse area adjacent to the lower
boundary in a region that we can call the {\it MT-BSS domain}
(highlighted in grey in Figure \ref{bss_models}).  

In Ferraro et al. (2009b), the presence of two distinct sequences of BSS has been
connected to the dynamical state of M~30, in particular to the fact
that this cluster might have recently ($1-2$ Gyr ago) experienced the
collapse of the core.  The dynamical state
of NGC~362 is quite debated, and controversial results are found in
the literature (Fischer et al. 1993; McLaughlin \&
van der Marel 2005).  The density profile cusp ($\alpha\sim-0.2$, see \cite{dalex13})
is shallower than typically observed in PCC
clusters and could indicate that NGC~362 is on the verge or is
currently experiencing the collapse of the core (Vesperini \& Trenti
2010).  The advanced dynamical age of NGC~362 is also suggested by its
monotonic BSS radial distribution.  In fact, in the "{\it dynamical
  clock}" classification (Ferraro et al. 2012), NGC~362 belongs (with M~30) to the
family of the highly dynamically-evolved clusters ({\it Family III}).

On the basis of this observational evidence we can argue that also in
the case of NGC~362 the presence of a double BSS sequence could be
connected to the advanced dynamical state of the cluster.  As in the
case of M~30, the fact that we observe two distinct sequences, and in
particular a well defined blue one, implies that the event that
triggered the formation of the double sequence is recent and
short-lived. If this event is connected with the dynamical evolution
of the system, it could likely be the collapse of the core (or its
initial phase). Indeed, during the collapse, the central density
rapidly increases, also enhancing the probability of gravitational
encounters (Meylan \& Heggie 1997): thus, blue BSS could be formed by
direct collisions boosted by the high densities reached in the core,
while the red BSS population could have been incremented by binary
systems brought to the mass-transfer regime by hardening processes
induced by gravitational encounters (McMillan, Hut \& Makino 1990;
Hurley et al. 2008). 
Within such a scenario the properties of the blue BSS suggest that the core
collapse occurred very recently ($\sim0.2$ Gyr ago) and over a quite
short time scale, of the order of the current core relaxation time
($\sim 10^8$ yr; Harris 1996).


\clearpage{\pagestyle{empty}\cleardoublepage}

%




\clearpage{\pagestyle{empty}\cleardoublepage}







\end{document}